\documentclass[acmsmall,acmtiis]{acmart}
\def\BibTeX{{\rm B\kern-.05em{\sc i\kern-.025em b}\kern-.08emT\kern-.1667em\lower.7ex\hbox{E}\kern-.125emX}}

\usepackage{graphicx}
\usepackage{xcolor,colortbl}

\usepackage{enumitem}
\usepackage{multirow}
\usepackage{hhline}

\usepackage{makecell}

\usepackage{todonotes}
\usepackage{pifont}

\usepackage{amsmath}

\DeclareFontFamily{U}{mathb}{}
\DeclareFontShape{U}{mathb}{m}{n}{
  <-5.5> mathb5
  <5.5-6.5> mathb6
  <6.5-7.5> mathb7
  <7.5-8.5> mathb8
  <8.5-9.5> mathb9
  <9.5-11.5> mathb10
  <11.5-> mathb12
}{}
\DeclareSymbolFont{mathb}{U}{mathb}{m}{n}
\DeclareMathSymbol{\ulsh}{3}{mathb}{"E8}
\DeclareMathSymbol{\ursh}{3}{mathb}{"E9}
\DeclareMathSymbol{\dlsh}{3}{mathb}{"EA}
\DeclareMathSymbol{\drsh}{3}{mathb}{"EB}

\definecolor{notSureColor}{RGB}{147, 9, 229}
\colorlet{ColorforSina}{green!5!orange!95!}
\definecolor{ColorOfNil}{rgb}{0.6, 0.4, 0.8}

\begin{document}

%
\title[A Multidisciplinary Survey and Framework for Explainable AI]{A Multidisciplinary Survey and Framework for Design and Evaluation of Explainable AI Systems}

\author{Sina Mohseni}
\email{sina.mohseni@tamu.edu}
\author{Niloofar Zarei}
\email{n.zarei.3001@tamu.edu}
\affiliation{
  \institution{Texas A\&M University}
  \city{College Station}
  \state{Texas}
}

\author{Eric D. Ragan}
\email{eragan@ufl.edu}
\affiliation{
  \institution{University of Florida}
  \city{Gainesville}
  \state{Florida}
}

%
\renewcommand{\shortauthors}{Mohseni et al.}

\begin{abstract}
The need for interpretable and accountable intelligent systems grows along with the prevalence of artificial intelligence applications used in everyday life. 
Explainable AI systems are intended to self-explain the reasoning behind system decisions and predictions. 
Researchers from different disciplines work together to define, design, and evaluate explainable systems. 
However, scholars from different disciplines focus on different objectives and fairly independent topics of Explainable AI research, which poses challenges for identifying appropriate design and evaluation methodology and consolidating knowledge across efforts. 
To this end, this paper presents a survey and framework intended to share knowledge and experiences of Explainable AI design and evaluation methods across multiple disciplines. 
Aiming to support diverse design goals and evaluation methods in XAI research, after a thorough review of Explainable AI related papers in the fields of machine learning, visualization, and human-computer interaction, we present a categorization of Explainable AI design goals and evaluation methods.
Our categorization presents the mapping between design goals for different Explainable AI user groups and their evaluation methods.  
From our findings, we develop a framework with step-by-step design guidelines paired with evaluation methods to close the iterative design and evaluation cycles in multidisciplinary Explainable AI teams. 
Further, we provide summarized ready-to-use tables of evaluation methods and recommendations for different goals in Explainable AI research.
\end{abstract}

%
%

%

\setcopyright{acmcopyright}
\acmJournal{TIIS}
\acmYear{2020} \acmVolume{1} \acmNumber{1} \acmArticle{1} \acmMonth{1} \acmPrice{15.00}
\acmDOI{10.1145/3387166}

\maketitle

\section{Introduction}

Impressive applications of Artificial Intelligence (AI) and machine learning have become prevalent in our time.
Tech giants like Google, Facebook, and Amazon have collected and analyzed enough personal data through smartphones, personal assistant devices, and social media that can model individuals better than other people. 
Recent negative interference of social media bots in political elections~\cite{woolley2016automating, howard2016bots} were yet another sign of how susceptible our lives are to the misuse of artificial intelligence and big data~\cite{o2016weapons}. 
In these circumstances, despite tech giants and the thirst for more advanced systems, others suggest holding off on fully unleashing AI for critical applications until they can be better understood by those who will rely on them.
The demand for predictable and accountable AI grows as tasks with higher sensitivity and social impact are more commonly entrusted to AI services.
Hence, algorithm transparency is an essential factor in holding organizations responsible and accountable for their products, services, and communication of information.

\textit{Explainable Artificial Intelligence} (XAI) systems are a possible solution towards accountable AI, making it possible by explaining AI decision-making processes and logic for end users~\cite{gunning2017explainable}.
Specifically, explainable algorithms can enable control and oversight in case of adverse or unwanted effects, such as biased decision-making or social discrimination. 
An XAI system can be defined as a self-explanatory intelligent system that describes the reasoning behind its decisions and predictions. 
The AI explanations (either on-demand explanations or in the form of model description) could benefit users in many ways such as improving safety and fairness when relying on AI decisions.

While the increasing impact of advanced black-box machine learning systems in the big-data era has attracted much attention from different communities, interpretability of intelligent systems has also been studied in numerous contexts~\cite{poulin2006visual, gregor1999explanations}.
The study of personalized agents, recommendation systems, and critical decision-making tasks (e.g., medical analysis, powergrid control) has added to the importance of machine-learning explanation and AI transparency for end-users. 
For instance, as a step towards this goal, the legal right to explanations has been established in the European Union General Data Protection Regulation (GDPR) commission.
While the current state of regulations is mainly focused on user data protection and privacy, it is expected to cover more algorithmic transparency and explanations requirements from AI systems~\cite{goodman2016eu}.

Clearly, addressing such a broad array of  definitions and expectations for XAI requires multidisciplinary research efforts, as existing communities have different requirements and often have drastically different priorities and areas of specialization. 
For instance, research in the domain of machine learning seeks to design new interpretable models and explain black -box models with ad-hoc explainers. 
Along the same line but with different approaches, researchers in visual analytics design and study tools and methods for data and domain experts to visualize complex black-box models and study interactions to manipulate machine learning models. 
In contrast, research in human-computer interaction (HCI) focuses on end-user needs such as user trust and understanding of machine generated explanations.
Psychology research also studies the fundamentals of human understanding, interpretability, and the structure of explanations.

Looking at the broad spectrum of research on XAI, it is evident that scholars from different disciplines have different goals in mind.  
Even though different aspects of XAI research are following the general goals of AI interpretability, researchers in each discipline use different measures and metrics to evaluate the XAI goals.
For example, numerical analytic methods are employed in machine learning fields to evaluate computational interpretability, while human interpretability and human-subjects evaluations are more commonly the primary goals in HCI and visualization communities.
In this regard, although there seems to be a mismatch in specific objectives for designing and evaluating explainability and interpretability, a convergence in goals is beneficial for achieving the full potential of XAI.
To this end, this paper presents a survey and framework intended to share knowledge and experiences of XAI design and evaluation methods across multiple disciplines.
To support the diverse design goals and evaluation methods in XAI research, after a thorough review of XAI related papers in the fields of machine learning, visualization, and HCI, we present a categorization of interpretable machine learning design goals and evaluation methods and show a mapping between design goals for different XAI user groups and their evaluation methods. 
From our findings, we develop a framework with step-by-step design guidelines paired with evaluation methods to close the iterative design and evaluation loops in multidisciplinary teams.
Further, we provide summarized ready-to-use evaluation methods for different goals in XAI research.
Lastly, we review recommendations for XAI design and evaluation drawn from our literature review.
\section{Background}

Nowadays, algorithms analyze user data and affect decision-making processes for millions of people on matters like employment, insurance rates, loan rates, and even criminal justice~\cite{chouldechova2017fair}.
However, these algorithms that serve critical roles in many industries have their own disadvantages that can result in discrimination~\cite{datta2015automated, sweeney2013discrimination}, and unfair decision-making~\cite{o2016weapons}. 
For instance, recently, news feed and targeted advertising algorithms in social media have attracted much attention for aggravating the lack of information diversity in social media~\cite{bozdag2015breaking}.
A significant part of the trouble could be because algorithmic decision-making systems---unlike recommender systems---do not allow their users to choose between the recommended items, but instead, present the most relevant content or option themselves. 
To address this, Heer~\cite{heer2019agency} suggests the use of shared representations of tasks that are augmented with both machine learning models and user knowledge to reduce negative effects of immature AI autonomous systems. 
They present case studies of interactive systems that integrate proactive computational support into interactive systems.

Bellotti and Edwards~\cite{bellotti2001intelligibility} argue that intelligent context-aware systems should not act on our behalf.
They suggest user control over the system as a principle to support the accountability of a system and its users. 
Transparency can provide essential information for decision-making that is hidden to the end-users and prevents blind faith~\cite{zarsky2016trouble}. 
The key benefits of algorithmic transparency and interpretability include: user awareness~\cite{ananny2018seeing}; bias and discrimination detection~\cite{diakopoulos2014algorithmic, sweeney2013discrimination}; interpretable behavior of intelligent systems~\cite{lim2009assessing}; and accountability for users~\cite{diakopoulos2017enabling}. 
Furthermore, considering the growing body of examples of discrimination and other legal aspects of algorithmic decision making, researchers are demanding and investigating transparency and accountability of AI under the law to mitigate adverse effects of algorithmic decision making~\cite{doshi2017accountability, mittelstadt2016automation, turilli2009ethics}. 
In this section, we review research background related to XAI systems from a broad and multidisciplinary perspective.
At the end, we relate the summaries and positions derived through our survey to other work in the field.

\subsection{Auditing Inexplicable AI}

Researchers audit algorithms to study bias and discrimination in algorithmic decision making~\cite{sandvig2014auditing} and study the users' awareness of the effects of these algorithms~\cite{eslami2015always}.
\textit{Auditing} of algorithms is a mechanism for investigating algorithms' functionality to detect bias and other unwanted algorithm behaviors without the need to know about its specific design details.
Auditing methods focus on problematic effects on the results of algorithmic decision-making systems.
To audit an algorithm, researchers feed new inputs to the algorithm and review system output and behavior. 
Researchers generate new data and user accounts with the help of scripts, bots~\cite{datta2015automated}, and crowdsourcing~\cite{hannak2013measuring} to emulate real data and real users in the auditing process.
For bias detection among multiple algorithms, cross-platform auditing can detect if an algorithm behaves differently from another algorithm.
A recent example of cross-platform auditing is a work by Eslami et al.~\cite{eslami2017careful}, in which they analyzed user reviews in three hotel booking websites to study user awareness of bias in online rating algorithms.
These examples demonstrate that auditing is a valuable yet time-intensive process that could not be scaled easily to large numbers of algorithms.
This calls for new research for more effective solutions toward algorithmic transparency.

\begin{figure}[t]
\centering
  \includegraphics[width=0.9\columnwidth]{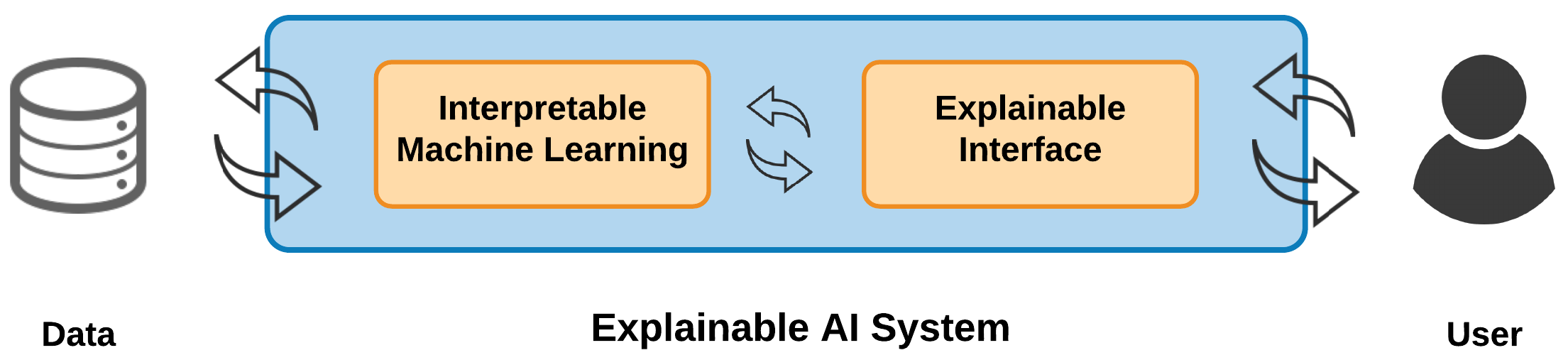} 
  \caption{
  The user interacts with the explainable interface to send queries to the interpretable machine learning and receive model prediction and explanations. The interpretable model interacts with the data to generate explanation or new new prediction for the user query. 
  }
  \label{fig:xai-system-model}
\end{figure}

\subsection{Explainable AI}

Along with the methods mentioned above for supporting transparency, machine learning explanations have also become a common approach to achieve transparency in many applications such as social media, e-commerce, and data-driven management of human workers~\cite{tang2012etrust,tintarev2011designing,lee2015working}.
The XAI system, as illustrated in Figure~\ref{fig:xai-system-model}, is able to generate explanations and describe the reasoning behind machine-learning decisions and predictions.
Machine-learning explanations enable users to understand how the data is processed. They aim to bring awareness to possible bias and system malfunctions.
For example, to measure user perception of justice in intelligent decision making, Binns et al.~\cite{Binns:2018:RHP:3173574.3173951} studied explanations in systems for everyday tasks such as determining car insurance rates and loan application approvals.
Their results highlight the importance of machine learning explanations in users' comprehension and trust in algorithmic decision-making systems.
In a similar work studying knowledge of social media algorithms, Radar et al.~\cite{rader2018explanations} ran a crowdsourced study to see how different types of explanations affect users' beliefs on news feed algorithmic transparency in a social media platform.
In their study, they measured users' awareness, correctness, and accountability to evaluate algorithmic transparency.
They found that all explanations caused users to become more aware of the system's behavior.
Stumpf et al.~\cite{stumpf2009interacting} designed experiments to investigate meaningful explanations and interactions to hold users accountable by machine learning algorithms.
They show explanations as a potential method for supporting richer human-computer collaboration to share intelligence.

The recent advancements and trends for explainable AI research demand a wide range of goals for algorithmic transparency which calls for research across varied application areas.
To this end, our review encourages a cross-discipline perspective of intelligibility and transparency goals.

\subsection{Related Surveys and Guidelines}

In recent years, there have been surveys and position papers suggesting research directions and highlighting challenges in interpretable machine learning research \cite{herman2017promise, lipton2016mythos, doshi2017towards}.
Although our review is limited to computer science literature, here we summarize several of the most relevant peer-reviewed surveys related to the topic of XAI across active disciplines including social science.
While all surveys, models, and guidelines in this section add value to the XAI research, to the best of our knowledge, there is no existing comprehensive survey and framework for evaluation methods of explainable machine learning systems.

\subsubsection{Social Science Surveys}

Research in the social sciences is particularly important for XAI systems to understand how people generate, communicate, and understand explanations by taking into account each others' thinking, cognitive biases, and social expectations in the process of explaining.
Hoffman, Mueller, and Klein reviewed the key concepts of explanations for intelligent systems in a series of essays to identify how people formulate and accept explanations, ways to generate self-explanations, and identified purposes and patterns for causal reasoning~\cite{hoffman2017part1explaining,hoffman2017part2explaining,klein2018part3explaining}. 
They lastly focus on deep neural networks (DNN) to examine their theoretical and empirical findings on a machine learning algorithm~\cite{hoffman2018part4explaining}. 
In other work, they presented a conceptual model of the process of explaining in the XAI context~\cite{hoffman2018metrics}. 
Their framework includes specific steps and measures for the goodness of explanations, user satisfaction and understanding of explanations, users' trust and reliance on XAI systems, effects of curiosity on the search for explanations, and human-XAI system performance.

Miller~\cite{miller2017explanation} suggests a close collaboration between machine learning researchers in the space of XAI with social science would further refine the explainability of AI for people.
He discusses how understanding and replicating how people generate, select, and present explanations could improve human-XAI interactions. 
For instance, Miller reviews how people generate and select explanations that are involved with cognitive biases and social expectations.
Other papers reviewing social science aspects of XAI systems include studies on the role of algorithmic transparency and explanation in lawful AI~\cite{doshi2017accountability} and of fair and accountable algorithmic decision-making processes~\cite{lepri2017fair}.

\subsubsection{Human Computer Interactions Surveys}

Many HCI surveys discuss the limitations and challenges in AI transparency~\cite{weller2017challenges} and interactive machine learning~\cite{amershi2014power}.
Others suggest a set of theoretical and design principles to support intelligibility of intelligent system and accountability of human users (e.g.,~\cite{hook2000steps, bellotti2001intelligibility}).
In a recent survey, Abdul et al.~\cite{abdul2018trends} presented a thorough literature analysis to find XAI-related topics and relationships among these topics.
They used visualization of keywords, topic models, and citation networks to present a holistic view of research efforts in a wide range of XAI related domains; from privacy and fairness to intelligent agents and context-aware systems.
In another work, Wang et al.~\cite{Wang2019Theory} explored theoretical underpinnings of human decision-making and proposed a conceptual framework for building human-centered decision-theory-driven XAI systems. 
Their framework helps to choose better explanations to present, backed by reasoning theories, and human cognitive biases.
Focused on XAI interface design, Eiband et al.~\cite{Eiband2018practice} present a stage-based participatory process for integration of transparency in existing intelligent systems using explanations.
Another design framework is XAID from Zhu et al.~\cite{zhu2018explainable}, which presents a human-centered approach for facilitating game designers to co-create with machine learning techniques.
Their study investigates the usability of XAI algorithms in terms of how well they support game designers.

\subsubsection{Visual Analytics Surveys}
XAI-related surveys in the visualization domain follow visual analytics goals such as understanding and interacting with machine learning systems in different visual analytics applications~\cite{sacha2016human,endert2017state}. 
Choo and Liu~\cite{choo2018visual} reviewed challenges and opportunities for Visual Analytics for explainable deep learning design.
In a recent paper, Hohman et al.~\cite{hohman2018visual} provide an excellent review and categorization of visual analytics tools for deep learning applications.
They cover various data and visualization techniques that are being used in deep visual analytics applications. 
Also, Spinner et al.~\cite{spinner2019explainer} proposed a XAI pipeline which maps the XAI process to an iterative workflow in three stages: model understanding, diagnosis, and refinement.
To operationalize their framework, they designed explAIner, a visual analytics
system for interactive and interpretable machine learning that instantiates all steps of their pipeline.

\subsubsection{Machine Learning Surveys}
In the machine learning area, Guidotti et al.~\cite{guidotti2018survey} present a comprehensive review and classification of machine learning interpretability techniques. 
Also, Montavon et al.~\cite{montavon2017methods} focus on interpretability techniques for DNN models. 
On Convolutional Neural Network (CNN), Zhang et al.~\cite{zhang2018visual} reviews research on interpretability techniques in six directions including visualization of CNN representations, diagnosing techniques for CNNs, approaches for transforming CNN representations into interpretable graphs, building explainable models, and semantic-level learning based on model interpretability.
In another work, Gilpin et al.~\cite{gilpin2018explaining} reviews interpretability techniques in machine learning algorithms and categorizes evaluation approaches to bridge the gap between machine learning and HCI communities.
\newline

In complementing the existing work, our survey provides a multidisciplinary categorization of design goals and evaluation methods for XAI systems.
As a result of surveyed papers, we propose a framework that provides a step-by-step design and evaluation plan for a multidisciplinary team of designers for building real-world XAI systems.
Unlike Eiband et al.~\cite{Eiband2018practice}, we do not make the assumption of \textit{adding transparency} to \textit{an existing} intelligent interface and do not limit the evaluation of XAI systems to the users' mental model.
We instead characterize both design goals and evaluation methods and compile all in a unified framework for multidisciplinary teamwork.
Our design framework has similarities to  Wang et al's~\cite{Wang2019Theory} theoretical framework which supports our design goals (see Section \ref{sec:generalization-validation}).
Our multidisciplinary work extends their conceptual framework by 1) including the design of interpretability algorithms as part of the framework and 2) pairing evaluation methods with each design step to close the iterative design and evaluation loops.

\begin{figure}[t]
\centering
  \includegraphics[width=1.0\columnwidth]{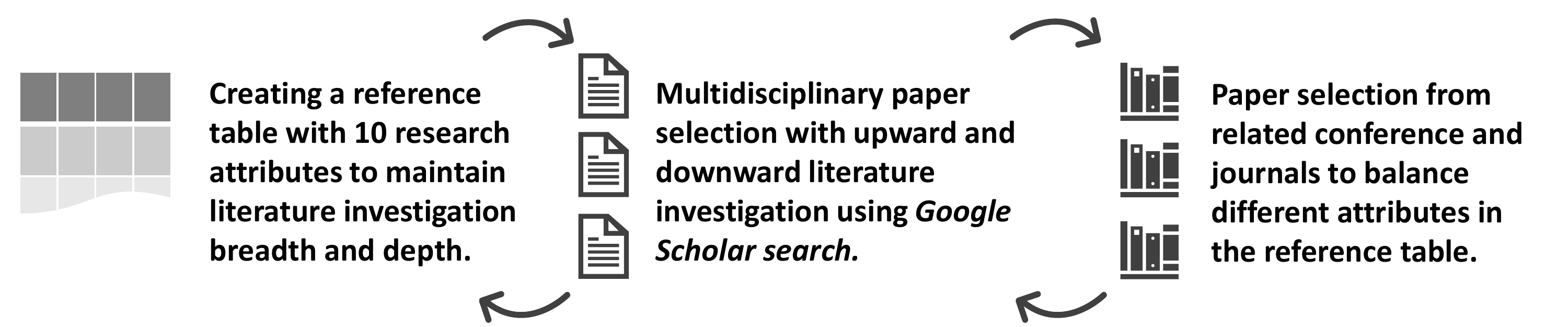}
  \caption{
  A diagram summarizing our iterative and multi-pass literature selection and review process to achieve desired literature investigation breadth and depth. 
  We started with 40 papers to create the reference table. 
  Then we added 80 papers by upward and downward literature investigation to improve review breath and depth. 
  Finally, we added another 80 papers from related conferences proceedings and journals to balance the reference table.
  }
  \label{tab:survey-method}
\end{figure}

\section{Survey Method}

We conducted a survey of the existing research literature to capture and organize the breadth of designs and goals for XAI evaluation.
We used a structured and iterative methodology to find XAI-relevant research and categorize the evaluation methods presented in research articles (summarized in Figure~\ref{tab:survey-method}).
In our iterative paper selection process, we started by selecting existing work from top computer science conferences and journals across the fields of HCI, visualization, and machine learning. 
However, since XAI is a quite fast growing topic, we also wanted to include \textit{arXiv} preprints and useful discussions in workshop papers. 
We started with 40 papers related to XAI topics across three research fields including but not limited to research on interpretable machine learning techniques, deep learning visualization, interactive model visualization, machine explanations in intelligent agents and context-aware systems, explainable user interfaces, explanatory debugging, and algorithmic transparency and fairness.

We then used selective coding to identify 10 main research attributes in those papers.
The main attributes we identified include: 
research discipline (social science, HCI, visualization, or machine learning), 
paper type (interface design, algorithm design, or evaluation paper), 
application domain (machine learning interpretability, algorithmic fairness, recommendation systems, transparency of intelligent systems, intelligent interactive systems and agents, explainable intelligent systems and agents, human explanations, or human trust), 
machine learning model (e.g., deep learning, decision trees, SVM), 
data modality (image, text, tabular data), 
explanation type (e.g., graphical, textual, data visualization), 
design goal (e.g., model debugging, user reliance, bias mitigation), 
evaluation type (e.g., qualitative, computational, quantitative with human-subjects),
targeted user (AI novices, data experts, AI experts),
and evaluation measure (e.g., user trust, task performance, user mental model).

In the second round of collecting XAI literature, we conducted an upward and downward literature investigation using the \textit{Google Scholar} search engine to add 80 more papers to our reference table. 
We narrowed down our search by XAI related topics and keywords including but not limited to: interpretability, explainability, intelligibility, transparency, algorithmic decision-making, fairness, trust, mental model, and debugging in machine learning and intelligent systems.
With this information, we performed axial coding to organize the literature and started discussions on our proposed design and evaluation categorization.

Finally, to maintain reasonable literature coverage and balance the number of papers for each of our categories of design goals and evaluation measures, we added another 80 papers to our reference table.
The conferences from which we selected XAI related paper were including but not limited to: CHI, IUI, HCOMP, SIGDIAL, UbiComp, AIES, VIS, ICWSM, IJCAI, KDD, AAAI, CVPR, and NeurIPS conferences. The journals included: Trends in cognitive science, Transactions on Cognitive and Developmental Systems, Cognition Journal, Transactions on Interactive Intelligent Systems, International Journal of Human-Computer Studies, Transactions on Visualization and Computer Graphics, and Transactions on Neural Networks and Learning Systems.

Following a review of 226 papers, our categorization of XAI design goals and evaluation methods is supported by references from papers performing design or evaluation of XAI systems. 
Our reference table\footnote{https://github.com/SinaMohseni/Awesome-XAI-Evaluation} is available online to the research community to provide further insight beyond our discussions in this document. 
Table~\ref{tab:main-table} shows a digest of our surveyed papers that contains 42 papers with both design and evaluation of XAI system. 
Later in the Section~\ref{sec:e_measure_section}, we provide a series of tables to organize different evaluation methods from research papers with example references for each, documenting our in-depth analysis of 69 papers in total.

\section{XAI Terminology}

To familiarize the readers with common XAI concepts and terminologies that are repeatedly referenced in this review, the following four subsections summarize high-level characterizations of model explanations.
Many related surveys (e.g.,~\cite{adadi2018peeking,Daniel2019challenge}) and reports (e.g.,~\cite{clinciu2019survey,tomsett2018interpretable}) also provide useful compilations of terminology and concepts in comprehensive reports. 
For instance, Abdul et al.~\cite{abdul2018trends} present a citation graph from diverse domains related to explanations, including intelligible intelligent systems, context-aware systems, and software learnability. 
Later, Arrieta et al.~\cite{arrieta2019explainable} present a thorough review of XAI concepts and taxonomies and arrives at the concept of \textit{Responsible AI} as a manifold of multiple AI principles including model fairness, explainability, and privacy.
Similarly, the concept of~\textit{Safe AI} has been reviewed by Amodei et al.~\cite{amodei2016concrete}, which is an interest in safety-critical intelligent applications such as autonomous vehicles~\cite{mohseni2020practical}.
Table~\ref{tab:treminology-table} presents descriptions for 14 common terms related to this survey's topic and organizes their relation to \textit{Intelligible Systems} and \textit{Transparent AI} topics.
We consider Transparent AI systems as the AI-based class of Intelligible Systems.
Therefore, properties and goals previously established for Intelligible Systems are ideally transferable to Transparent AI systems.
However, challenges and limitations for achieving transparency in complex machine learning algorithms raise issues (e.g., ensuring the fairness of an algorithm) that were not necessarily problematic in intelligible rule-based systems but now require closer attention from research communities.

The descriptions presented in Table~\ref{tab:treminology-table} are meant to be an introduction to these terms, though exact definitions and interpretations can depend on usage context and research discipline. 
Consequently, researchers from different disciplines often use these terms interchangeably, disregarding  differences in meaning~\cite{adadi2018peeking}. 
Perhaps the two generic terms of \textit{Black-box Model} and \textit{Transparent Model} are in the center of XAI terminology ambiguity. 
The black-box term refers to complex machine learning models that are not human-interpretable~\cite{lipton2016mythos} as opposed to transparent models which are simple enough to be human-interpretable~\cite{arrieta2019explainable}. 
We find it more accurate and consistent to separate the transparency of an XAI system (as described in Figure~\ref{fig:xai-system-model}) from the interpretability of its internal machine learning models. 
Specifically, Table~\ref{tab:treminology-table} shows that Transparent AI could be achieved by either \textit{Interpretable AI} or \textit{Explainable AI} approaches. 
Other examples of terminology ambiguity include the terms \textit{Interpretability} and \textit{Explainability} that are often used as synonyms in the field of machine learning. 
For example the phrase ``interpretable machine learning technique'' often refers to ad-hoc techniques for generating explanations for non-interpretable models such as DNNs~\cite{molnar2018interpretable}. 
Another example is the occasional case of using the terms \textit{Transparent System} and \textit{Explainable System} interchangeably in HCI research~\cite{Eiband2018practice}, while others clarify that explainability is not equivalent to transparency because it does not require knowing the flow of the bits in the AI decision-making process~\cite{doshi2017accountability}.

\begin{table}[]
\caption{Table of common terminology related to Intelligible Systems and Transparent AI.
Higher-level main concepts are shown in gray, while related terms for the main concepts are listed below and categorized as a desired outcome, property, or practical approach.
Explainable AI is one particular practical approach for intelligible systems to enable improve transparency.
Note that definitions and interpretations can vary across the literature, and this table is meant to serve as a quick reference.}
\label{tab:treminology-table}
\resizebox{\textwidth}{!}{%
\begin{tabular}{ccl}
\hline
\textbf{Concept} & \textbf{Category} & \textbf{Description} \\ \hline
\rowcolor[HTML]{C0C0C0} 
\textbf{Intelligible System} & Main Concept & \begin{tabular}[c]{@{}l@{}}A system that is understandable and predictable for its users\\ through transparency or explanations~\cite{abdul2018trends,bellotti2001intelligibility,Daniel2019challenge}.\end{tabular} \\ \hline
\textbf{\begin{tabular}[c]{@{}c@{}} Understandability \\  (Intelligibility) \end{tabular}} &  & \begin{tabular}[c]{@{}l@{}} Intelligible systems support user understanding\\  of system's underlying functions ~\cite{arrieta2019explainable,lim2011improving}.\end{tabular} \\ \cline{1-1} \cline{3-3} 
\textbf{Predictability} & \multirow{-3}{*}{\begin{tabular}[c]{@{}c@{}}Desired\\ Properties\end{tabular}} & \begin{tabular}[c]{@{}l@{}} Intelligibility supports building a mental model of the system\\  that enables user to predict system behavior~\cite{Daniel2019challenge}.\end{tabular} \\ \hline
\textbf{Trustworthiness} &  & \begin{tabular}[c]{@{}l@{}} Enabling positive user attitude toward the system that\\  emerges from knowledge, experience, and emotion~\cite{hoffman2013trust,hoffman2018metrics}.\end{tabular} \\ \cline{1-1} \cline{3-3} 
\textbf{Reliability} &  & \begin{tabular}[c]{@{}l@{}}Supporting user trust to rely and follow \\ system's advice for higher performance~\cite{hoffman2013trust,hoffman2018metrics}.\end{tabular} \\ \cline{1-1} \cline{3-3} 
\textbf{Safety} & \multirow{-5}{*}{\begin{tabular}[c]{@{}c@{}}Desired \\ Outcomes\end{tabular}} & \begin{tabular}[c]{@{}l@{}}Improving safety by reducing user unintended \\  misuse due to misperception and unawareness~\cite{mohseni2020practical}.\end{tabular} \\ \hline
\rowcolor[HTML]{C0C0C0} 
\textbf{Transparent AI} & Main Concept  & \begin{tabular}[c]{@{}l@{}}An AI-based system that provides information\\  about its decision-making processes~\cite{lipton2016mythos,clinciu2019survey}.\end{tabular} \\ \hline
\textbf{Interpretable AI} &  & \begin{tabular}[c]{@{}l@{}}Inherently human-interpretable models due to \\ their low complexity of machine learning algorithms~\cite{molnar2018interpretable}.\end{tabular} \\ \cline{1-1} \cline{3-3} 
\textbf{Explainable AI} & \multirow{-3}{*}{\begin{tabular}[c]{@{}c@{}}Practical \\ Approaches\end{tabular}} & \begin{tabular}[c]{@{}l@{}}Supporting user understanding of complex models \\ by providing explanations for predictions~\cite{Wang2019Theory}.\end{tabular} \\ \hline
\textbf{Interpretability} &  & \begin{tabular}[c]{@{}l@{}}The ability to support user understanding and comprehension\\ of the model decision making process and predictions~\cite{lipton2016mythos,arrieta2019explainable}.\end{tabular} \\ \cline{1-1} \cline{3-3} 
\textbf{Explainability} & \multirow{-3}{*}{\begin{tabular}[c]{@{}c@{}}Desired\\ Properties\end{tabular}} & \begin{tabular}[c]{@{}l@{}}The ability to explain underlying model and its reasoning \\ with accurate and user comprehensible explanations~\cite{lipton2016mythos,arrieta2019explainable}.\end{tabular} \\ \hline
\textbf{Accountable AI} &  & \begin{tabular}[c]{@{}l@{}}Allowing for auditing and documentation to hold organizations\\ accountable for their AI-based products and services~\cite{lepri2017fair,doshi2017accountability}.\end{tabular} \\ \cline{1-1} \cline{3-3} 
\textbf{Fair AI} & \multirow{-3}{*}{\begin{tabular}[c]{@{}c@{}}Desired \\ Outcomes\end{tabular}} & \begin{tabular}[c]{@{}l@{}}Enabling ethical and fairness analysis of models \\ and data used in decision-making processes~\cite{lepri2017fair,arrieta2019explainable}.\end{tabular} \\ \hline  
\end{tabular}%
}
\end{table}

\subsection{Global and Local Explanations}
\label{sec:global-and-local}

One way to classify explanations is by their interpretation scale.
For instance, an explanation could be as thorough as describing the entire machine learning model.
Alternatively, it could only partially explain the model, or it could be limited to explaining an individual input instance. 
\textit{Global Explanation} (or \textit{Model Explanation}) is an explanation type that describes how the overall machine learning model works.
Model visualization~\cite{liu2014topicpanorama, liu2017towards} and decision rules~\cite{lakkaraju2016interpretable} are examples of explanations falling in this category.
In other cases, interpretable approximations of complex models serve as the model explanation. 
Tree regularization~\cite{wu2018beyond} is a recent example of regularized complex model to learn tree-like decision boundaries. 
Model complexity and explanation design are the main factors used to choose between different types of global explanations.

In contrast, \textit{Local Explanations} (or \textit{Instance Explanations}) aim to explain the relationship between specific input-output pairs or the reasoning behind the results for an individual user query.
This type of explanation is thought to be less overwhelming for novices, and it can be suited for investigating edge cases for the model or debugging data.
Local explanations often make use of saliency methods~\cite{baehrens2010explain, zeiler2014visualizing} or local approximation of the main model~\cite{ribeiro2016should,ribeiro2018anchors}.
Saliency methods (also as known as attribution maps or sensitivity maps) use different approaches (e.g., perturbation-based methods, gradient-based methods) to show what features in the input strongly influence the model's prediction. 
Local approximation of the model, on the other hand, trains an interpretable model (learned from the main model) to locally represent the complex model's behavior.

\subsection{Interpretable Models vs. Ad-hoc Explainers}
\label{sec:ad-hoc-explainers}

The human interpretability of a machine learning model is inversely proportional to the model's size and complexity.
Complex models (e.g., deep neural networks) with high performance and robustness in real-world applications are not interpretable by human users due to their large variable space.
Linear regression models or decision trees offer better interpretability but have limited performance on high-dimensional data, whereas a random forest model (ensemble of hundreds of decision trees) can have much higher performance but is less understandable.
This trade-off between model interpretability and performance led researchers to design ad-hoc methods to explain any black-box machine learning algorithm such as deep neural networks.
Ad-hoc explainers (e.g., \cite{ribeiro2016should, lundberg2017unified}) are independent algorithms that can describe model predictions by explaining ``why'' a certain decision has been made instead of describing the whole model. 
However, there are limitations in explaining black-box models with ad-hoc explainers, such as the uncertainty of the fidelity of the explainer itself. 
We will discuss more about the fidelity of explanations in Section \ref{sec:explanation-truthfulness}.
Furthermore, although ad-hoc explainers generally describe ``why'' a prediction is made, these methods lack in explaining ``how'' the decision is made.

\subsection{What to Explain}
\label{sec:what_to_explain}

When users face a complex intelligent system, they may demand different types of explanatory information and each explanation type may require its own design.
Here we review six common types of explanations used in XAI system designs.
\\

\noindent \textbf{How Explanations} demonstrate a holistic representation of the machine learning algorithm to explain \textit{How} the model works. 
For visual representations, model graphs~\cite{lakkaraju2016interpretable} and decision boundaries~\cite{maaten2008visualizing} are common design examples for \textit{How} explanations.
However, research shows users may also be able to develop a mental model of the algorithm based on a collection of explanations from multiple individual instances~\cite{lombrozo2009explanation}.

\noindent \textbf{Why Explanations} describe \textit{Why} a prediction is made for a particular input.
Such explanations aim to communicate what features in the input data~\cite{ribeiro2016should} or what logic in the model~\cite{ribeiro2018anchors,lakkaraju2016interpretable} has led to a given prediction by the algorithm. 
This type of explanation can have either model agnostic~\cite{ribeiro2016should,lundberg2017unified} or model dependent~\cite{selvaraju2017grad} solutions.

\noindent \textbf{Why-not Explanations} help users to understand the reasons why a specific output was not in the output of the system~\cite{vermeulen2010pervasivecrystal}.
\textit{Why-not} explanations (also called \textit{Contrastive Explanations}) characterize the reasons for differences between a model prediction and the user's expected outcome.
Feature importance (or feature attribution) is commonly used as an interpretability technique for \textit{Why} and \textit{Why-not} explanations.

\noindent \textbf{What-If Explanations} involve demonstration of how different algorithmic and data changes affect model output given new inputs~\cite{cai2019effects}, manipulation of inputs~\cite{lim2009and}, or changing model parameters~\cite{kocielnik2019will}. 
Different what-if scenarios may be automatically recommended by the system or can be chosen for exploration through interactive user control. 
Domains with high-dimensional data (e.g., image and text) and complex machine learning models (e.g., DNNs) have fewer parameters for users to directly tune and examine trained model compared to simpler data (e.g., low-dimensional tabular data) and models.

\noindent \textbf{How-to Explanations} spell out hypothetical adjustments to the input or model that would result in a different output~\cite{lim2009and,lim2019these}, such as a user-specified output of interest. 
Techniques to generate~\textit{How-to} (or \textit{Counterfactual}) explanations are ad-hoc and model-agnostic considering that model structure and internal values are not a part of the explanation~\cite{wachter2017counterfactual}.
Such methods can work interactively with the user's curiosity and partial conception of the system to allow an evolving mental model of the system through iterative testing.

\noindent \textbf{What-else Explanations} present users with similar instances of input that generate the same or similar outputs from the model. 
Also called \textit{Explanation by Example}, \textit{What-else} explanations pick samples from the model's training dataset that are similar to the original input in the model representation space~\cite{cai2019human}. 
Although very popular and easy to achieve, research shows example-based explanations could be misleading when training datasets lack uniform distribution of the data~\cite{kim2016examples}.

\subsection{How to Explain}
\label{sec:how_to_explain}

In all types of machine learning explanations, the goal is to reveal new information about the underlying system.
In this survey, we mainly focus on human-understandable explanations, though we note that research on interpretable machine learning has also studied other purposes such as knowledge transfer, object localization, and error detection~\cite{olah2018the,fong2017interpretable}.

Explanations can be designed using a variety of formats for different user groups~\cite{yu2018user}. \textit{Visual Explanations} use visual elements to describe the reasoning behind the machine learning models. 
Attention maps and visual saliency in the form of saliency heatmaps~\cite{zeiler2014visualizing, simonyan2013deep} are examples of visual explanations that are widely used in machine learning literature.
Verbal Explanations describe the machine's model or reasoning with words, phrases, or natural language. 
Verbal explanations are popular in applications like question-answering explanations and decision lists~\cite{lakkaraju2016interpretable}. 
This form of explanation has also been implemented in recommendation systems~\cite{Berkovsky:2017:RUT:3025171.3025209,herlocker2000explaining} and robotics~\cite{rosenthal2016verbalization}.
Explainable interfaces commonly make use of multiple modalities (e.g., visual, verbal, and numerical elements) for explanations to support user understanding~\cite{myers2006answering}.
\textit{Analytic Explanation} is another approach to view and explore the data and the machine learning models representations~\cite{hohman2018visual}. 
Analytic explanations commonly rely on numerical metrics and data visualizations.
Visual analytics tools also allow researchers to review model structures, relations, and their parameters in complex deep models. 
Heatmap visualizations~\cite{strobelt2018lstmvis}, graphs and networks~\cite{goodall2018situ}, and hierarchical (decision trees) visualizations are commonly used to visualize analytic explanations for interpretable algorithms. 
Recently, Hohman et al.~\cite{hohman2019telegam} implemented a combination of visualization and verbalization to communicate or summarize key aspects of a model.

From a different perspective, Chromik et al.~\cite{chromik2019dark} extends the idea of ``dark patterns'' from interactive user interface design~\cite{gray2018dark} into machine learning explanations. 
They review possible ways that phrasing of explanations and their implementation in the interface could deceive users for the benefit of other parties.
They review negative effects such as lack of user attention to explanations, formation of an incorrect mental model, and even algorithmic anxiety~\cite{jhaver2018algorithmic} could be among the consequences of such deceptive presentations and interactions of machine learning explanations.

\section{Categorization of XAI Design Goals and Evaluation Methods}
\label{sec:categorization_section}

\begin{figure}[t]
\centering
  \includegraphics[width=0.9\columnwidth]{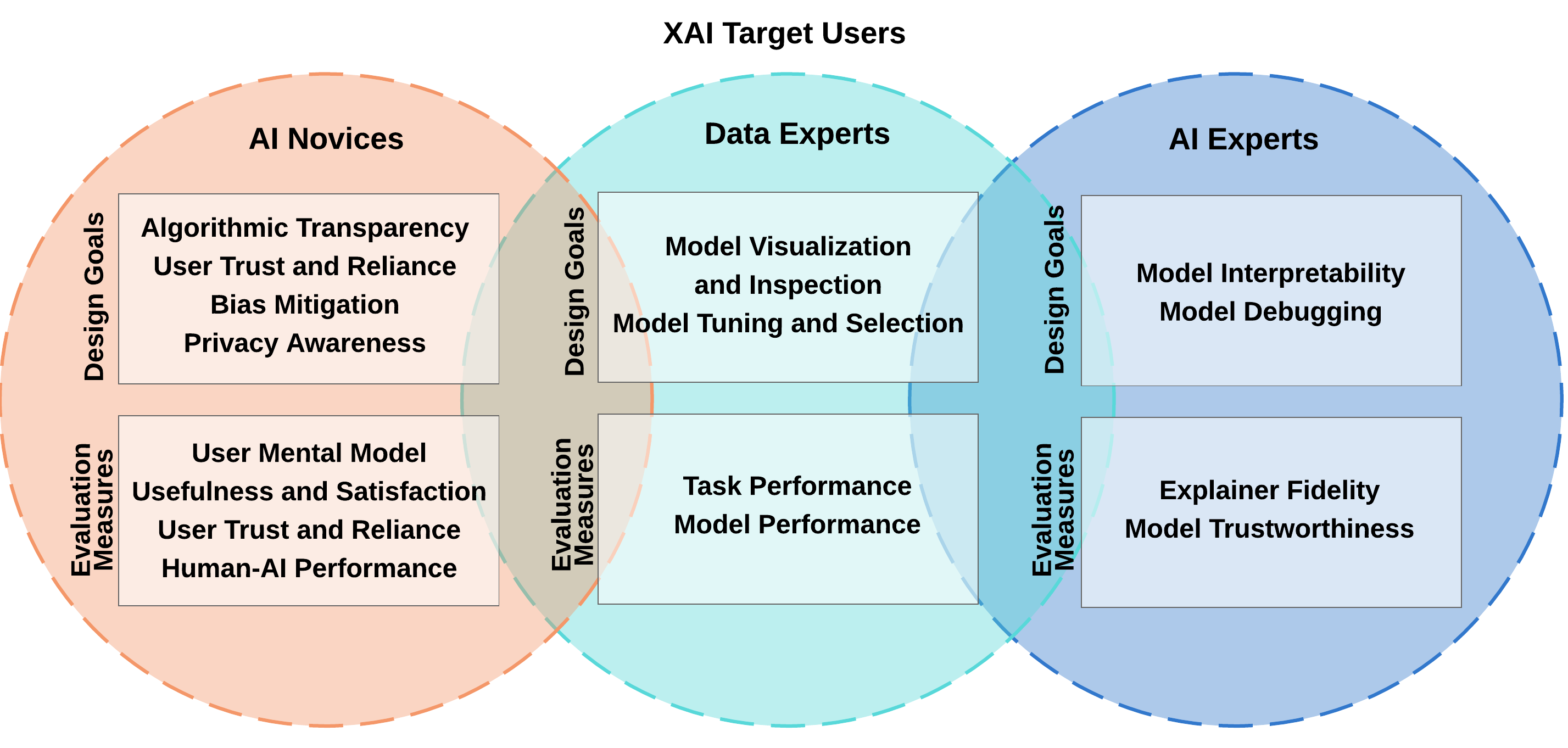} 
  \caption{
  A summary of our categorization of XAI design goals and evaluation measures between user groups. \textbf{Top:} Different system design goals for each user group. \textbf{Bottom:} Common evaluation measures used in each user group. 
  Notice that similar XAI goals for different user groups require different research objectives, design methods, and implementation paths.
  }
  \label{fig:3D-categorization}
\end{figure}

While an ideal XAI system should be able to answer all user queries and meet all XAI concept goals \cite{gunning2017explainable}, individual research efforts focus on designing and studying XAI systems with respect to specific interpretability goals and specific users.
Similarly, evaluating the explanations can demonstrate and verify the effectiveness of the explainable systems for their intended goals.

After careful review and analysis of XAI goals and their evaluation methods in the literature, we recognized the following two attributes to be most significant for our purposes of interdisciplinary organization of XAI design and evaluation methods:

\begin{itemize}

\item \textbf{Design Goals}. The first attribute in our categorization is the design goal for interpretable algorithms and explainable interfaces in XAI research.
We obtain XAI design goals from multiple research disciplines: machine learning, data visualization, and HCI. 
To better understand the differences between various goals for XAI, we organize XAI design goals with their three user groups: AI novices (i.e., general AI product end-user), data experts (experts in data analytics and domain experts), and AI experts (machine learning model designers). 

\item \textbf{Evaluation Measures}. We review evaluation methods and discuss measures used to evaluate machine learning explanations.
The measures include user mental model, user trust and reliance, explanation usefulness and satisfaction, human-machine task performance, and computational measures.
In our review, we will pay more attention to evaluation measures of XAI as the authors believe that this category is relatively less explored.

\end{itemize}

Figure~\ref{fig:3D-categorization} presents the pairing between XAI design goals and their evaluation measures. 
Note that user groups is used as an auxiliary dimension to emphasize on the importance of end users for system goals. 
The overlap between XAI user groups shows similarities in the design and evaluation methods between different targeted user groups. 
However, the similar XAI goals in different user groups require different research objectives, design methods, and implementation paths.
To help summarize our characterization along with example literature, Table~\ref{tab:main-table} presents a cross-reference table of XAI evaluation literature to emphasize the importance of design goals, evaluation measures, and user types. 
We first review details of research focusing on XAI design goals in Section~\ref{sec:d_goal_section} including eight goals organized by their user groups. 
We then review evaluation measures and methods in Section~\ref{sec:e_measure_section} including six main measures and their methods collected from the surveyed literature.

\begin{table}[]
\centering
\caption{Tabular summary of our XAI evaluation dimensions of measures and targeted user types. The table includes 42 papers that represent a subset of the surveyed literature organized by the two dimensions.}
\label{tab:main-table}
\resizebox{\textwidth}{!}{%
\begin{tabular}{c|c|c|c|c|c|c|c|c|c|c|c|c|c}
 & \multicolumn{8}{c|}{\textbf{Design Goals}} & \multicolumn{5}{c|}{} \\
 & \multicolumn{4}{c|}{\textbf{Novice Users}} & \multicolumn{2}{c|}{\textbf{Data Experts}} & \multicolumn{2}{c|}{\textbf{AI Experts}} & \multicolumn{5}{c|}{\multirow{-2}{*}{\textbf{Evaluation Measures}}} \\
\multirow{-3}{*}{\textbf{Work}} & \textbf{\rotatebox{90}{\begin{tabular}[c]{@{}c@{}}G1: Algorithmic Transparency\end{tabular}}} & \textbf{\rotatebox{90}{G2: User  Trust and Reliance}} & \textbf{\rotatebox{90}{G3: Bias Mitigation}} & \textbf{\rotatebox{90}{G4: Privacy Awareness}} & \textbf{\rotatebox{90}{\begin{tabular}[c]{@{}c@{}}G5: Model Visualization \\ and Inspection\end{tabular}}} & \textbf{\rotatebox{90}{\begin{tabular}[c]{@{}c@{}}G6: Model Tuning \\ and Selection\end{tabular}}} & \textbf{\rotatebox{90}{\begin{tabular}[c]{@{}c@{}} \vspace*{-8pt} \\ G7: Model Interpretability \vspace{1.2mm} \end{tabular}}} & \textbf{\rotatebox{90}{G8: Model Debugging}} & \textbf{\rotatebox{90}{M1: Mental Model}} & \textbf{\rotatebox{90}{\begin{tabular}[c]{@{}c@{}}M2: Usefulness and \\ Satisfaction\end{tabular}}} & \textbf{\rotatebox{90}{\begin{tabular}[c]{@{}c@{}}M3: User Trust \\ and Reliance \end{tabular}}} & \textbf{\rotatebox{90}{\begin{tabular}[c]{@{}c@{}}M4: Human-AI \\ Task Performance \end{tabular}}} & \multicolumn{1}{c|}{\textbf{\rotatebox{90}{M5: Computational Measures}}} \\ \hline
\rowcolor[HTML]{EFEFEF} 
Herlocker et al. 2000 \cite{herlocker2000explaining} &  & \ding{117} &  &  &  &  &  &  &  & \ding{117} & \ding{117} & \ding{117} &  \\
Kulesza et al. 2012 \cite{Kulesza2012Tell} & \ding{117} &  &  &  &  &  &  &  & \ding{117} & \ding{117} &  & \ding{117} &  \\
\rowcolor[HTML]{EFEFEF} 
Lim et al. 2009 \cite{lim2009assessing} & \ding{117} &  &  &  &  &  &  &  & \ding{117} & \ding{117} &  &  &  \\
Stumpf et al. 2018 \cite{stumpf2018explaining} & \ding{117} & \ding{117} &  &  &  &  &  &  & \ding{117} &  & \ding{117} &  &  \\
\rowcolor[HTML]{EFEFEF} 
Bilgic et al. 2005\cite{bilgic2005explaining} &  & \ding{117} &  &  &  &  &  &  &  & \ding{117} & \ding{117} &  &  \\
Bunt et al. 2012 \cite{bunt2012explanations} & \ding{117} &  &  &  &  &  &  &  &  & \ding{117} &  &  &  \\
\rowcolor[HTML]{EFEFEF} 
Gedikli et al. 2014 \cite{gedikli2014should} &  & \ding{117} &  &  &  &  &  &  &  & \ding{117} &  &  &  \\
Kulesza et al. 2013 \cite{kulesza2013too} & \ding{117} & \ding{117} &  &  &  &  &  &  & \ding{117} &  & \ding{117} &  &  \\
\rowcolor[HTML]{EFEFEF} 
Lim et al. 2009 \cite{lim2009and} & \ding{117} & \ding{117} &  &  &  &  &  &  & \ding{117} & \ding{117} & \ding{117} & \ding{117} &  \\
Lage et al. 2019 \cite{lage2019human} & \ding{117} &  &  &  &  &  &  &  &  & \ding{117} &  & \ding{117} &  \\
\rowcolor[HTML]{EFEFEF} 
Schmid et al. 2016 \cite{schmid2016does} & \ding{117} &  &  &  &  &  &  &  &  &  &  & \ding{117} &  \\
Berkovsky et al. 2017 \cite{Berkovsky:2017:RUT:3025171.3025209} &  & \ding{117} &  &  &  &  &  &  &  & \ding{117} & \ding{117} &  &  \\
\rowcolor[HTML]{EFEFEF} 
Glass et al. 2008 \cite{glass2008toward} &  & \ding{117} &  &  &  &  &  &  &  & \ding{117} & \ding{117} &  &  \\
Haynes et al. 2009 \cite{haynes2009designs} &  & \ding{117} &  &  &  &  &  &  &  & \ding{117} & \ding{117} &  &  \\
\rowcolor[HTML]{EFEFEF} 
Holliday et al. 2016 \cite{holliday2016user} & \ding{117} & \ding{117} &  &  &  &  &  &  & \ding{117} &  & \ding{117} &  &  \\
Nothdurft et al. 2014 \cite{nothdurft2014probabilistic} & \ding{117} & \ding{117} &  &  &  &  &  &  & \ding{117} &  & \ding{117} &  &  \\
\rowcolor[HTML]{EFEFEF} 
Pu and Chen et al. 2006 \cite{pu2006trust} &  & \ding{117} &  &  &  &  &  &  &  &  & \ding{117} & \ding{117} &  \\
Bussone et al. 2015 \cite{bussone2015role} & \ding{117} & \ding{117} &  &  &  &  &  &  &  &  & \ding{117} &  &  \\
\rowcolor[HTML]{EFEFEF} 
Groce et al. 2014 \cite{groce2014you} & \ding{117} &  &  &  &  &  &  &  & \ding{117} &  &  & \ding{117} &  \\
Myers et al. 2006 \cite{myers2006answering} & \ding{117} &  &  &  &  &  &  &  & \ding{117} &  &  & \ding{117} &  \\
\rowcolor[HTML]{EFEFEF} 
Binns et al. 2018 \cite{Binns:2018:RHP:3173574.3173951} & \ding{117} &  & \ding{117} &  &  &  &  &  & \ding{117} &  &  &  &  \\
Lee et al. 2019 \cite{lee2019procedural} & \ding{117} &  & \ding{117} &  &  &  &  &  & \ding{117} & \ding{117} &  &  &  \\
\rowcolor[HTML]{EFEFEF} 
Rader et al. 2018 \cite{rader2018explanations} & \ding{117} &  &  & \ding{117} &  &  &  &  & \ding{117} & \ding{117} &  &  &  \\
Datta et al. 2015 \cite{datta2015automated} &  &  &  & \ding{117} &  &  &  &  &  &  &  &  & \ding{117} \\
\rowcolor[HTML]{EFEFEF} 
Kulesza et al. 2015 \cite{kulesza2015principles} & \ding{117} &  &  &  & \ding{117} & \ding{117} &  &  & \ding{117} &  &  & \ding{117} &  \\
Kulesza et al. 2010 \cite{kulesza2010explanatory} & \ding{117} &  &  &  & \ding{117} & \ding{117} &  &  & \ding{117} &  &  & \ding{117} &  \\
\rowcolor[HTML]{EFEFEF} 
Krause et al. 2016 \cite{krause2016interacting} &  &  &  &  & \ding{117} & \ding{117} &  &  &  &  &  & \ding{117} &  \\
Krause et al. 2017 \cite{Krause2017workflow} &  &  &  &  & \ding{117} & \ding{117} &  &  &  &  &  & \ding{117} &  \\
\rowcolor[HTML]{EFEFEF} 
Liu et al. 2014 \cite{liu2014topicpanorama} &  &  &  &  & \ding{117} &  &  &  &  &  &  & \ding{117} &  \\
Ribeiro et al. 2016 \cite{ribeiro2016should} &  &  &  &  &  &  & \ding{117} &  & \ding{117} &  & \ding{117} & \ding{117} & \ding{117} \\
\rowcolor[HTML]{EFEFEF} 
Ribeiro et al. 2018 \cite{ribeiro2018anchors} &  &  &  &  &  &  & \ding{117} &  & \ding{117} &  & \ding{117} & \ding{117} & \ding{117} \\
Ross et al. 2017 \cite{ross2017improving} &  &  &  &  &  &  & \ding{117} &  &  &  &  &  & \ding{117} \\
\rowcolor[HTML]{EFEFEF} 
Adebayo et al. 2018 \cite{adebayo2018sanity} &  &  &  &  &  &  & \ding{117} &  &  &  &  &  & \ding{117} \\
Samek et al. 2017 \cite{samek2017evaluating} &  &  &  &  &  &  & \ding{117} &  &  &  &  &  & \ding{117} \\
\rowcolor[HTML]{EFEFEF} 
Zeiler et al. 2014 \cite{zeiler2014visualizing} &  &  &  &  &  &  & \ding{117} &  &  &  &  & \ding{117} & \ding{117} \\
\rowcolor[HTML]{FFFFFF} 
Lakkaraju et al. 2016 \cite{lakkaraju2016interpretable} &  &  &  &  &  &  & \ding{117} &  &  &  &  & \ding{117} &  \\
\rowcolor[HTML]{EFEFEF} 
Kahng et al. 2018 \cite{kahng2018cti} &  &  &  &  &  &  &  & \ding{117} &  & \ding{117} &  & \ding{117} &  \\
Liu et al. 2018 \cite{liu2018analyzing} &  &  &  &  &  &  &  & \ding{117} &  & \ding{117} &  & \ding{117} &  \\
\rowcolor[HTML]{EFEFEF} 
Liu 2017 et al. 2009 \cite{liu2017towards} &  &  &  &  &  &  &  & \ding{117} &  & \ding{117} &  & \ding{117} &  \\
Ming et al. 2017 \cite{ming2017understanding} &  &  &  &  &  &  &  & \ding{117} &  & \ding{117} &  & \ding{117} &  \\
\rowcolor[HTML]{EFEFEF} 
Pezzotti et al. 2018 \cite{pezzotti2018deepeyes} &  &  &  &  &  &  &  & \ding{117} &  & \ding{117} &  & \ding{117} &  \\
Strobelt et al. 2018 \cite{strobelt2018lstmvis} &  &  &  &  &  &  &  & \ding{117} &  & \ding{117} &  & \ding{117} & 
\end{tabular}%
}
\end{table}

\section{XAI Design Goals}
~\label{sec:d_goal_section}

Research efforts have explored many goals for XAI systems.
Doshi-Velez and Kim ~\cite{doshi2017towards} reviewed multiple high-level priorities for XAI systems with examples including safety, ethics, user reliance, and scientific understanding. 
Later, Arrieta et al.~\cite{arrieta2019explainable} presented a thorough review of XAI opportunities in different application domains.
Accordingly, different design choices such as explanation type, scope, and level of detail will be affected by the application domain, design goal, and user type.
For example, while machine learning experts might prefer highly-detailed visualizations of deep models to help them optimize and diagnose algorithms, end-users of daily-used AI products do not expect fully detailed explanations for every query from a personalized agent.
Therefore, XAI systems are expected to provide the right type of explanations for the right group of users, meaning it will be most efficient to design an XAI system according to the user's needs and levels of expertise. 

To this end, we distinguish XAI design goals based on the designated end-user and evaluation subjects, which we categorize into three general groups of AI experts, data experts, and AI novices. 
We emphasize that this separation of groups is presented primarily for organizational convenience, as goals are not mutually exclusive across groups, and  specific priorities are case dependent for any particular project.
The XAI design goals also extend to the broader goal of \textit{Responsible AI} by improving transparency and explainability of intelligent systems. 
Note that although there are overlaps in the methods used to achieve these goals, the research objectives and design approaches are substantially different among distinct research fields and their user groups. 
For instance, even though leveraging interpretable models to reduce machine learning model bias is a research goal for AI experts, bias mitigation is also a design goal for AI novices to avoid adverse effects of algorithmic decision-making in their respective domain settings.
However, interpretability techniques for AI experts and bias mitigation tools for AI novice require different design methods and elements.
In the following subsections, we review eight design goals for XAI systems organized by their user groups.

\subsection{AI Novices}

\textit{AI novices} refer to end-users who use AI products in daily life but have no (or very little) expertise on machine learning systems.
These include end-users of intelligent applications like personalized agents (e.g., home assistant devices), social media, and e-commerce websites.
In most smart systems, machine learning algorithms serve as internal functions and APIs to enable specific features embedded in intelligent and context-aware interfaces.
Previous research shows intuitive interface and interaction design can enhance users' experience with the system through improving end-users' comprehension and reliance on the intelligent systems~\cite{muir1987trust}.
In this regard, creating human-understandable and yet accurate representations of complicated machine learning explanations for novice end-users is a challenging design trade-off in XAI systems. 
Note that although there are overlaps among goals for \textit{AI Novices} and AI experts who build interpretable algorithms, each user group requires a different set of design methods and objectives that are being studied in different research communities.

The main design goals for AI novice end-users of XAI system can be itemized as the following:
\newline

\noindent \textbf{G1: Algorithmic Transparency:} An immediate goal for a XAI system -- in comparison to an inexplicable intelligent system -- is to help end-users understand how the intelligent system works. 
Machine learning explanations improve users' mental model of the underlying intelligent algorithms by providing comprehensible transparency for the complex intelligent algorithms~\cite{weller2017challenges}.
Further, transparency of a XAI system can improve user experience through better user understanding of model output~\cite{lim2011improving}, thus improving user interactions with the system~\cite{kulesza2015principles}.
\newline

\noindent \textbf{G2: User Trust and Reliance:} XAI system can improve end-users trust in the intelligent algorithm by providing explanations.
A XAI system lets users assess system reliability and calibrate their perception of system accuracy.
As a result, users' trust in the algorithm leads to their reliance on the system.
Example applications where XAI aims to improve user reliance through its transparent design include recommendation systems~\cite{Berkovsky:2017:RUT:3025171.3025209}, autonomous systems~\cite{wiegand2019drive}, and critical decision making systems~\cite{bussone2015role} .
\newline

\noindent \textbf{G3: Bias Mitigation:} Unfair and biased algorithmic decision-making is a critical side effect of intelligent systems. 
Bias in machine learning has many sources, including biased training data and feature learning that could result in discrimination in algorithmic decision-making~\cite{mehrabi2019survey}.
Machine learning explanations can help end-users to inspect if the intelligent systems are biased in their decision-making.
Examples of cases in which XAI is used for bias mitigation and fairness assessment are criminal risk assessment~\cite{lee2019procedural,Binns:2018:RHP:3173574.3173951} and loan and insurance rate prediction~\cite{chen2019fairness}.
It is worth mentioning that there is  overlap between the biased decision-making mitigation goal for AI novices and the goal of dataset bias for AI experts (Section~\ref{sec:data-experts}), which results in shared implementation techniques.
However, the two distinct user groups require their own sets of XAI design goals and processes.
\newline

\noindent \textbf{G4: Privacy Awareness:} Another goal in designing XAI systems is to provide a means for end-users to assess their data privacy. 
Machine learning explanations can disclose to end-users what user data is being used in algorithmic decision-making.
Examples of AI application examples in which privacy awareness is primarily important include personalized advertisements using users' online advertisement~\cite{datta2015automated} and personalized news feed in social media~\cite{rader2018explanations,eslami2015always}.
\newline

In addition to the major XAI goals, interactive visualization tools have also been developed to help AI novices to learn machine learning concepts and models by interacting with simplified data and model representations. 
Examples of these educative tools include TensorFlow PlayGround~\cite{smilkov2017direct} for teaching elementary neural networks concepts and Adversarial Playground~\cite{norton2017adversarial} for learning concept of adversarial examples in DNNs. 
These minor goals cover XAI system objectives that have limited extent compared to main goals.

\subsection{Data Experts}
\label{sec:data-experts}
\textit{Data experts} include data scientists and domain experts who use machine learning for analysis, decision-making, or research. 
Visual analysis tools can support interpretable machine learning in many ways, such as visualizing the network architecture of a trained model and training process of machine learning models.
Researchers have implemented various visualization designs and interaction techniques to understand better and improve machine learning models.

Data experts analyze data in specialized forms and domains, such as cybersecurity~\cite{goodall2018situ,best20147}, medicine~\cite{caruana2015intelligible,krause2016interacting}, text~\cite{liu2014topicpanorama, liu2016uncertainty}, and satellite image analysis~\cite{robinson2017deep}. 
These users might be experts of certain domain areas or experts in general areas of data science, but in our categorization, we consider users in the \textit{data experts} category to generally lack expertise in the technical specifics of the machine learning algorithms.
Instead, this group of users often utilize intelligent data analysis tools or visual analytics systems to obtain insights from the data. 
Notice that there are overlaps between XAI goals in different disciplines and visual analytics tools designed for \textit{Data Experts} could be used by both model designers and data analysts. 
However, design needs and approaches for these XAI systems may be different across research communities.
The main design goals for data experts users of a XAI system are as follows:
\newline

\noindent \textbf{G5: Model Visualization and Inspection:}
Similar to AI novices, data experts also benefit from machine learning interpretability to inspect model uncertainty and trustworthiness~\cite{sacha2016role}.
For instance, machine-learning explanations help data experts to visualize models~\cite{hohman2019summit} and inspect for problems like bias~\cite{ahn2019fairsight}. 
Another important aspect of model visualization and inspection for domain experts is to identify and analyze failure cases of machine learning models and systems~\cite{ming2018rulematrix}.
Therefore, the main challenge for data-analysis and decision-support systems is to improve model transparency via visualization and interaction techniques for domain experts~\cite{yu2018user}.
\newline

\noindent \textbf{G6: Model Tuning and Selection:}
Visual analytics approaches can help data experts to tune machine learning parameters for their specific data in an interactive visual fashion~\cite{liu2014topicpanorama}.
The interpretability element in XAI visual analytic systems increase data experts' ability to compare multiple models~\cite{alexander2015task} and select the right model for the targeted data.
As an example, Du et al.~\cite{du2019eventaction} present EventAction, an event sequence recommendation approach that allows the users to interactively select records that share their desired attribute values.
In the case of tuning DNN networks, visual analytics tools enhance designers' ability to modify networks~\cite{pezzotti2018deepeyes}, improve training~\cite{liu2018analyzing}, and to compare different networks~\cite{wongsuphasawat2017visualizing}.

\subsection{AI Experts}
In our categorization, \textit{AI experts} are machine learning scientists and engineers who design machine learning algorithms and interpretability techniques for XAI systems. 
Machine learning interpretability techniques either provide model interpretation or instance explanations. 
Examples of model interpretation techniques include inherently interpretable models~\cite{wang2015falling}, deep model simplification~\cite{wu2018beyond}, and visualization of model internals~\cite{yosinski2015understanding}.
Instance explanations techniques, however, present feature importance for individual instances such as saliency map in image data and attention in textual data~\cite{das2017human}.
AI engineers also benefit from visualization and visual analytics tools to interactively inspect model internal variables~\cite{liu2018analyzing} to detect architecture and training flaws or monitor and control the training process~\cite{kahng2018cti}, which indicates possible overlaps among design goals.
We list main design goals for AI Experts into two following items:
\newline

\noindent \textbf{G7: Model Interpretability:} Model interpretability is often a primary XAI Goal for AI experts.
Model interpretability allows getting new insights into how deep models learn patterns from data~\cite{olah2018the}. 
In this regard, various interpretability techniques for different domains have been proposed to satisfy the need for explanation~\cite{selvaraju2017grad,kim2018interpretability}.
For example, Yosinski et al.~\cite{yosinski2015understanding} created an interactive toolbox to explore CNN's activation layers in real-time that gives an intuition about ``how the CNN works'' to the user.
\newline

\noindent \textbf{G8: Model Debugging:}
AI researchers use interpretability techniques in different ways to improve model architecture and training process. 
For example, Zeiler and Fergus~\cite{zeiler2014visualizing} present a use case in which visualization of filters and feature maps in CNN leads to revising training hyper-parameters and, therefore, model performance improvement.
In another work, Ribeiro et al.~\cite{ribeiro2016should} used model instance explanations and human review of explanations to improve model performance through feature engineering.
\newline

Other than main XAI goals for AI experts, machine learning explanations are used for other purposes including detecting dataset bias~\cite{zhang2018examining}, adversarial attack detection~\cite{fong2017interpretable}, and model failure prediction~\cite{mohseni2019predicting}. 
Also, visual saliency maps and attention mechanisms have been used as weakly supervised object localization~\cite{simonyan2013deep}, multiple object recognition~\cite{ba2014multiple}, and knowledge transfer~\cite{li2019attention} techniques.

\section{XAI Evaluation Measures}
\label{sec:e_measure_section}

Evaluation measures for XAI systems is another important factor in the design process of XAI systems. 
Explanations are designed to answer different interpretability goals, and hence different measures are needed to verify explanation validity for the intended purpose. 
For example, experimental design with human-subject studies is a common approach to perform evaluations with AI novice end-users. 
Various controlled in-lab and online crowdsourced studies have been used for XAI evaluation. 
Also, case studies aim to collect domain expert users' feedback while performing high-level cognitive tasks with analytics tools.
By contrast, computational measures are designed to evaluate the accuracy and completeness of explanations from interpretable algorithms.

In this section, we review and categorize the main evaluation measures for XAI systems and algorithms.
Table~\ref{tab:main-table} shows a list of five evaluation measures associated with their design goals.
Additionally, we provide summarized and ready-to-use XAI evaluation measures and methods extracted from the literature in Tables~\ref{tab:mental-table}-\ref{tab:computational-table}.
\newline

\subsection{M1: Mental Model}


\begin{table}[]
\centering
\caption{Evaluation measures and methods used in studying user mental models in XAI systems}
\label{tab:mental-table}
\begin{tabular}{ll}
\hline
\textbf{Mental Model Measures} & \textbf{Evaluation Methods} \\ \hline
\multirow{2}{*}{User Understanding of Model} & Interview (\cite{costanza2014doing}) and Self-explanation (\cite{dodge2018should, Penney:2018:TFU:3172944.3172946, Binns:2018:RHP:3173574.3173951}) \\ \cline{2-2} 
 & Likert-scale Questionnaire (\cite{lombrozo2009explanation, rader2015understanding, lim2009assessing, kulesza2013too, kim2018interpretability, lakkaraju2016interpretable}) \\ \hline
Model Output Prediction & User Prediction of Model Output (\cite{kay2016ish, ribeiro2016should, ribeiro2018anchors}) \\ \hline
Model Failure Prediction & User Prediction of Model Failure (\cite{bansal2019beyond,nushi2018towards}) \\ \hline
\end{tabular}
\end{table}

Following cognitive psychology theories, a mental model is a representation of how users understands a system. 
Researchers in HCI study users' mental models to determine their understanding of intelligent systems in various applications.
For example, Costanza et al.~\cite{costanza2014doing} studied how users understand a smart grid system, and Kay et al.~\cite{kay2016ish} studied how users understand and adapt to uncertainty in machine learning prediction of bus arrival times.

In the context of XAI, explanations help users to create a mental model of \textit{how the AI works}. 
Machine learning explanation is a way to help the users in building a more accurate mental model.
Studying users' mental models of XAI systems can help verify explanation effectiveness in describing an algorithm's decision-making process.
Table~\ref{tab:mental-table} summarizes different evaluation methods used to measure users' mental model of machine learning models. 

Psychology research in human-AI interactions has also explored structure, types, and functions of explanations to find essential ingredients of ideal explanation for better user understanding and more accurate mental models~\cite{keil2006explanation, lombrozo2006structure}. 
For instance, Lombrozo~\cite{lombrozo2009explanation} studied how different types of explanations can help structure conceptual representation.
In order to find out how an intelligent system should explain its behavior for non-experts, research on machine learning explanations has studied how users interpret intelligent agents~\cite{dodge2018should, Penney:2018:TFU:3172944.3172946} and algorithms~\cite{rader2015understanding} to find out what users expect from machine explanations.
Related to this, Lim and Dey~\cite{lim2009assessing} elicit types of explanations that users might expect in four real-world applications. They specifically study what types of explanations users demand in different scenarios such as system recommendation, critical events, and unexpected system behavior.
In measuring user mental model through model failure prediction, Bansal et al.~\cite{bansal2019beyond} designed a game in which participants receive monetary incentives based on their final performance score. 
Although experiments were done on a simple three-dimensional task, their results indicate a decrease in users' ability to predict model failure as data and model get more complicated.

A useful way of studying users comprehension of intelligent systems is to directly ask them about the intelligent system's decision-making process. 
Analyzing users' interviews, think-alouds, and self-explanations provides valuable information about the users' thought processes and mental models~\cite{kulesza2010explanatory}.
On studying user comprehension, Kulesza et al.~\cite{kulesza2013too} studied the impact of explanation soundness and completeness on fidelity of end-users mental model in a music recommendation interface.
Their results found that explanation completeness (broadness) had a more significant effect on user understanding of the agent compared to explanation soundness.
In another example, Binns et al.~\cite{Binns:2018:RHP:3173574.3173951} studied the relation between machine explanations and users' perception of justice in algorithmic decision-making with different sets of explanation styles. 
User attention and expectations may also be considered during the interpretable interface design cycles for intelligent systems~\cite{stumpf2018explaining}.

Interest in developing and evaluating human-understandable explanations has also led to interpretable models and ad-hoc explainers to measure mental models.
For example, Ribeiro et al.~\cite{ribeiro2016should} evaluated users' understanding of the machine learning algorithm with visual explanations.
They showed how explanations mitigate human overestimation of the accuracy of an image classifier and help users choose a better classifier based on the explanations.
In a follow-up work, they compared the global explanations
of a classifier model with the instance explanations
of the same model and found global explanations were more effective solutions for finding the model weaknesses ~\cite{ribeiro2018anchors}. 
In another paper, Kim et al.~\cite{kim2018interpretability} conducted a crowdsourced study to evaluate feature-based explanation understandability for end-users.
Addressing understanding of model representations, Lakkaraju et al.~\cite{lakkaraju2016interpretable} presented interpretable decision sets, an interpretable classification model, and measured users' mental models with different metrics such as user accuracy on predicting machine output and length of users' self-explanations.

\subsection{M2: Explanation Usefulness and Satisfaction}
\label{user_satisfaction}

End-user satisfaction and usefulness of machine explanation are also of importance when evaluating explanations in intelligent systems~\cite{bilgic2005explaining}.
Researchers use different subjective and objective measures for understandability, usefulness, and sufficiency of details to assess explanatory value for users~\cite{miller2017explanation}.
Although there are implicit methods to measure user satisfaction~\cite{hoffman2017theory}, a considerable part of the literature follows qualitative evaluation of satisfaction in explanations, such as questionnaires and interviews.
For example, Gedikli et al.~\cite{gedikli2014should} evaluated ten different explanation types with user ratings of explanation satisfaction and transparency.
Their results showed a strong relationship between user satisfaction and perceived transparency.
Similarly, Lim et al.~\cite{lim2009and} explore explanation usefulness and efficiency in their interpretable context-aware system by presenting different types of explanations such as ``why'', ``why not'' and ``what if'' explanation types and measuring users response time.

\begin{table}[]
\centering
\caption{User satisfaction measures and study methods used in measuring user satisfaction and usefulness of explanations in XAI studies.}
\label{tab:satisfaction-table}
\begin{tabular}{ll}
\toprule
\textbf{Satisfaction Measures} & \textbf{Evaluation Methods} \\ \midrule
\multirow{3}{*}{User Satisfaction} & 
Interview and Self-report (\cite{lim2009assessing, gedikli2014should,lim2009and,bunt2012explanations}) \\ \cmidrule(l){2-2} 
 & Likert-scale Questionnaire (\cite{coppers2018intellingo,lage2019human,lim2009assessing,gedikli2014should,lim2009and}) \\ \cmidrule(l){2-2} 
 & 
 Expert Case Study (\cite{kahng2018cti,strobelt2018lstmvis,liu2016uncertainty,Krause2014,liu2017towards}) \\ \midrule
\multirow{2}{*}{Explanation Usefulness} & Engagement with Explanations (\cite{coppers2018intellingo}) \\ \cmidrule(l){2-2} 
 & Task Duration and Cognitive Load (\cite{lim2009and,lage2019human,gedikli2014should}) \\ \bottomrule
\end{tabular}
\end{table}


Another line of research studies whether intelligible systems are always appreciated by the users or it has a conditional value.
An early work from Lim and Dey~\cite{lim2009assessing} studied user understanding and satisfaction of different explanation types in four real-world context-aware applications.
Their findings show that, when considering scenarios involved with criticality, users want more information explaining the decision making process and experience higher levels of satisfaction after receiving these explanations.
Similarly, Bunt et al.~\cite{bunt2012explanations} considered whether explanations are always necessary for users in every intelligent system.
Their results show that, in some cases, the cost of viewing explanations in diary entries like Amazon and YouTube recommendations could outweigh their benefits.
To study the impact of explanation complexity on users' comprehension, Lage et al.~\cite{lage2019human} studied how explanation length and complexity affect users' response time, accuracy, and subjective satisfaction. 
They also observed that increasing explanation complexity resulted in lowered subjective user satisfaction.
In a recent study, Coppers et al.~\cite{coppers2018intellingo} also show that adding intelligibility does not necessarily improve user experience in a study with expert translators.
Their experiment suggests that an intelligible system is preferred by experts when the additional explanations are not part of the translators readily available knowledge.
In another work, Curran et al. ~\cite{curran2012towards} measured users' understanding and preference of explanations in an image recognition task by ranking and coding user transcripts.
They provide three types of instance explanations for participants and show that although all explanations were coming from the same model, participants had different levels of trust in explanations' correctness, according to explanations clarity and understandability.

Table~\ref{tab:satisfaction-table} summarizes the study methods used to measure user satisfaction and usefulness of machine learning explanations.
Note that the primary goal of XAI system evaluations for domain and AI experts is through direct evaluation of user satisfaction of explanation design during the design cycle. 
For example, case studies and participatory design are common approaches for directly including expert users as part of the system design and evaluation processes.

\subsection{M3: User Trust and Reliance}
\label{user_trust_section}

User trust in an intelligent system is an affective and cognitive factor that influences positive or negative perceptions of a system~\cite{madsen2000measuring, hoffman2013trust}.
Initial user trust and the development of trust over time have been studied and presented with different terms such as \textit{swift} trust~\cite{meyerson1996swift}, \textit{default} trust \cite{merritt2013trust} and \textit{suspicious} trust \cite{bobko2014construct}.
Prior knowledge and beliefs are important in shaping the initial state of trust; however, trust and confidence can change in response to exploring and challenging the system with edge cases \cite{hoffman2014myths}.
Therefore, the user may have different feelings of trust and mistrust during different stages of experience with any given system.

\begin{table}[]
\centering
\caption{Evaluation measures and methods used in measuring user trust in XAI studies.}
\label{tab:trust-table}
\begin{tabular}{@{}ll@{}}
\toprule
\textbf{Trust Measures} & \textbf{Evaluation Methods} \\ \midrule
\multirow{2}{*}{Subjective Measures} & Self-explanation and Interview (\cite{cahour2009does,bussone2015role}) \\ \cmidrule(l){2-2} 
 & Likert-scale Questionnaire (\cite{cahour2009does,Berkovsky:2017:RUT:3025171.3025209,nourani2019effects,bussone2015role}) \\ \midrule
\multirow{3}{*}{Objective Measures} & User Perceived System Competence (\cite{yin2019understanding,pu2006trust,nourani2019effects}) \\ \cmidrule(l){2-2} 
 & User Compliance with System (\cite{eiband2019impact}) \\ \cmidrule(l){2-2} 
 & User Perceived Understandability (\cite{yin2019understanding,nothdurft2014probabilistic}) \\ \bottomrule
\end{tabular}
\end{table}


Researchers define and measure trust in different ways. 
User knowledge, technical competence, familiarity, confidence, beliefs, faith, emotions, and personal attachments are common terms used to analyze and investigate trust~\cite{madsen2000measuring, jian2000foundations}. 
For these outcomes, user trust and reliance can be measured by explicitly asking about user opinions during and after working with a system, which can be done through interviews and questionnaires. 
For example, Ming et al.~\cite{yin2019understanding} studied the importance of model accuracy on user trust. 
Their findings show that user trust in the system was affected by both the system's stated accuracy and users' perceived accuracy over time.
Similarly, Nourani et al.~\cite{nourani2019effects} explored how explanation inclusion and level of meaningfulness would affect the user's perception of accuracy. 
Their controlled experiment results show that whether explanations are human-meaningful can significantly affect perception of system accuracy independent of the actual accuracy observed from system usage.
Additionally, trust assessment scales could be specific to the systems application context and XAI design purposes. 
For instance, multiple scales would assess user opinion on systems reliability, predictability, and safety separately. 
Related to this, a detailed trust measurement setup is presentation in the paper by Cahour and Forzy~\cite{cahour2009does}, which measures user trust with multiple trust scales (constructs of trust), video recording, and self-confrontation interviews to evaluate three modes of system presentation.
Also, to better understand factors that influence trust in adaptive agents, Glass et al.~\cite{glass2008toward} studied which types of questions users would like to be able to ask an adaptive assistant.
Others have looked at changes to user awareness over time by displaying system confidence and uncertainty of the machine learning outputs in applications with different degrees of criticality~\cite{antifakos2005towards,kay2016ish}.

Multiple efforts have studied the impact of XAI on developing justified trust in users in different domains.
For instance, Pu and Chen~\cite{pu2006trust} proposed an organizational framework for generating explanations and measured perceived competence and user's intention to return as the measures for user trust. 
Another example compared user trust with explanations for different goals like transparency and justification explanation~\cite{nothdurft2014probabilistic}.
They considered perceived understandability to measure user trust and show that transparent explanations can help reduce the negative effects of trust loss in unexpected situations.

Studying user trust in real-world applications, Berkovsky et al.~\cite{Berkovsky:2017:RUT:3025171.3025209} evaluated trust with various recommendation interfaces and content selection strategies.
They measured user reliance on a movie recommender system with six distinct constructs of trust.
Also on recommender algorithms, Eiband et al. \cite{eiband2019impact} repeats the Langer et al.'s experiment~\cite{langer1978mindlessness} on the role of ``placebic'' explanations (i.e., explanations that convey no information) in mindlessness of user behavior.
They studied if providing placebic explanations would increase user reliance on the recommender system. 
Their results suggest that future work on explanations for intelligent systems may consider using placebic explanations as a baseline for comparison with machine learning generated explanations.
Also concerned with expert user's trust, Bussone et al. \cite{bussone2015role} measured trust by Likert-scale and think-alouds and found that explanations of facts lead to higher user trust and reliance in a clinical decision-support system.
Table~\ref{tab:trust-table} summarizes a list of subjective and objective evaluation methods to measure user trust in the machine learning systems and their explanations. 

Many studies evaluate user trust as a static property. 
However, it is essential to take user's experience and learning over time into account when working with complex AI systems. 
Collecting repeated measures over time can help in understanding and analyzing the trend of users' developing trust with the progression of experience.
For instance, in their study, Holliday et al.~\cite{holliday2016user} evaluated trust and reliance in multiple stages of working with an explainable text-mining system.
They showed that the level of user trust in the system varied over time as the user gained more experience and familiarity with the system.

We note that although our literature review did not find a direct measurement of trust to be commonly prioritized in analysis tools for data and machine learning experts, users' reliance on tools and the tendency to continue using tools are often considered as a part of the evaluation pipeline during case studies.
In other words, our summarization is not meant to claim that data experts do not consider trust, but rather we did not find it to be a core outcome explicitly measured in the literature for this user group.

\subsection{M4: Human-AI Task Performance}
\label{task_performance_section}

A key goal of XAI is to help end-users to be more successful in their tasks involving machine learning systems~\cite{hook2000steps}.
Thus, human-AI task performance is a measure relevant to all three groups of user types.
For example, Lim et al.~\cite{lim2009and} measured users' performance in terms of success rate and task completion time to evaluate the impact of different types of explanations.
They use a generic interface that can be applied to various types of sensor-based context-aware systems, such as weather prediction.
Further, explanations can assist users in adjusting the intelligent system to their needs.
Kulesza et al.~\cite{Kulesza2012Tell} study of explanations for a music recommender agent found a positive effect of explanations on users' satisfaction with the agent's output, as well as on users' confidence in the system and their overall experience.

Another use case for machine learning explanations is to help users judge the correctness of system output~\cite{groce2014you, Krause2017workflow,stumpf2009interacting}.
Explanations also assist users in debugging interactive machine learning programs for their needs~\cite{kulesza2015principles, kulesza2010explanatory}.
In a study of end-users interacting with an email classifier system, Kulesza et al.~\cite{kulesza2015principles} measured classifier performance to show that explanatory debugging benefits user and machine performance. 
Similarly, Ribeiro et al.~\cite{ribeiro2016should} found users could detect and remove wrong explanations in text classification, resulting in training better classifiers with higher performance and explanations quality.
To support these goals, Myers et al.~\cite{myers2006answering} designed a framework that users can ask \textit{why} and \textit{why not} questions and expect explanations from the intelligent interfaces. 
Table~\ref{tab:performance-table} summarizes a list of evaluation methods to measure task performance in human-AI collaboration and model tuning scenarios.

\begin{table}[]
\centering
\caption{Evaluation measures and methods used in measuring human-machine task performance in XAI studies.}
\label{tab:performance-table}
\begin{tabular}{ll}
\hline
\textbf{Performance Measures} & \textbf{Evaluation Methods} \\ \hline
\multirow{3}{*}{User Performance} & Task Performance (\cite{kulesza2010explanatory,lim2009and,kahng2018cti,groce2014you}) \\ \cline{2-2} 
 & Task Throughput(\cite{kulesza2010explanatory,lim2009and,lakkaraju2016interpretable}) \\  \cline{2-2} 
 & Model Failure Prediction (\cite{groce2014you,Krause2017workflow,stumpf2009interacting}) \\ \hline
\multirow{2}{*}{Model Performance} & Model Accuracy (\cite{ribeiro2016should,kulesza2015principles,stumpf2009interacting,liu2017towards,pezzotti2018deepeyes}) \\ \cline{2-2} 
 & Model Tuning and Selection (\cite{liu2014topicpanorama}) \\ \hline
\end{tabular}
\end{table}


Visual analytics tools also help domain experts to better perform their tasks by providing model interpretations.
Visualizing model structure, details, and uncertainty in machine outputs can allow domain experts to diagnose models and adjust hyper-parameters to their specific data for better analysis.
Visual analytics research has explored the need for model interpretation in text~\cite{wise1995visualizing, hu2014interactive, liu2016uncertainty} and multimedia~\cite{choo2010ivisclassifier, bryan2013efficient} analysis tasks. This body of work demonstrates the importance of integrating user feedback to improve model results.
An example of a visual analytics tool for text analysis is TopicPanaroma~\cite{liu2014topicpanorama}, which models a textual corpus as a topic graph and incorporates machine learning and feature selection to allow users to modify the graph interactively.
In their evaluation procedure, they ran case studies with two domain experts: a public relations manager used the tool to find a set of tech-related patterns in news media, and a professor analyzed the impact of news media on the public during a health crisis. 
In analysis of streaming data, automated approaches are error-prone and require expert users to review model details and uncertainty for better decision making~\cite{rudolph2009finvis,best20147}.
For example, Goodall et al.~\cite{goodall2018situ} presented Situ, a visual analytics system for discovering suspicious behavior in cyber network data. 
The goal was to make anomaly detection results understandable for analysts, so they performed multiple case studies with cybersecurity experts to evaluate how the system could help users to improve their task performance.
Ahn and Lin~\cite{ahn2019fairsight} present a framework and visual analytic design to aid fair data-driven decision making. 
They proposed FairSight, a visual analytic system to achieve different notions of fairness in ranking decisions through visualizing, measuring, diagnosing, and mitigating biases.

Other than domain experts using visual analytics tools, machine learning experts also use visual analytics to find shortcomings in the model architecture or training flaws in DNNs to improve the classification and prediction performance~\cite{liu2017towards, pezzotti2018deepeyes}.
For instance, Kahng et al.~\cite{kahng2018cti} designed a  system to visualize instance-level and subset-level of neuron activation in a long-term investigation and development with machine learning engineers.
In their case studies, they interviewed three machine learning engineers and data scientists who used the tool and reported the key observations.
Similarly, Hohman et al.~\cite{hohman2019summit} present an interactive system that scalably summarizes and visualizes what features a DNN model has learned and how those features interact in instance predictions.
Their visual analytic system presents activation aggregation to discover important neurons and neuron-influence aggregation to identify interactions between important neurons.
In the case of recurrent neural networks (RNN), LSTMVis~\cite{strobelt2018lstmvis} and RNNVis~\cite{ming2017understanding} are tools to interpret RNN models for natural language processing tasks.
In another recent paper, Wang et al.~\cite{wang2019visual} presented DNN Genealogy, an interactive visualization tool that offers a visual summary of DNN representations.

Another critical role of visual analytics for machine learning experts is to visualize model training processes~\cite{zhong2017evolutionary}.
An example of a visual analytics tool for diagnosing the training process of a deep generative model is DGMTracker~\cite{liu2018analyzing}, which helps experts understand the training process by visually representing training dynamics.
An evaluation of DGMTracker was conducted in two case studies with experts to validate efficiency of the tool in supporting understanding of the training process and diagnosing a failed training process.

\subsection{M5: Computational Measures}
\label{sec:explanation-truthfulness}

Computational measures are common in the field of machine learning to evaluate interpretability techniques' correctness and completeness in terms of explaining what the model has learned. 
Herman~\cite{herman2017promise} notes that reliance on human evaluation of explanations may lead to persuasive explanations rather than transparent systems due to user preference for simplified explanations.
Therefore, this problem leads to the argument that explanations' fidelity to the black-box model should be evaluated by computational methods instead of human subject studies. 
Fidelity of an ad-hoc explainer refers to the correctness of the ad-hoc technique in generating the true explanations (e.g., correctness of a saliency map) for model predictions.
This leads to a series of computational methods to evaluate correctness of generated explanations, consistency of explanation results, and fidelity of ad-hoc interpretability techniques to the original black-box model~\cite{robnik2018perturbation}.

\begin{table}[]
\centering
\caption{Evaluation measures and methods used for evaluating fidelity of interpretability techniques and reliability of trained models. This set of evaluation methods is used by machine learning and data experts to eighter evaluate the correctness of interpretability methods or evaluate the training quality trained models beyond standard performance metrics.}
\label{tab:computational-table}
\begin{tabular}{@{}ll@{}}
\toprule
\textbf{Computational Measures} & \textbf{Evaluation Methods} \\ \midrule
\multirow{3}{*}{\begin{tabular}[c]{@{}l@{}}Explainer Fidelity\end{tabular}} & Simulated Experiments (\cite{ribeiro2018anchors, ribeiro2016should}) \\ \cmidrule(l){2-2} 
 & Sanity Check (\cite{olah2018the, zahavy2016graying, zintgraf2017visualizing, yosinski2015understanding,kindermans2017reliability, ross2017improving} \\ \cmidrule(l){2-2} 
 & Comparative Evaluation (\cite{samek2017evaluating, ijcai2017-371}) \\ \midrule
\multirow{2}{*}{Model Trustworthiness} & Debugging Model and Training (~\cite{zeiler2014visualizing}) \\ \cmidrule(l){2-2} 
 & Human-Grounded Evaluation (\cite{mohseni2018human, lundberg2017unified, schmidt2019quantifying, das2017human}) \\ \bottomrule
\end{tabular}
\end{table}


In many cases, machine learning researchers often consider consistency in explanation results, computational interpretability, and qualitative self-interpretation of results as evidence for explanation correctness~\cite{olah2018the, zahavy2016graying,zintgraf2017visualizing,yosinski2015understanding}.
For example, Zeiler and Fergus~\cite{zeiler2014visualizing} discuss fidelity of the visualization for CNN network by its validity in finding model weaknesses resulted in improved prediction results. 
In other cases, comparing a new explanation technique with existing state-of-the-art explanation techniques is is used to verify explanation quality~\cite{chu2018exact,shrikumar2017learning, lundberg2017unified}.
For instance, Ross et al.~\cite{ijcai2017-371} designed a set of empirical evaluations and compared their explanations' consistency and computational cost with the LIME technique~\cite{ribeiro2016should}.
In a comprehensive setup, Samek et al.~\cite{samek2017evaluating} proposed a framework for evaluating saliency explanations for image data that quantify the feature importance with respect to the classifier prediction. 
They compared three different saliency explanation techniques for image data (sensitivity-based~\cite{simonyan2013deep}, deconvolution ~\cite{zeiler2014visualizing}, and layer-wise relevance propagation~\cite{bach2015pixel}) and investigated the correlation between saliency map quality and network performance on different image datasets under input perturbation. 
On the contrary, Kindermans et al.~\cite{kindermans2017reliability} show interpretability techniques have inconsistencies on simple image transformations, hence their saliency maps can be misleading.
They define an input invariance property for reliability of explanations from saliency methods. 
To extend a similar idea, Adebayo et al.~\cite{adebayo2018sanity} propose three tests to measure adequacy of interpretability techniques for tasks that are sensitive to either data or the model itself.

Other evaluation methods include assessing explanation's fidelity in comparison to inherently interpretable models (e.g., linear regression and decision trees).  
For example, Ribeiro et al.~\cite{ribeiro2016should} compared explanations generated by the LIME ad-hoc explainer to explanations from an interpretable model. 
They created gold standard explanations directly from the interpretable models (sparse logistic regression and decision trees) and used these for comparisons in their study. 
A downside of this approach is that the evaluation is limited to generating a gold standard by an interpretable model.
User simulated evaluation is another method to perform computational evaluation of model explanations.
Ribeiro et al.~\cite{ribeiro2016should} simulated user trust in explanations and models by defining ``untrustworthy'' explanations and models. 
They tested a hypothesis on how real users would prefer more reliable explanations and choose better models. 
The authors later repeated similar user simulated evaluations in the \textit{Anchors} explanation approach~\cite{ribeiro2018anchors} to report simulated users' precision and coverage in finding the better classifier by only looking at explanations.

A different approach in quantifying explanations quality with human intuition has been taken by Schmidt and Biessmann~\cite{schmidt2019quantifying} by defining an explanation quality metric based on user task completion time and agreement of predictions.
Another example is the work of Lundberg and Lee~\cite{lundberg2017unified}, who compared the SHAP ad-hoc explainer model with LIME and DeepLIFT~\cite{shrikumar2017learning} with the assumption that good model explanations should be consistent with the explanations from humans who understand the model. 
Lertvittayakumjorn and Toni~\cite{Lertvittayakumjorn2019human} also present three user tasks to evaluate local explanation techniques for text classification through revealing model behavior to human users, justifying the predictions, and helping humans investigate uncertain predictions.
A similar idea has been implemented in~\cite{mohseni2018human} by feature-wise comparison of a ground-truth and model explanation. 
They provide a user annotated benchmark to evaluate machine learning instance explanations.
Later, Poerner et al.~\cite{poerner2018evaluating} use this benchmark as human annotated ground truth in comparison to small-context (word level) and large-context (sentence level) explanation evaluation. 
User annotated benchmarks can be valuable when considering human meaningfulness of explanations, though the discussion by Das et al.~\cite{das2017human} implies that machine learning models (visual question answering attention models in their case) do not seem to look at the same regions as humans. 
They introduce a human-attention dataset~\cite{VQA-HAT} (collection of mouse-tracking data) and evaluate attention maps generated by state-of-the-art models against human.

Interpretability techniques also enable quantitative measures for evaluating model trustworthiness (e.g., model fairness, reliability, and safety) through its explanations.
Trustworthiness of a model represents a set of domain specific goals such as fairness (by fair feature learning), reliability, and safety (by robust feature learning).
For example, Zhang et al.~\cite{zhang2018examining} present a case of using machine learning explanations to find representation learning flaws caused by potential biases in the training dataset. 
Their technique mines the relationships between pairs of attributes according to their inference patterns.
Further, Kim et al.~\cite{kim2018interpretability} presented quantitative testing of machine learning models by their explanations. 
In their concept activation vector technique, the model can be tested for specific concepts (e.g., image patterns) and a vector score shows if the model is biased toward that concept.
They later extended their concept-based global explanation of model representation learning for systematic discovery of concepts that are human meaningful and important for the model prediction~\cite{ghorbani2019towards}.
They used human subject experiments to evaluate learned concepts.
Table~\ref{tab:computational-table} summarizes a list of evaluation methods to measure fidelity of interpretability techniques and model trustworthiness with computational techniques.

\section{XAI Design and Evaluation Framework}

The variety of different XAI design goals (Section~\ref{sec:d_goal_section}) and evaluation methods (Section~\ref{sec:e_measure_section}) from our review suggests the need for diverse sets of techniques to build end-to-end XAI systems.
However, it is generally insufficient to take design practices and evaluation methods separately.
A holistic and more actionable vantage will require consideration of dependencies between design goals and evaluation methods and will inform when to choose between them during the design cycles. 
Previously, various models and guidelines for the design and evaluation of AI-infused interactive user interfaces~\cite{dudley2018review,amershi2019guidelines} and visual analytics systems~\cite{munzner2009nested} have been proposed to help designers through the design process.
However, challenges in generating useful machine learning explanations and presenting them through an appropriate interface that aligns with target outcomes call for a multidisciplinary workflow framework.

\begin{figure*}[t]
\centering
  \includegraphics[width=0.65\columnwidth]{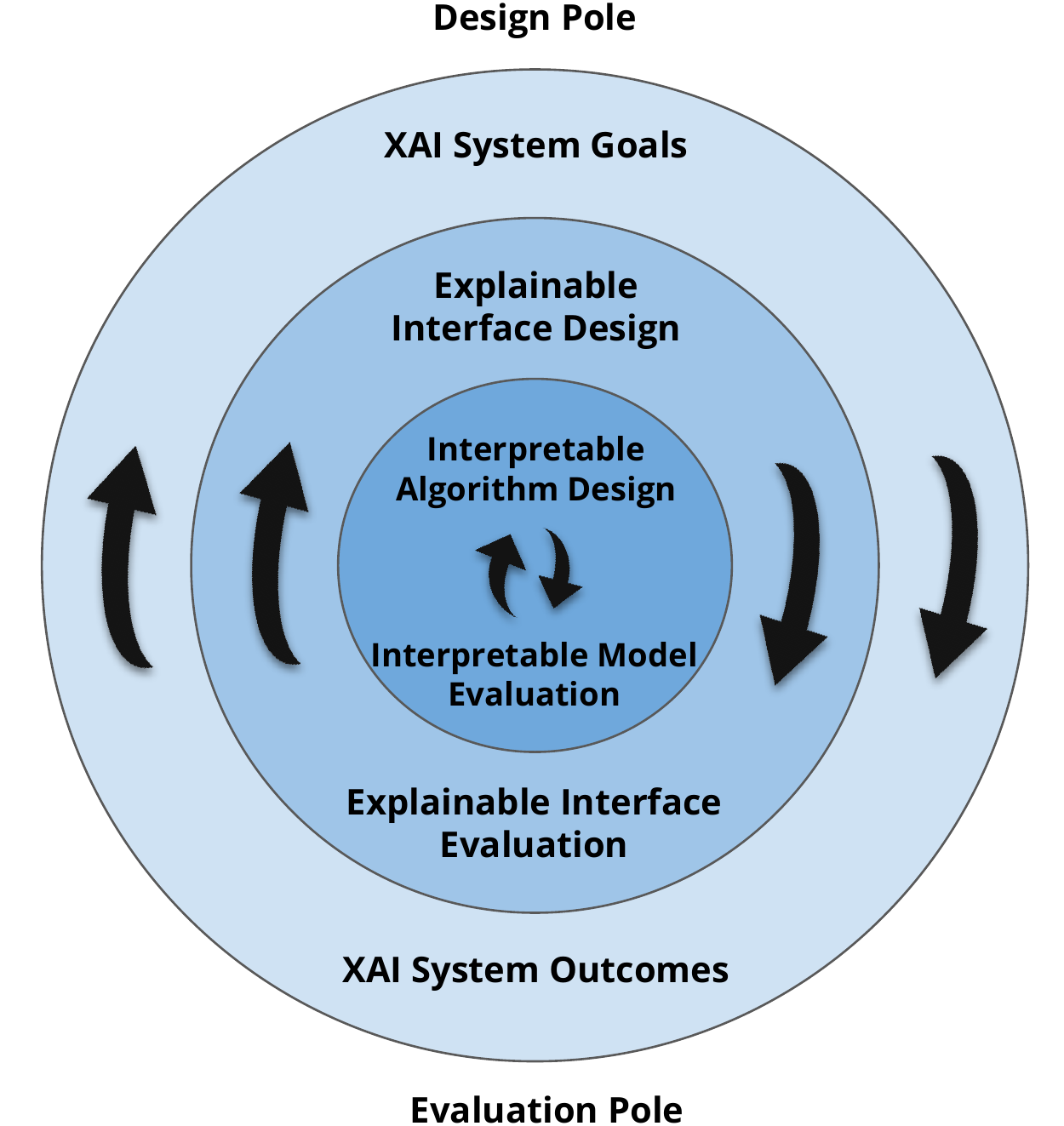}
  \caption{XAI Design and Evaluation Framework: our nested model for design and evaluation of explainable machine learning systems.
  The \textit{outer layer} demonstrates system-level design goals which are paired with evaluation of high-level XAI outcomes.
  The \textit{middle layer} shows explainable user interface and visualization design step paired with appropriate user understandability and satisfaction evaluation measures.
  The \textit{innermost layer} presents design and evaluation of trustworthy interpretable machine learning algorithms.
  }
  \label{fig:evaluation}
\end{figure*}

Thus, based on our analysis of prior work, we propose a design and evaluation framework for XAI systems.
The impetus for this framework is the desire to organize and relate the diverse set of existing design goals and evaluation methods in a unified model.
The framework is intended to give guidance on what evaluation measures are appropriate to use at which design stage of the XAI system. 
Figure~\ref{fig:evaluation} summarizes the framework as a nested model for end-to-end XAI system design and evaluation. 
The formulation of the model as layers relates to the core design goals and evaluation interests from the different research communities (as identified from the literature review) to help promote interdisciplinary progress in XAI research. 
The model is structured to support system design steps by starting from the outer layer (\textit{XAI System Goals}), then addressing end-user needs in the middle layer (\textit{Explainable Interface}), and finally focusing on underlying interpretable algorithms in the innermost layer (\textit{Interpretable Algorithms}). 
The nested model is organized with a \textit{Design Pole} focusing on design goals and choices, and an \textit{Evaluation Pole} presenting appropriate evaluation methods and measures for each layer.
Our framework suggests iterative cycles of design and evaluation to cover both algorithmic and human-related aspects of XAI systems.
In this section, we elaborate on details of the nested framework and provide guidelines on using it for multidisciplinary XAI system design.\\

\noindent \textbf{Case Study Example:} 
To showcase a practical example of using the framework, we also include a case study of a collaborative design and development effort for an XAI system. 
In the scenario of the case study, a multidisciplinary team of researchers designed a XAI system for fake news detection for non-expert (not AI experts or news analysts) daily newsreaders. 
The design team planned to add a \textit{XAI Assistant} feature to a news reading and sharing website to perform fake news detection. 
The system design consisted of a news reading interface equipped with the XAI news assistant (news assistant) to help the user identify fake news while reviewing news stories and articles. 
The presented case is the result of an ongoing research done over a one-year period by a team of eight university researchers with HCI, Visualization, and AI backgrounds. 
During the following subsections, each framework guideline is followed by an example of its application in our case study.

\subsection{XAI System Goals Layer}
\label{sec:xai-system-goal}
As designers in a multidisciplinary team have different roles and priorities in building a XAI system, we suggest beginning the system design cycle from the \textit{XAI Goal} layer (the outer layer of Figure~\ref{fig:evaluation}) to characterize design goal and system expectations. 
Specifically, this step involves identifying the~\textit{purpose for explanation} and choosing \textit{what to explain} for the targeted end-user and dedicated application. 
The iterative refinements between XAI goal (top pole) and system evaluation (bottom pole) present how the paired evaluation measures help to improve system design. 
We organize the following guidelines for the XAI goal layer.

At the beginning of the system design process, the team will need to specify explainability requirements for each framework layer based on the evaluation metrics. 
The explainability requirements are intended to satisfy overall system goals defined by user (or customer) needs, and sometimes regulations, laws, and safety standards. 
Later, the evaluation step in each design cycle will have the team revisit the initial XAI system requirements. 
The sufficiency of the evaluation results in comparison to the initial design requirements serves as a key indicator of whether to stop or continue design iteration.
However, since many subjective measures are used in the process, it is important to choose an appropriate evaluation baseline (see Section~\ref{sec:eval-baseline}) to track progress during design cycles.
\newline

\noindent \textbf{Guideline 1: Determine XAI System Goals: } 
Identifying and establishing clear goals and expectations from XAI system is the first step in the design process.
XAI Design goals could be driven by many motivations like improving user experience on an existing system, advancing scientific findings~\cite{krause2016interacting, Lex2012}, or adhering to new regulations~\cite{tikkinen2018eu}.
In Section~\ref{sec:d_goal_section} we reviewed eight main goals (G1-G8) for XAI systems.
Also, ordering the priority of goals in cases with multiple design goals can be beneficial in the next steps of the process (see Guideline 2).
Given the fact that different XAI user types and applications are interested in various design goals, it is important to establish these goals early in the design process to identify and align with appropriate design principles.
A pitfall in this stage is to pick XAI goals without considering the end-user group, algorithmic limitations, and user preferences in the context of the application.
Overshooting XAI goals could hurt evaluation results moving forward in the design process.
\vspace{0.7em}
\begin{center}
\framebox{
\parbox{0.9\textwidth}{\textit{Application in Case Study:} In the first step of our case study with a news curation application, the team started with identifying the main goals and expectations for the XAI news assistant.
The design focused on novice end-users without any particular expertise.
The XAI design goal was to improve user reliance and mental model of news predictions through explainable design.
The team hypothesized that end-users would trust and rely on the fake news detection assistant, given that the new XAI is capable of providing explanations for each news story. 
Also, the team hoped that users would be able to use the explanations to learn model weaknesses and strengths to provide feedback to the developer team.}}\par
\end{center}
\vspace{1em}

\noindent \textbf{Guideline 2: Decide What to Explain: } 
The second step in the XAI system design is to identify ``what to explain'' to the user in order to achieve the initial XAI goals (see Guideline 1) of the system.
We reviewed multiple machine learning interpretability techniques and explanation types in Sections~\ref{sec:global-and-local},~\ref{sec:ad-hoc-explainers}, and ~\ref{sec:what_to_explain} which can provide different types of information to the user.
Although theory-driven design frameworks discuss explanation mechanisms driven by human reasoning semantics~\cite{lim2019these}, user-centered methods to identify useful explanations such as online surveys, interviews, and user observations (e.g.,~\cite{mennicken2014today,bussone2015role}) to understand \textit{when} and \textit{what} needs to be explained for the users to understand better and trust intelligent systems.
Preliminary experiments are valuable in the early steps of the design cycle to identify and narrow down explanation options for the user in order to satisfy design goals.
A typical approach for evaluating the effectiveness and usefulness of explanation choice in user-centric experiments is to compare the user's mental model of the system with and without explanation components.
On this subject, Lim and Dey~\cite{lim2009assessing} conducted experiments to discover what type of information users are interested in different real-world context-aware application scenarios.
Stumpf et al.~\cite{stumpf2018explaining} also performed end-user interviews to identify user perceptions and expectations from an interpretable interface as well as finding main decision points where users may need explanations.
In another work, Haynes et al.~\cite{haynes2009designs} provide a review and studies incorporating different explanations (operational, ontological, mechanistic, and design rationale explanations) in intelligent systems. 
Similarly, visual analytics design involves expert interviews and focus groups in the design path to identify design goals~\cite{munzner2009nested}.

The design process in this step involves algorithmic implementation constraints like ``what can be explained'' to the user.
For example, global explanations from a DNN may not feasible and comprehensible due to the large number of variables in the graph. 
Additionally, research shows instance explanations from a DNN lack completeness and may fail to present salient features in cases~\cite{adebayo2018sanity}.
Such constraints and decision points could be solved through focused groups, brainstorming, and interviews between model designers and interface designers in the team.
Therefore, a design pitfall for explanation choices is not to take limitations of interpretability techniques into account.

\vspace{0.7em}
\begin{center}
\framebox{
\parbox{0.9\textwidth}{\textit{Application in Case Study:} In our scenario, efficient news curation required fake news detection with the help of our XAI assistant.
In the analysis of what the system should explain, the design team decided to identify candidate useful and impactful explanation options.
We started with reviewing machine learning research on false information (e.g., rumor, hoax, fake news, clickbait) detection as well as HCI research on news feeds and news search systems to identify key attributes for news veracity checking~\cite{mohseni2019open}. 
Given the non-expert target end-users, explanatory information needed to limit technical details.
Next, the user interface designers and machine learning designers in the team discussed candidate explanation choices and algorithmic constraints in interpretability techniques.
That is, some options for what to explain may not be entirely possible given the interpretability of existing models, and the team needed to consider whether alternative learning techniques could provide better explanations or if the design team would need to figure out meaningful ways to explain the information that was available from the model.}}\par
\end{center}
\vspace{1em}

\noindent \textbf{Guideline 3: Evaluate System Outcomes: } 
Evaluation of XAI system outcomes is the final step in the evaluation process. 
Figure~\ref{fig:evaluation} shows how the final system outcome evaluation is paired with the initial design goals in the outer layer of our framework. 
The main goal of this stage is to quantitatively and qualitatively assess the effectiveness of the XAI system for the initially established system-level XAI goals. 
Clearly, evaluation of final system outcomes could be influenced by the design of the explainable user interface (intermediate layer) and the design of interpretable algorithms (innermost layer).
For example, evaluating a newborn interpretable machine learning algorithm's output using human subjects through a weak in-lab or crowdsourced user study may not be meaningful or productive for XAI system outcomes if core computational changes are still in progress and could ultimately change the entire model interpretability and explanation format later.
Also, changes in the targeted user could affect evaluation results at this stage.
For example, a system designed for novices may not satisfy the needs of an expert user and hence would not improve performance as expected.
Evaluation measures in this layer depend on the design goals, application domain, and targeted users. 
Example evaluation measures for final system outcomes include user trust~\cite{pu2006trust} and reliance on the system~\cite{Berkovsky:2017:RUT:3025171.3025209}, human-machine task performance~\cite{bansal2019beyond}, user awareness~\cite{kay2016ish}, and user understanding of their personal data~\cite{rader2018explanations}.
An effective process for evaluation of high-level XAI outcomes is to break down the evaluation goal into multiple well-defined measures and metrics.
This way, the team can perform evaluation studies on different steps using valid methods in controlled setup.
For example, in the evaluation of XAI systems for trustworthiness, several factors of human trust could be measured during and after a period of user experience with the XAI system. 
In addition, computational measures (Section~\ref{sec:explanation-truthfulness}) are used to examine the fidelity of interpretability methods and trustworthiness of the model with objective metrics.
A possible pitfall in evaluation of the XAI system outcomes is performing the evaluation without considering the model trustworthiness and explanations' correctness from the interpretable model layer (see Guideline 7) and explanation understandability and usefulness from the user interface layer (see Guideline 5).

\vspace{0.7em}
\begin{center}
\framebox{
\parbox{0.9\textwidth}{\textit{Application in Case Study:} In our case study with news review and curation, we needed to evaluate our XAI news assistant with non-expert users who would gather news stories while flagging fake news articles.
In the evaluation step, the team ran multiple large-scale human-subject studies with novice participants recruited through Amazon Mechanical Turk to work with our news reading system. 
Note that both the explainable interface and interpretable algorithm passed multiple design and testing iterations before this evaluation step. 
Major decisions for this evaluation was how to structure the duration and complexity of the user task while appropriately testing the system's full range of functionality.
The task was designed with questions built in to help collect subjective data in addition to the objective user performance data.
Multiple evaluation measures were chosen for system outcomes, including: subjective user trust in the news assistant, user agreement rate with the news assistant, veracity of user-shared news stories, and user accuracy in guessing the news assistant output.
Both qualitative and quantitative analysis of user feedback and interaction data were valuable to the evaluation of system outcomes.
The results and analysis from these evaluations helped the team to understand the effectiveness of the XAI elements (in both the algorithm and the interface) for the initial system goals.}}\par
\end{center}
\vspace{1em}

\subsection{User Interface Design Layer}
\label{sec:interface-design-layer}
The middle layer of our framework is concerned with designing and evaluating an explainable interface or visualization for the user to interact with XAI system. 
Interface design for explanations consists of presenting model explanations from interpretable algorithms to end-users in terms of their \textit{explanation format} and \textit{interaction design}. 
The importance of this layer is to satisfy design requirements and needs to be determined in the XAI system design layer (see Guideline 2).
An elegant translation of machine-generated explanations (e.g., verbal, numeric, or visual explanation) needs carefully designed human-understandable and satisfying explanations in the user interface.
In Section~\ref{sec:how_to_explain}, we reviewed multiple types of explanation formats for integrating XAI elements into the user interface. 
The iterative movement between \textit{Design pole} and \textit{Evaluation pole} in this layer presents design refinement in pursuit a desired goal state.
\newline

\noindent \textbf{Guideline 4: Decide How to Explain: }
Identifying candidate explanation formats for the targeted system and user group is the first step to deliver machine learning explanations to end-users.
The design process can account for different levels of complexity, length, presentation state (e.g., permanent or on-demand), and interactivity options depending on the application and user type. 
The explanations format in the interface is particularly important to improve user understanding of underlying algorithms. 
Studies show that while detailed and complex interactive representations may aim to communicate the explanations to the expert users, AI-novice users of XAI system prefer more simplified explanation and representation interfaces~\cite{lage2019human}.
User satisfaction of interface design is also another critical factor in user engagement of the interface components~\cite{muir1987trust}.
Additionally, interaction design for explainable interfaces can allow a user to communicate with the system to adjust explanations and could better support user inspection of the system~\cite{kulesza2010explanatory}.
Research of intelligent interface design presents multiple design methods such as wireframing and low-fidelity prototyping (e.g.,~\cite{mennicken2014today,bussone2015role}) that could also be adapted to the explainable interface design. 
Also, existing design guidelines and best-practice knowledge for AI-infused interfaces (e.g., \cite{amershi2019guidelines}) and visualizations (e.g., ~\cite{meyer2015nested}) could be used in this stage to leverage similar systems for explainable interface design. 
Aside from model explanations, providing prediction uncertainty also has been identified as an important factor for both general end-users and data expert users~\cite{sacha2016role}. 
For example, Kay et al.~\cite{kay2016ish} presented the full design cycle for an uncertainty visualization interface in a bus arrival time application. 
Their design process included surveying to identify usage requirements, developing alternative layouts, running user testing, and final evaluation of user understanding of machine learning output.

\vspace{0.7em}
\begin{center}
\framebox{
\parbox{0.9\textwidth}{\textit{Application in Case Study:} To determine \textit{how} to explain news classification results to non-expert end users, the user interface design team started the process by reviewing the initial system goals and explanation types.
The team then continued with multiple interface sketches that matched the intended application and user tasks. 
During the initial design steps, the team tried to keep a balance between interface complexity and explanation usefulness by choosing among available explanation types from our interpretable machine learning algorithms.
Next, mock-ups from the top three designs were implemented for testing with a small number of participants. 
Each mock-up had a different arrangement of data, user task flow, and explanation format for the news assistant interface.  
Our human-subject experiments in this stage were based on user observations and post-usage interviews to collect qualitative feedback regarding participant understanding and subjective satisfaction of explanation components and interface arrangements. 
Interviews resulted in the selection of the most comprehensible and conclusive design among the available options to continue with (see Guidelines 5).}}\par
\end{center}
\vspace{1em}

\noindent \textbf{Guideline 5: Evaluate Explanation Usefulness: }
This mid-layer evaluation step can be used along with various measures to help assess user understanding of the XAI underlying intelligent algorithms. 
A series of user-centered evaluations of explainable interface with multiple goals and granularity levels could be performed to measure: 

\begin{enumerate}
    \item User understanding of explanation.
    \item User satisfaction of explanation.
    \item User mental model of the intelligent system.
\end{enumerate}

Evaluations in the middle layer are particularly important due to the impact on XAI system outcomes (outer layer) and being affected by interpretable model outputs (inner-most layer). 
Specifically, evaluation measures in this stage can inform how well users understand the interpretable system, however, the design validity at this step also may be reflected by higher-level XAI outcomes (i.e., outer-layer evaluation) such as user trust and task performance. 
Note that user understanding of an XAI system could be limited to parts of the system rather than the entire system; similarly, understanding may be limited to a subspace of scenarios rather than all possible scenarios.

The three evaluation measures introduced for this step could be used on multiple iterative cycles to improve overall explainable interface design.
For example, Saket et al.~\cite{saket2017evaluating} studies users understanding of visualization encoding and effectiveness of interactive graphical encoding for end-user. 
On the other hand, user satisfaction of explanation type and format depends on factors such as targeted application criticality and user-preferred cognitive load~\cite{doshi2017towards}. 
Evaluating user mental model is also an effective way to measure usefulness of explainable interfaces.
Tables~\ref{tab:mental-table} and ~\ref{tab:satisfaction-table} present a list of measures for evaluating explainable interfaces in this step.
The choice of baseline is another important factor in evaluating explainable interfaces. 
Typically, a combination of qualitative and quantitative analysis are used to measure effects of explanation components (in comparison to non-explainable system) or to compare multiple explanations types. 
However, the choice of placebic explanations has been proposed as the evaluation baseline for more accurate measurement of explanation content~\cite{eiband2019impact}. 
In the case of expert review, evaluation of a domain expert's mental model commonly involves comparison with the AI expert's mental model and description of ``how the model works''. 
Section~\ref{sec:eval-baseline} reviews common choices of ground-truth baselines in XAI evaluation studies.
With all approaches, updates in explanation components of the interface require assessment of their impact on user experience and understandability. 
However, the metrics and depth of evaluation vary during the evaluation cycles as the team narrows down specific needs.
Finally, a possible evaluation pitfall for explainable interfaces is going after broad measures of XAI outcomes (See Guideline 3) rather than focusing on a narrower scope of explanation components and interactions.

\vspace{0.7em}
\begin{center}
\framebox{
\parbox{0.9\textwidth}{\textit{Application in Case Study:}
In our case study, interface designers started evaluation of candidate explanation components by a series of small studies with a repeated-measures design so that the same study participant could experience different explanation designs in one session. 
Next, we analyzed quantitative and qualitative data collected from the end-users to choose candidate designs and routes to further improve the interface for explainable components.
Discussions with the machine learning team also helped to find sources of limitations in the interpretability technique that could possibly affect user satisfaction.
After the initial cycles of revision, we collected a round of external and internal expert reviews to update the study methodology and data collection details according to project progress.}}\par
\end{center}
\vspace{1em}

\subsection{Interpretable Algorithm Design Layer}

The innermost layer of our framework involves designing interpretable algorithms that are able to generate explanations for the users.
The last design step in our XAI system framework is the choice of interpretability technique (design pole) to generate the outlined explanation types.
However, evaluating the generated explanation (evaluation pole) is the first evaluation step before human-subject evaluations in the explainable interface.
Ideally, the interpretability techniques should generate explanations in accordance with the requirements in the explainable interface design step (see Guideline 4); however, the choice of interpretability technique depends on  domain and carries implementation limitations. 
For example, while shallow models are desired for their high interpretability, these models typically do not perform well in cases of complex and high dimensional data like image and text.
On the other hand, highly accurate predictions in black-box models (e.g., deep neural networks and random forest models) require post-processing and ad-hoc algorithms to generate explanations. 
The ad-hoc approach also has limitations on both choice of explanation type and need for completeness~\cite{adebayo2018sanity} and fidelity~\cite{ribeiro2016should} validation compared to the original model.
This shows not only machine learning designers should consider the trade-off between model interpretability and performance but also should consider the fidelity of the ad-hoc explainer to black-box model. 
We suggest two following design and evaluation guidelines for this layer: 
\newline

\textbf{Guideline 6: Design Interpretability Technique: }
Designing interpretable decision-making algorithms starts with the choice of machine learning model.
Shallow machine learning models (e.g., linear models and decision trees) have intrinsic interpretability due to low number of variables and model simplicity.
For more complex models (e.g., random forest and DNN), ad-hoc explainer technique (see Section~\ref{sec:ad-hoc-explainers}) are needed to generate explanations.
However, the choice of machine learning model (i.e., shallow vs. deep) is bounded by model's performance on data domain.
Secondly, ad-hoc explainer techniques have certain limitations in their explanation type.
The importance of choosing the right combination of model and explainer is in their impact on providing useful  (See Guideline 4) and trustworthy explanations for end-users.

Machine learning research has proposed various ad-hoc explainers to generate ``Why'' explanations (e.g., feature attribution~\cite{kim2018interpretability,lundberg2017unified}), ``How'' explanations (e.g., rules list~\cite{wang2015falling,letham2015interpretable}), ``What else'' explanation (e.g., similar training instance~\cite{maaten2008visualizing,kim2016examples}), and ``What if'' (e.g., sensitivity analysis~\cite{zeiler2014visualizing}) explanation types.
However, despite substantial research in interpretable machine learning techniques, a core issue in model explanations is the difference between machine learning model's decision-making logic and human sense-making as the receiver~\cite{kim2017human,zhang2019dissonance}.

\vspace{0.7em}
\begin{center}
\framebox{
\parbox{0.9\textwidth}{\textit{Application in Case Study:} In our fake news detection case study, the explainable interface design team had previously discussed candidate explanation choices with the machine learning design team (see Guidelines 2 and 4). 
Therefore, a final review of model-generated explanations and an assessment of implementation limitations were performed in this step.
For example, removing noise-like features from saliency maps, normalizing attributions scores, and resolving contradicting explanations between an ensemble of models were primary implementation bottlenecks that were resolved in this step.
Specifically, as a decision point for trade-offs between clarity and faithfulness of explanations, the team decided on using heuristic filters to eliminate features with a very low attribution score for the sake of presentation simplicity.}}\par
\end{center}
\vspace{1em}

\textbf{Guideline 7: Evaluate Model Trustworthiness: }
Evaluating the interpretable machine learning is the first evaluation step in our framework due to its impact on outer layer evaluation measure.
The high significance of this evaluation step stems from the possibility that any unreliability of interpretability at this inner layer will propagate to all other outer layers.
Such unintended error propagation may lead to problematic outer-layer design decisions as well as misleading evaluation results.
We discuss two main evaluation goals for the innermost layer:

\begin{enumerate}
    \item Evaluating model trustworthiness.
    \item Evaluating ad-hoc explainer fidelity.
\end{enumerate}

The first evaluation goal aims to utilize interpretability techniques as a debugging tool to analyze the model's trustworthiness on learning concepts beyond general performance measures~\cite{kim2018interpretability}.
Examples of model trustworthiness validation include evaluating model reliability in financial risk assessment~\cite{florez2015enhancing},  model fairness in social influencing applications~\cite{zhang2018examining}, and model safety for its intended functionality~\cite{bojarski2018visualbackprop}. 
Researchers have also proposed various regularization techniques for enhancing trustworthy feature learning in machine learning models~\cite{ijcai2017-371,hendricks2018women}. 
Next, the second evaluation goal targets fidelity of ad-hoc explainer techniques to the black-box model. 
Research shows that different ad-hoc interpretability techniques have inconsistencies and can be misleading~\cite{adebayo2018sanity}. 
Evaluating explanation trustworthiness can verify explainer fidelity in terms of how well it represents the black-box model (see Section ~\ref{sec:explanation-truthfulness}).

\vspace{0.7em}
\begin{center}
\framebox{
\parbox{0.9\textwidth}{\textit{Application in Case Study:} 
In our case study, we paid careful attention to qualitative reviewing of the model explanations after each design iteration.
Our initial qualitative review of model explanations led to dataset cleaning through a heuristic search aimed at the removal of mislabeled examples and unrelated news articles. An improvement to model performance was achieved after dataset cleaning. 
Then, after the first round of human-subject evaluation of the explainable interface (see Guideline 5), the team identified negative effects of keyword explanations with low attention scores from end-users. 
The team decided on using a lower threshold for visualizing attention maps to reduce clutter and ``noisy explanations'' for end-users.
Finally, after one round of XAI outcome evaluation (see Guideline 3), analysis of users' mental models revealed that a dataset imbalance between the ``fake news'' and ``true news'' was causing a bias for the model in that the model was usually more confident in predicting fake news over true news.}}\par
\end{center}
\vspace{1em}

\section{Discussion}

In our review, we discussed multiple XAI design goals and evaluation measures appropriate for various targeted user types. 
Table~\ref{tab:main-table} presents the categorization of selected existing design and evaluation methods that organizes literature along three perspectives: \textit{design goals}, \textit{evaluation methods}, and the \textit{targeted users} of the XAI system.
Our categorization revealed the necessity of an interdisciplinary effort for designing and evaluating XAI systems. 
To address these issues, we proposed a design and evaluation framework that connects design goals and evaluation methods for end-to-end XAI systems design, as presented through a model (Figure~\ref{fig:evaluation}) and guidelines.
In this section, we discuss further considerations for XAI designers to benefit from the body of knowledge of XAI system design and evaluation.
The following recommendations support and promote different layers of the proposed design and evaluation framework as well.

\subsection{Pairing Design Goals with Evaluation Methods}

It is essential to use appropriate measures for evaluating the effectiveness of design elements. 
A common pitfall in choosing evaluation measures in XAI systems is that the same evaluation measure is sometimes used for multiple design goals. 
A simple solution to address this issue is to distinguish between measurements by using multiple scales to capture different attributes in each evaluation target. 
For example, the concept of \textit{user trust} consists of multiple constructs~\cite{cahour2009does} that could be measured with separate scales in questionnaires and interviews (see Section~\ref{user_trust_section}). 
User satisfaction measurements could also be designed for various attributes such as understandability of explanations, usefulness of explanations, and sufficiency of details~\cite{hoffman2018metrics} to target specific explanation qualities (see Section~\ref{user_satisfaction}).

An efficient way to pair design goals with appropriate evaluation measures is to balance different design methods and evaluation types in iterative cycles of design.
Managing the trade-offs between qualitative and quantitative methods in the design process can allow designers to take advantage of different approaches, as needed.
For example, while focus groups and interviews provide more detailed and in-depth feedback on the users' mental model~\cite{Kulesza2012Tell}, remote measurements are highly valuable due to the scalability of the collected data even though they provide less detail for drawing conclusions~\cite{lage2019human}.
Thus, one successful approach could be to start with multiple small-scale prototyping and formative studies collecting qualitative measures at the earlier stages of the design (e.g., for XAI system goals layer in the framework) and continue with larger-scale studies and quantitative measures in the later stages (e.g., for interpretable model and interface evaluations in the framework).

\subsection{Overlap Among Design Goals}

In our categorization of XAI systems, we chose two main dimensions to organize XAI systems by their \textit{Design Goals} and \textit{Evaluation Measures} in Section~\ref{sec:categorization_section}. 
The XAI design goals (G1--G8) were based on the goals extracted from the surveyed papers, 
and since the XAI design goals are primarily derived from their targeted user groups, we note that overlaps among goals do exist across disciplines. 
For instance, there is overlap of the goals of \textit{G1: Algorithmic Transparency} for novice users in HCI research, \textit{G5: Model Visualization} for data experts in visual analytics, and \textit{G7: Interpretability Techniques} for AI experts in machine learning research.  
While overlapping, these similar goals are studied with different objectives across the three research disciplines leading to diverse sets of design requirements and implementation paths. 
For example, designing XAI systems for AI novices requires processes and steps to build human-centered explainable interfaces to communicate model explanations to the end-users, whereas designing new interpretability techniques for AI experts has a different set of computational requirements. 
Another example of overlap in XAI goals is between the goal for \textit{G6: Model Visualization and Inspection} for data experts and \textit{G8: Model Debugging} for AI experts, in which different sets of tools and requirements are used to address different research objectives. 

To address the overlap between XAI goal among research disciplines, we used the XAI \textit{User Groups} as an auxiliary dimension to organize XAI goals in this cross-disciplinary topic (Section~\ref{sec:d_goal_section}) and emphasize the diversity of diverse research objectives.  
The three user groups were chosen to organize research objectives and efforts into HCI (for AI novices), visual analytics (for data experts), and machine learning (for AI experts) research fields. 
Additionally, as described in the framework, the three user groups prioritize design objectives in the design process for the XAI system rather than absolute separation of design goals. 
For example, the objectives and priorities in XAI system design for algorithmic bias mitigation for domain experts in a law firm are certainly different from those of model training and tuning tools for AI experts. 
However, by following the multidisciplinary design framework, a design team can translate XAI system goals into design objectives for explainable interface and machine learning techniques to improve the design process in different layers. 
Therefore, in the above example, the design team can focus on diverse interface design and interpretability technique objectives to achieve the primary XAI goal of bias mitigation for the domain experts. 
Note that the specifics of any particular system will determine the priorities of different objectives. 

\subsection{System Evaluation Over Time}

An important aspect in evaluating complex AI and XAI systems is to take the user learning into account.
Learnability is even more critical when measuring mental models and user trust in the system. 
A user learns and gets more familiar with the system over time with continued interaction with the system.
This brings the importance of repeated temporal data capture (in contrast to static measurements) for XAI evaluations.
Holliday et al.~\cite{holliday2016user} present an example of multiple trust assessments during the user study.
They measured user trust at regular intervals during the study to capture changes in user trust as the user interacts more with the system.
Their results indicates an XAI system outperformed a non-XAI counterpart in maintaining user trust over time. 
Time-based measurements, also referred to as \textit{dynamic measurements}, allows designers to monitor explanation usability and effectiveness in various contexts and situations~\cite{doyle2008measuring, schaffernicht2011comprehensive}. 
For instance, Zhang et al.~\cite{zhang2020effect} explore the effect of intelligent system explanations in user trust calibration.
In their experiments, they observe significant effect on calibration of trust when model prediction confidence score was shown to participants. 
In another example, a study by Nourani et al.\cite{nourani2020time} controlled whether users' early experiences with an explainable activity recognition system had better or worse model outputs, and the first impressions significantly affected both task performance and user confidence in understanding how the system works.
In a study with a news review task, Mohseni et al.~\cite{mohseni2020trust} identified different user profiles for changes in trust over time (trust dynamics) while working with the assistance of an explainable fake news detector. 
Their analysis of results revealed a significant effect of machine learning explanations on user trust dynamics.

Long-term evaluation of XAI systems can also allow designers to estimate valuable user experience factors such as over-trust and under-trust on the system.
User-perceived system accuracy~\cite{kulesza2010explanatory} and transparency~\cite{rader2015understanding} are examples of long-term measures for explanation usability that depend on building user trust in the system's interpretability.
As more information is provided by explanations over time, reasoning and mental strategies may change as users create new hypotheses about system functionality.
Therefore, it is essential to also consider users' mental models and trust in extended studies to evaluate all aspects of the XAI system.

Another use case of long-term measurements is to evaluate the effects of intelligent system's non-uniform behaviors in real-world scenarios. 
This means, although in a controlled study setup, a balanced set of input examples will present the system to the user, in real-world scenarios, users may face alterations in system performance in long-term interaction with the system.
Long-term measurements will identify user's unjust trust in the system due to a limited or biased set of interactions with the system.
For example, in the context of autonomous vehicles, Kraus et al.~\cite{kraus2019more} presented a model of trust calibration and presented studies on trust dynamics in the early phases of user interaction with the system. 
Their results indicate the effects of error-free automation in steady increase of user trust as well as the effects of user a priori information in eliminating the decrease of trust in case of system malfunction.

\subsection{Evaluation Ground Truth} 
\label{sec:eval-baseline}

Research on XAI systems study various goals with different measures across multiple domains.
The breadth of XAI research makes it challenging to interpret and transfer findings from one task and domain to another. 
Knowing key factors for interpreting implications of evaluation results is  essential to aggregate findings across  domains and disciplines. 
An important factor in understanding XAI evaluation results and comparing results among multiple studies is the choice of ground truth. 
In the following, we review common choices of ground truth for both human-subject and computational evaluation methods.

Human-subject experiments often take the form of controlled studies to examine the effects of machine learning explanations on a control group in comparison to a baseline group. 
In these setups, the choice of the baseline could affect results implications and significance. 
Our review of papers in the space of XAI evaluation shows the majority of study designs use a \textit{no explanation} condition as the baseline condition to measure the effectiveness of model explanations in an explanation group. 
Examples for the baseline include approaches that remove model explanations related components and features form the interface in the baseline condition~\cite{kocielnik2019will,nourani2019effects}.
In other work, Poursabzi et al.~\cite{poursabzi2018manipulating} also included a \textit{no AI} baseline to measure participants' performance without the help of model predictions. 
Another way is to compare the effects of explanation type or complexity between study conditions without the \textit{no explanation} baseline. 
For instance, Lage et al.~\cite{lage2019human} present a study to evaluate the effects of explanation complexity on participants' comprehension and performance.
They used linear and logistic regression to estimate the effects of explanation complexity on participants' normalized response time, response accuracy, and subjective task difficulty rating.

Though the above mentioned studies are controlled experiments, there may still be unaccounted human behavioral implications due to differences in the complex process of explaining worthy of consideration. 
Langer et al.~\cite{langer1978mindlessness} present an experiment on ``placebic'' explanations that shows people's mindless behavior when facing explanations for actions.
In a simple setup, their study showed that when asking a request, inclusion of explanations and justifications increased user's willingness to comply even if the explanations convey no meaningful information.  
Recently, Eiband et al. \cite{eiband2019impact} proposed using \textit{placebic explanations} instead of a \textit{no explanation} condition as the baseline for XAI human subject studies. 
Therefore, using non-informative or even randomly generated explanations as the baseline condition could potentially counteract a participant's positive tendency toward explanations and improve study results.

Considering other approaches, a commonly accepted computational technique for quantitatively evaluating instance explanations is to create a ground truth based on the input features that semantically contribute to the target class. 
For example, image segmentation maps (annotations of objects in images) are used to evaluate model generated saliency maps in weakly supervised object localization tasks~\cite{li2018tell}. 
Mohseni et al.~\cite{mohseni2018human} proposed a multi-layer \textit{Human-Attention} baseline for feature-level evaluation of machine learning explanations. 
Their \textit{Human-Attention} baseline provides a human-grounded feature attribution map with a higher level of granularity compared to object segmentation maps. 
Similarly, feature-level annotations have been used as the explanation ground truth in the text classification domain~\cite{8970999}. 
Other less accurate means of feature attribution like bounding box in images datasets have been used for quantitative evaluation of saliency maps.
For instance, Du et al.~\cite{du2018towards} evaluated saliency maps generated from a CNN model by calculating pixel-wise IOU (intersection over union) of model-explanation bounding boxes and ground truth bounding boxes.

\subsection{Role of User Interactions in XAI} 

Another important consideration in designing XAI systems is how to leverage user interactions to better support system understandability.
The benefits of interactive system design have been previously explored in the topic of interactive machine learning~\cite{amershi2014power,amershi2019guidelines} for novice end-users. 
AI and data experts also benefit from interactive visual tools to improve model and task performance~\cite{endert2017state}.  
In this section, we discuss multiple examples of interaction design that support user understanding of the underlying black-box model.

Focusing on interactive design for AI-based systems for AI novices, Amershi et al.~\cite{amershi2014power} reviewed multiple case studies that demonstrate the effectiveness of interactivity with a tight coupling between the algorithm and the user. 
They emphasize how interactive machine learning processes allow the users to instantly examine the impact of their actions and adapt their next queries to improve outcomes. 
Such interactions allow users to test various inputs and learn about the model by creating~\textit{What-If} explanations~\cite{Wang2019Theory}. 
Particularly, user-led cyles of trial and error help novices to understand how the machine learning model works and how to steer the model to improve results. 
In the context of XAI, Jongejan and Holbrook~\cite{cai2019effects} present a study in which users draw images to see whether an image recognition algorithm can correctly recognize the intended sketch. 
Their system and study allows for interactive trial-and-error to explore how the algorithm works. 
In addition, their system provides example-based explanations in cases where the algorithm fails to correctly classify drawings.
Another approach is to allow users to control or tune algorithmic parameters to achieve better results. 
For example, Kocielnik et al.~\cite{kocielnik2019will} present a study in which users were able to freely control detection sensitivity in an AI assistant. 
Their results showed a significant effect on user perception of control and acceptance.

Visual analytics tools also support model understanding for expert users through interaction with algorithms.
Examples including allowing data scientists and model experts to interactively explore model representations~\cite{hohman2019summit}, analyze model training processes~\cite{liu2018analyzing}, and detect learning biases~\cite{cabrera2019fairvis}. 
Also, embedded interaction techniques can support the exploration of very large deep learning networks.
For instance, Hohman et al.~\cite{hohman2019summit} present multiple interactive features to select and filter of neurons and zoom and pan in feature representations to support AI experts in interpreting and reviewing trained models.

\subsection{Generalization and Extension of the Framework}
\label{sec:generalization-validation}

Our framework is extendable and compatible with existing AI-infused interface and interaction design guidelines.
For example, Amershi et al.~\cite{amershi2019guidelines} propose 18 design guidelines for human-AI interaction design. 
Their guidelines are based on a review of a large number of AI-related design recommendation sources. 
They systematically validated guidelines through multiple rounds of evaluations with 49 design practitioners in 20 AI-infused products. 
Their design guidelines provide further details within the user interface design layer of our framework (Section~\ref{sec:interface-design-layer}) to guide the development of appropriate user interactions with model output and interactions. 
In other work, Dudley and Kristensson~\cite{dudley2018review} present a review and characterization of user interface design principles for interactive machine learning systems. 
They propose a structural breakdown of interactive machine learning systems and present six principles to support system design. 
This work also benefits our framework by contributing practices of interactive machine learning design to the XAI system goals layer (Section~\ref{sec:xai-system-goal}) and the user interface design layer (Section~\ref{sec:interface-design-layer}) 
From the standpoint of evaluation methods, Mueller and Klein~\cite{mueller2011improving} discuss how common usability tests cannot address intelligent tools where software replicates human intelligence. 
They suggest new solutions are needed to allow the users to experience an AI-based tool's strengths and weaknesses. 
Likewise, our nested framework points out the potential for error propagation from the inner layers (e.g., interpretable algorithms design) to the outer layers (e.g., system outcomes) in the XAI system evaluation pole. 
The iterative back-and-forth between layers in the nested model encourages expert review of system outcomes, user-centered evaluation of the explainable interface, and computational evaluation of machine learning algorithms.

\subsection{Limitations of the Framework}
Our framework provides a basis for XAI system design in interdisciplinary teamwork and we have described our case study example to validate and improve the framework. 
The presented case study serves as a practical example of using our framework in a multidisciplinary collaborative XAI design and development effort. 
Our use case is the result of a year-long (and ongoing) research done by a team of eight university researchers with diverse backgrounds.
The lessons learned and pitfalls in our end-to-end implementation case study are incorporated in the presented design guidelines.
However, no framework is perfect or entirely comprehensive.
We acknowledge that the validity and usefulness of a framework are to be proven in practice with further case studies. 
In our future work, we plan to run multiple validation case studies to examine practicality and usefulness of this framework.

Moreover, this framework has a common limitation of many multidisciplinary design frameworks of being light on specific details at each step. 
Rather than contributing detailed guidelines for each framework layer, the framework is intended to pave the path for efficient collaboration among and within different teams, which is essential for XAI system design given the inherently interdisciplinary nature of this field.
This higher level of freedom allows for extendability with other design guidelines (see the discussion in Section~\ref{sec:generalization-validation}) to integrate with more tailored approaches for specific domains.
Additionally, the diversity of design goals and evaluation methods at each layer can help maintain the balance of attention from the design team to different aspects of XAI system.

\section{Conclusion}

We reviewed XAI-related research to organize multiple XAI design goals and evaluation measures. 
Table~\ref{tab:main-table} presents our categorization of selected existing design and evaluation methods that organizes literature along three perspectives: \textit{design goals}, \textit{evaluation methods}, and the \textit{targeted users} of the XAI system. 
We provide summarized ready-to-use tables of evaluation methods and recommendations for different goals in XAI research. 
Our categorization revealed the necessity of an interdisciplinary effort for designing and evaluating XAI systems. 
We want to draw attention to related resources in the social sciences that can facilitate the extent of social and cognitive aspects of explanations.
To address these issues, we proposed a design and evaluation framework that connects design goals and evaluation methods for end-to-end XAI systems design, as presented through a model and a series of guidelines.
We hope our framework drives further discussion about the interplay between design and evaluation of explainable artificial intelligent systems.
Although the presented framework is organized to provide a high-level guideline for a multidisciplinary effort to build XAI systems, it is not meant to offer all aspects of interface and interaction design and development of interpretable machine learning techniques.
Lastly, we briefly discussed additional considerations for XAI designers to benefit from the body of knowledge of XAI system design and evaluation.

\section{Acknowledgments}

The authors would like to thank anonymous reviewers for their helpful comments on earlier versions of this manuscript.
The work in this paper is supported by the DARPA XAI program under N66001-17-2-4031 and by NSF award 1900767.
The views and conclusions in this paper are those of the authors and should not be interpreted as representing any funding agencies.

\bibliographystyle{ACM-Reference-Format}
\bibliography{XAI-bibliography}


\begin{thebibliography}{226}


\ifx \showCODEN    \undefined \def \showCODEN     #1{\unskip}     \fi
\ifx \showDOI      \undefined \def \showDOI       #1{#1}\fi
\ifx \showISBNx    \undefined \def \showISBNx     #1{\unskip}     \fi
\ifx \showISBNxiii \undefined \def \showISBNxiii  #1{\unskip}     \fi
\ifx \showISSN     \undefined \def \showISSN      #1{\unskip}     \fi
\ifx \showLCCN     \undefined \def \showLCCN      #1{\unskip}     \fi
\ifx \shownote     \undefined \def \shownote      #1{#1}          \fi
\ifx \showarticletitle \undefined \def \showarticletitle #1{#1}   \fi
\ifx \showURL      \undefined \def \showURL       {\relax}        \fi
\providecommand\bibfield[2]{#2}
\providecommand\bibinfo[2]{#2}
\providecommand\natexlab[1]{#1}
\providecommand\showeprint[2][]{arXiv:#2}

\bibitem[\protect\citeauthoryear{Abdul, Vermeulen, Wang, Lim, and
  Kankanhalli}{Abdul et~al\mbox{.}}{2018}]%
        {abdul2018trends}
\bibfield{author}{\bibinfo{person}{Ashraf Abdul}, \bibinfo{person}{Jo
  Vermeulen}, \bibinfo{person}{Danding Wang}, \bibinfo{person}{Brian~Y Lim},
  {and} \bibinfo{person}{Mohan Kankanhalli}.} \bibinfo{year}{2018}\natexlab{}.
\newblock \showarticletitle{Trends and trajectories for explainable,
  accountable and intelligible systems: An hci research agenda}. In
  \bibinfo{booktitle}{\emph{Proceedings of the 2018 CHI Conference on Human
  Factors in Computing Systems}}. ACM, \bibinfo{pages}{582}.
\newblock


\bibitem[\protect\citeauthoryear{Adadi and Berrada}{Adadi and Berrada}{2018}]%
        {adadi2018peeking}
\bibfield{author}{\bibinfo{person}{Amina Adadi} {and} \bibinfo{person}{Mohammed
  Berrada}.} \bibinfo{year}{2018}\natexlab{}.
\newblock \showarticletitle{Peeking inside the black-box: A survey on
  Explainable Artificial Intelligence (XAI)}.
\newblock \bibinfo{journal}{\emph{IEEE Access}}  \bibinfo{volume}{6}
  (\bibinfo{year}{2018}), \bibinfo{pages}{52138--52160}.
\newblock


\bibitem[\protect\citeauthoryear{Adebayo, Gilmer, Muelly, Goodfellow, Hardt,
  and Kim}{Adebayo et~al\mbox{.}}{2018}]%
        {adebayo2018sanity}
\bibfield{author}{\bibinfo{person}{Julius Adebayo}, \bibinfo{person}{Justin
  Gilmer}, \bibinfo{person}{Michael Muelly}, \bibinfo{person}{Ian Goodfellow},
  \bibinfo{person}{Moritz Hardt}, {and} \bibinfo{person}{Been Kim}.}
  \bibinfo{year}{2018}\natexlab{}.
\newblock \showarticletitle{Sanity checks for saliency maps}. In
  \bibinfo{booktitle}{\emph{Advances in Neural Information Processing
  Systems}}. \bibinfo{pages}{9505--9515}.
\newblock


\bibitem[\protect\citeauthoryear{Ahn and Lin}{Ahn and Lin}{2019}]%
        {ahn2019fairsight}
\bibfield{author}{\bibinfo{person}{Yongsu Ahn} {and} \bibinfo{person}{Yu-Ru
  Lin}.} \bibinfo{year}{2019}\natexlab{}.
\newblock \showarticletitle{Fairsight: visual analytics for fairness in
  decision making}.
\newblock \bibinfo{journal}{\emph{IEEE Transactions on Visualization and
  Computer Graphics}} (\bibinfo{year}{2019}).
\newblock


\bibitem[\protect\citeauthoryear{Alexander and Gleicher}{Alexander and
  Gleicher}{2015}]%
        {alexander2015task}
\bibfield{author}{\bibinfo{person}{Eric Alexander} {and}
  \bibinfo{person}{Michael Gleicher}.} \bibinfo{year}{2015}\natexlab{}.
\newblock \showarticletitle{Task-driven comparison of topic models}.
\newblock \bibinfo{journal}{\emph{IEEE Transactions on Visualization and
  Computer Graphics}} \bibinfo{volume}{22}, \bibinfo{number}{1}
  (\bibinfo{year}{2015}), \bibinfo{pages}{320--329}.
\newblock


\bibitem[\protect\citeauthoryear{Amershi, Cakmak, Knox, and Kulesza}{Amershi
  et~al\mbox{.}}{2014}]%
        {amershi2014power}
\bibfield{author}{\bibinfo{person}{Saleema Amershi}, \bibinfo{person}{Maya
  Cakmak}, \bibinfo{person}{William~Bradley Knox}, {and} \bibinfo{person}{Todd
  Kulesza}.} \bibinfo{year}{2014}\natexlab{}.
\newblock \showarticletitle{Power to the people: The role of humans in
  interactive machine learning}.
\newblock \bibinfo{journal}{\emph{AI Magazine}} \bibinfo{volume}{35},
  \bibinfo{number}{4} (\bibinfo{year}{2014}), \bibinfo{pages}{105--120}.
\newblock


\bibitem[\protect\citeauthoryear{Amershi, Weld, Vorvoreanu, Fourney, Nushi,
  Collisson, Suh, Iqbal, Bennett, Inkpen, et~al\mbox{.}}{Amershi
  et~al\mbox{.}}{2019}]%
        {amershi2019guidelines}
\bibfield{author}{\bibinfo{person}{Saleema Amershi}, \bibinfo{person}{Dan
  Weld}, \bibinfo{person}{Mihaela Vorvoreanu}, \bibinfo{person}{Adam Fourney},
  \bibinfo{person}{Besmira Nushi}, \bibinfo{person}{Penny Collisson},
  \bibinfo{person}{Jina Suh}, \bibinfo{person}{Shamsi Iqbal},
  \bibinfo{person}{Paul~N Bennett}, \bibinfo{person}{Kori Inkpen},
  {et~al\mbox{.}}} \bibinfo{year}{2019}\natexlab{}.
\newblock \showarticletitle{Guidelines for human-AI interaction}. In
  \bibinfo{booktitle}{\emph{Proceedings of the 2019 CHI Conference on Human
  Factors in Computing Systems}}. ACM, \bibinfo{pages}{3}.
\newblock


\bibitem[\protect\citeauthoryear{Amodei, Olah, Steinhardt, Christiano,
  Schulman, and Man{\'e}}{Amodei et~al\mbox{.}}{2016}]%
        {amodei2016concrete}
\bibfield{author}{\bibinfo{person}{Dario Amodei}, \bibinfo{person}{Chris Olah},
  \bibinfo{person}{Jacob Steinhardt}, \bibinfo{person}{Paul Christiano},
  \bibinfo{person}{John Schulman}, {and} \bibinfo{person}{Dan Man{\'e}}.}
  \bibinfo{year}{2016}\natexlab{}.
\newblock \showarticletitle{Concrete problems in AI safety}.
\newblock \bibinfo{journal}{\emph{arXiv preprint arXiv:1606.06565}}
  (\bibinfo{year}{2016}).
\newblock


\bibitem[\protect\citeauthoryear{Ananny and Crawford}{Ananny and
  Crawford}{2018}]%
        {ananny2018seeing}
\bibfield{author}{\bibinfo{person}{Mike Ananny} {and} \bibinfo{person}{Kate
  Crawford}.} \bibinfo{year}{2018}\natexlab{}.
\newblock \showarticletitle{Seeing without knowing: Limitations of the
  transparency ideal and its application to algorithmic accountability}.
\newblock \bibinfo{journal}{\emph{New Media \& Society}} \bibinfo{volume}{20},
  \bibinfo{number}{3} (\bibinfo{year}{2018}), \bibinfo{pages}{973--989}.
\newblock


\bibitem[\protect\citeauthoryear{Antifakos, Kern, Schiele, and
  Schwaninger}{Antifakos et~al\mbox{.}}{2005}]%
        {antifakos2005towards}
\bibfield{author}{\bibinfo{person}{Stavros Antifakos}, \bibinfo{person}{Nicky
  Kern}, \bibinfo{person}{Bernt Schiele}, {and} \bibinfo{person}{Adrian
  Schwaninger}.} \bibinfo{year}{2005}\natexlab{}.
\newblock \showarticletitle{Towards improving trust in context-aware systems by
  displaying system confidence}. In \bibinfo{booktitle}{\emph{Proceedings of
  the 7th International Conference on Human Computer Interaction with Mobile
  Devices \& Services}}. ACM, \bibinfo{pages}{9--14}.
\newblock


\bibitem[\protect\citeauthoryear{Arrieta, D{\'\i}az-Rodr{\'\i}guez, Del~Ser,
  Bennetot, Tabik, Barbado, Garc{\'\i}a, Gil-L{\'o}pez, Molina, Benjamins,
  et~al\mbox{.}}{Arrieta et~al\mbox{.}}{2020}]%
        {arrieta2019explainable}
\bibfield{author}{\bibinfo{person}{Alejandro~Barredo Arrieta},
  \bibinfo{person}{Natalia D{\'\i}az-Rodr{\'\i}guez}, \bibinfo{person}{Javier
  Del~Ser}, \bibinfo{person}{Adrien Bennetot}, \bibinfo{person}{Siham Tabik},
  \bibinfo{person}{Alberto Barbado}, \bibinfo{person}{Salvador Garc{\'\i}a},
  \bibinfo{person}{Sergio Gil-L{\'o}pez}, \bibinfo{person}{Daniel Molina},
  \bibinfo{person}{Richard Benjamins}, {et~al\mbox{.}}}
  \bibinfo{year}{2020}\natexlab{}.
\newblock \showarticletitle{Explainable artificial intelligence (XAI):
  Concepts, taxonomies, opportunities and challenges toward responsible AI}.
\newblock \bibinfo{journal}{\emph{Information Fusion}}  \bibinfo{volume}{58}
  (\bibinfo{year}{2020}), \bibinfo{pages}{82--115}.
\newblock


\bibitem[\protect\citeauthoryear{Ba, Mnih, and Kavukcuoglu}{Ba
  et~al\mbox{.}}{2014}]%
        {ba2014multiple}
\bibfield{author}{\bibinfo{person}{Jimmy Ba}, \bibinfo{person}{Volodymyr Mnih},
  {and} \bibinfo{person}{Koray Kavukcuoglu}.} \bibinfo{year}{2014}\natexlab{}.
\newblock \showarticletitle{Multiple object recognition with visual attention}.
\newblock \bibinfo{journal}{\emph{arXiv preprint arXiv:1412.7755}}
  (\bibinfo{year}{2014}).
\newblock


\bibitem[\protect\citeauthoryear{Bach, Binder, Montavon, Klauschen, M{\"u}ller,
  and Samek}{Bach et~al\mbox{.}}{2015}]%
        {bach2015pixel}
\bibfield{author}{\bibinfo{person}{Sebastian Bach}, \bibinfo{person}{Alexander
  Binder}, \bibinfo{person}{Gr{\'e}goire Montavon}, \bibinfo{person}{Frederick
  Klauschen}, \bibinfo{person}{Klaus-Robert M{\"u}ller}, {and}
  \bibinfo{person}{Wojciech Samek}.} \bibinfo{year}{2015}\natexlab{}.
\newblock \showarticletitle{On pixel-wise explanations for non-linear
  classifier decisions by layer-wise relevance propagation}.
\newblock \bibinfo{journal}{\emph{PloS one}} \bibinfo{volume}{10},
  \bibinfo{number}{7} (\bibinfo{year}{2015}), \bibinfo{pages}{e0130140}.
\newblock


\bibitem[\protect\citeauthoryear{Baehrens, Schroeter, Harmeling, Kawanabe,
  Hansen, and M{\"u}ller}{Baehrens et~al\mbox{.}}{2010}]%
        {baehrens2010explain}
\bibfield{author}{\bibinfo{person}{David Baehrens}, \bibinfo{person}{Timon
  Schroeter}, \bibinfo{person}{Stefan Harmeling}, \bibinfo{person}{Motoaki
  Kawanabe}, \bibinfo{person}{Katja Hansen}, {and}
  \bibinfo{person}{Klaus-Robert M{\"u}ller}.} \bibinfo{year}{2010}\natexlab{}.
\newblock \showarticletitle{How to explain individual classification
  decisions}.
\newblock \bibinfo{journal}{\emph{Journal of Machine Learning Research}}
  \bibinfo{volume}{11}, \bibinfo{number}{Jun} (\bibinfo{year}{2010}),
  \bibinfo{pages}{1803--1831}.
\newblock


\bibitem[\protect\citeauthoryear{Bansal, Nushi, Kamar, Lasecki, Weld, and
  Horvitz}{Bansal et~al\mbox{.}}{2019}]%
        {bansal2019beyond}
\bibfield{author}{\bibinfo{person}{Gagan Bansal}, \bibinfo{person}{Besmira
  Nushi}, \bibinfo{person}{Ece Kamar}, \bibinfo{person}{Walter~S Lasecki},
  \bibinfo{person}{Daniel~S Weld}, {and} \bibinfo{person}{Eric Horvitz}.}
  \bibinfo{year}{2019}\natexlab{}.
\newblock \showarticletitle{Beyond Accuracy: The Role of Mental Models in
  Human-AI Team Performance}. In \bibinfo{booktitle}{\emph{Proceedings of the
  AAAI Conference on Human Computation and Crowdsourcing}},
  Vol.~\bibinfo{volume}{7}. \bibinfo{pages}{2--11}.
\newblock


\bibitem[\protect\citeauthoryear{Bellotti and Edwards}{Bellotti and
  Edwards}{2001}]%
        {bellotti2001intelligibility}
\bibfield{author}{\bibinfo{person}{Victoria Bellotti} {and}
  \bibinfo{person}{Keith Edwards}.} \bibinfo{year}{2001}\natexlab{}.
\newblock \showarticletitle{Intelligibility and accountability: human
  considerations in context-aware systems}.
\newblock \bibinfo{journal}{\emph{Human--Computer Interaction}}
  \bibinfo{volume}{16}, \bibinfo{number}{2-4} (\bibinfo{year}{2001}),
  \bibinfo{pages}{193--212}.
\newblock


\bibitem[\protect\citeauthoryear{Berkovsky, Taib, and Conway}{Berkovsky
  et~al\mbox{.}}{2017}]%
        {Berkovsky:2017:RUT:3025171.3025209}
\bibfield{author}{\bibinfo{person}{Shlomo Berkovsky}, \bibinfo{person}{Ronnie
  Taib}, {and} \bibinfo{person}{Dan Conway}.} \bibinfo{year}{2017}\natexlab{}.
\newblock \showarticletitle{How to Recommend?: User Trust Factors in Movie
  Recommender Systems}. In \bibinfo{booktitle}{\emph{Proceedings of the 22nd
  International Conference on Intelligent User Interfaces}}
  \emph{(\bibinfo{series}{IUI '17})}. \bibinfo{publisher}{ACM},
  \bibinfo{address}{New York, NY, USA}, \bibinfo{pages}{287--300}.
\newblock
\showISBNx{978-1-4503-4348-0}
\urldef\tempurl%
\url{https://doi.org/10.1145/3025171.3025209}
\showDOI{\tempurl}


\bibitem[\protect\citeauthoryear{Best, Endert, and Kidwell}{Best
  et~al\mbox{.}}{2014}]%
        {best20147}
\bibfield{author}{\bibinfo{person}{Daniel~M Best}, \bibinfo{person}{Alex
  Endert}, {and} \bibinfo{person}{Daniel Kidwell}.}
  \bibinfo{year}{2014}\natexlab{}.
\newblock \showarticletitle{7 key challenges for visualization in cyber network
  defense}. In \bibinfo{booktitle}{\emph{Proceedings of the Eleventh Workshop
  on Visualization for Cyber Security}}. ACM, \bibinfo{pages}{33--40}.
\newblock


\bibitem[\protect\citeauthoryear{Bilgic and Mooney}{Bilgic and Mooney}{2005}]%
        {bilgic2005explaining}
\bibfield{author}{\bibinfo{person}{Mustafa Bilgic} {and}
  \bibinfo{person}{Raymond~J Mooney}.} \bibinfo{year}{2005}\natexlab{}.
\newblock \showarticletitle{Explaining recommendations: Satisfaction vs.
  promotion}. In \bibinfo{booktitle}{\emph{Beyond Personalization Workshop,
  IUI}}, Vol.~\bibinfo{volume}{5}. \bibinfo{pages}{153}.
\newblock


\bibitem[\protect\citeauthoryear{Binns, Van~Kleek, Veale, Lyngs, Zhao, and
  Shadbolt}{Binns et~al\mbox{.}}{2018}]%
        {Binns:2018:RHP:3173574.3173951}
\bibfield{author}{\bibinfo{person}{Reuben Binns}, \bibinfo{person}{Max
  Van~Kleek}, \bibinfo{person}{Michael Veale}, \bibinfo{person}{Ulrik Lyngs},
  \bibinfo{person}{Jun Zhao}, {and} \bibinfo{person}{Nigel Shadbolt}.}
  \bibinfo{year}{2018}\natexlab{}.
\newblock \showarticletitle{``It's Reducing a Human Being to a Percentage'':
  Perceptions of Justice in Algorithmic Decisions}. In
  \bibinfo{booktitle}{\emph{Proceedings of the 2018 CHI Conference on Human
  Factors in Computing Systems}}. ACM, \bibinfo{pages}{377}.
\newblock


\bibitem[\protect\citeauthoryear{Bobko, Barelka, and Hirshfield}{Bobko
  et~al\mbox{.}}{2014}]%
        {bobko2014construct}
\bibfield{author}{\bibinfo{person}{Philip Bobko}, \bibinfo{person}{Alex~J
  Barelka}, {and} \bibinfo{person}{Leanne~M Hirshfield}.}
  \bibinfo{year}{2014}\natexlab{}.
\newblock \showarticletitle{The construct of state-level suspicion: A model and
  research agenda for automated and information technology (IT) contexts}.
\newblock \bibinfo{journal}{\emph{Human Factors}} \bibinfo{volume}{56},
  \bibinfo{number}{3} (\bibinfo{year}{2014}), \bibinfo{pages}{489--508}.
\newblock


\bibitem[\protect\citeauthoryear{Bojarski, Choromanska, Choromanski, Firner,
  Ackel, Muller, Yeres, and Zieba}{Bojarski et~al\mbox{.}}{2018}]%
        {bojarski2018visualbackprop}
\bibfield{author}{\bibinfo{person}{Mariusz Bojarski}, \bibinfo{person}{Anna
  Choromanska}, \bibinfo{person}{Krzysztof Choromanski},
  \bibinfo{person}{Bernhard Firner}, \bibinfo{person}{Larry~J Ackel},
  \bibinfo{person}{Urs Muller}, \bibinfo{person}{Phil Yeres}, {and}
  \bibinfo{person}{Karol Zieba}.} \bibinfo{year}{2018}\natexlab{}.
\newblock \showarticletitle{Visualbackprop: Efficient visualization of cnns for
  autonomous driving}. In \bibinfo{booktitle}{\emph{2018 IEEE International
  Conference on Robotics and Automation (ICRA)}}. IEEE, \bibinfo{pages}{1--8}.
\newblock


\bibitem[\protect\citeauthoryear{Bozdag and van~den Hoven}{Bozdag and van~den
  Hoven}{2015}]%
        {bozdag2015breaking}
\bibfield{author}{\bibinfo{person}{Engin Bozdag} {and} \bibinfo{person}{Jeroen
  van~den Hoven}.} \bibinfo{year}{2015}\natexlab{}.
\newblock \showarticletitle{Breaking the filter bubble: democracy and design}.
\newblock \bibinfo{journal}{\emph{Ethics and Information Technology}}
  \bibinfo{volume}{17}, \bibinfo{number}{4} (\bibinfo{year}{2015}),
  \bibinfo{pages}{249--265}.
\newblock


\bibitem[\protect\citeauthoryear{Bryan and Mysore}{Bryan and Mysore}{2013}]%
        {bryan2013efficient}
\bibfield{author}{\bibinfo{person}{Nicholas Bryan} {and}
  \bibinfo{person}{Gautham Mysore}.} \bibinfo{year}{2013}\natexlab{}.
\newblock \showarticletitle{An efficient posterior regularized latent variable
  model for interactive sound source separation}. In
  \bibinfo{booktitle}{\emph{International Conference on Machine Learning}}.
  \bibinfo{pages}{208--216}.
\newblock


\bibitem[\protect\citeauthoryear{Bunt, Lount, and Lauzon}{Bunt
  et~al\mbox{.}}{2012}]%
        {bunt2012explanations}
\bibfield{author}{\bibinfo{person}{Andrea Bunt}, \bibinfo{person}{Matthew
  Lount}, {and} \bibinfo{person}{Catherine Lauzon}.}
  \bibinfo{year}{2012}\natexlab{}.
\newblock \showarticletitle{Are explanations always important?: a study of
  deployed, low-cost intelligent interactive systems}. In
  \bibinfo{booktitle}{\emph{Proceedings of the 2012 ACM International
  Conference on Intelligent User Interfaces}}. ACM, \bibinfo{pages}{169--178}.
\newblock


\bibitem[\protect\citeauthoryear{Bussone, Stumpf, and O'Sullivan}{Bussone
  et~al\mbox{.}}{2015}]%
        {bussone2015role}
\bibfield{author}{\bibinfo{person}{Adrian Bussone}, \bibinfo{person}{Simone
  Stumpf}, {and} \bibinfo{person}{Dympna O'Sullivan}.}
  \bibinfo{year}{2015}\natexlab{}.
\newblock \showarticletitle{The role of explanations on trust and reliance in
  clinical decision support systems}. In
  \bibinfo{booktitle}{\emph{International Conference on Healthcare Informatics
  (ICHI)}}. IEEE, \bibinfo{pages}{160--169}.
\newblock


\bibitem[\protect\citeauthoryear{Cabrera, Epperson, Hohman, Kahng, Morgenstern,
  and Chau}{Cabrera et~al\mbox{.}}{2019}]%
        {cabrera2019fairvis}
\bibfield{author}{\bibinfo{person}{Angel Cabrera}, \bibinfo{person}{Will
  Epperson}, \bibinfo{person}{Fred Hohman}, \bibinfo{person}{Minsuk Kahng},
  \bibinfo{person}{Jamie Morgenstern}, {and} \bibinfo{person}{Duen~Horng
  Chau}.} \bibinfo{year}{2019}\natexlab{}.
\newblock \showarticletitle{FairVis: visual analytics for discovering
  intersectional bias in machine learning}.
\newblock \bibinfo{journal}{\emph{IEEE Conference on Visual Analytics Science
  and Technology (VAST)}} (\bibinfo{year}{2019}).
\newblock


\bibitem[\protect\citeauthoryear{Cahour and Forzy}{Cahour and Forzy}{2009}]%
        {cahour2009does}
\bibfield{author}{\bibinfo{person}{B{\'e}atrice Cahour} {and}
  \bibinfo{person}{Jean-Fran{\c{c}}ois Forzy}.}
  \bibinfo{year}{2009}\natexlab{}.
\newblock \showarticletitle{Does projection into use improve trust and
  exploration? An example with a cruise control system}.
\newblock \bibinfo{journal}{\emph{Safety Science}} \bibinfo{volume}{47},
  \bibinfo{number}{9} (\bibinfo{year}{2009}), \bibinfo{pages}{1260--1270}.
\newblock


\bibitem[\protect\citeauthoryear{Cai, Jongejan, and Holbrook}{Cai
  et~al\mbox{.}}{2019a}]%
        {cai2019effects}
\bibfield{author}{\bibinfo{person}{Carrie~J Cai}, \bibinfo{person}{Jonas
  Jongejan}, {and} \bibinfo{person}{Jess Holbrook}.}
  \bibinfo{year}{2019}\natexlab{a}.
\newblock \showarticletitle{The effects of example-based explanations in a
  machine learning interface}. In \bibinfo{booktitle}{\emph{Proceedings of the
  24th International Conference on Intelligent User Interfaces}}.
  \bibinfo{pages}{258--262}.
\newblock


\bibitem[\protect\citeauthoryear{Cai, Reif, Hegde, Hipp, Kim, Smilkov,
  Wattenberg, Viegas, Corrado, Stumpe, et~al\mbox{.}}{Cai
  et~al\mbox{.}}{2019b}]%
        {cai2019human}
\bibfield{author}{\bibinfo{person}{Carrie~J Cai}, \bibinfo{person}{Emily Reif},
  \bibinfo{person}{Narayan Hegde}, \bibinfo{person}{Jason Hipp},
  \bibinfo{person}{Been Kim}, \bibinfo{person}{Daniel Smilkov},
  \bibinfo{person}{Martin Wattenberg}, \bibinfo{person}{Fernanda Viegas},
  \bibinfo{person}{Greg~S Corrado}, \bibinfo{person}{Martin~C Stumpe},
  {et~al\mbox{.}}} \bibinfo{year}{2019}\natexlab{b}.
\newblock \showarticletitle{Human-centered tools for coping with imperfect
  algorithms during medical decision-making}. In
  \bibinfo{booktitle}{\emph{Proceedings of the 2019 CHI Conference on Human
  Factors in Computing Systems}}. \bibinfo{pages}{1--14}.
\newblock


\bibitem[\protect\citeauthoryear{Caruana, Lou, Gehrke, Koch, Sturm, and
  Elhadad}{Caruana et~al\mbox{.}}{2015}]%
        {caruana2015intelligible}
\bibfield{author}{\bibinfo{person}{Rich Caruana}, \bibinfo{person}{Yin Lou},
  \bibinfo{person}{Johannes Gehrke}, \bibinfo{person}{Paul Koch},
  \bibinfo{person}{Marc Sturm}, {and} \bibinfo{person}{Noemie Elhadad}.}
  \bibinfo{year}{2015}\natexlab{}.
\newblock \showarticletitle{Intelligible models for healthcare: Predicting
  pneumonia risk and hospital 30-day readmission}. In
  \bibinfo{booktitle}{\emph{Proceedings of the 21th ACM SIGKDD International
  Conference on Knowledge Discovery and Data Mining}}. ACM,
  \bibinfo{pages}{1721--1730}.
\newblock


\bibitem[\protect\citeauthoryear{Chen, Kallus, Mao, Svacha, and Udell}{Chen
  et~al\mbox{.}}{2019}]%
        {chen2019fairness}
\bibfield{author}{\bibinfo{person}{Jiahao Chen}, \bibinfo{person}{Nathan
  Kallus}, \bibinfo{person}{Xiaojie Mao}, \bibinfo{person}{Geoffry Svacha},
  {and} \bibinfo{person}{Madeleine Udell}.} \bibinfo{year}{2019}\natexlab{}.
\newblock \showarticletitle{Fairness under unawareness: Assessing disparity
  when protected class is unobserved}. In \bibinfo{booktitle}{\emph{Proceedings
  of the Conference on Fairness, Accountability, and Transparency}}. ACM,
  \bibinfo{pages}{339--348}.
\newblock


\bibitem[\protect\citeauthoryear{Choo, Lee, Kihm, and Park}{Choo
  et~al\mbox{.}}{2010}]%
        {choo2010ivisclassifier}
\bibfield{author}{\bibinfo{person}{Jaegul Choo}, \bibinfo{person}{Hanseung
  Lee}, \bibinfo{person}{Jaeyeon Kihm}, {and} \bibinfo{person}{Haesun Park}.}
  \bibinfo{year}{2010}\natexlab{}.
\newblock \showarticletitle{iVisClassifier: An interactive visual analytics
  system for classification based on supervised dimension reduction}. In
  \bibinfo{booktitle}{\emph{Visual Analytics Science and Technology (VAST),
  2010 IEEE Symposium on}}. IEEE, \bibinfo{pages}{27--34}.
\newblock


\bibitem[\protect\citeauthoryear{Choo and Liu}{Choo and Liu}{2018}]%
        {choo2018visual}
\bibfield{author}{\bibinfo{person}{Jaegul Choo} {and} \bibinfo{person}{Shixia
  Liu}.} \bibinfo{year}{2018}\natexlab{}.
\newblock \showarticletitle{Visual analytics for explainable deep learning}.
\newblock \bibinfo{journal}{\emph{IEEE Computer Graphics and Applications}}
  \bibinfo{volume}{38}, \bibinfo{number}{4} (\bibinfo{year}{2018}),
  \bibinfo{pages}{84--92}.
\newblock


\bibitem[\protect\citeauthoryear{Chouldechova}{Chouldechova}{2017}]%
        {chouldechova2017fair}
\bibfield{author}{\bibinfo{person}{Alexandra Chouldechova}.}
  \bibinfo{year}{2017}\natexlab{}.
\newblock \showarticletitle{Fair prediction with disparate impact: A study of
  bias in recidivism prediction instruments}.
\newblock \bibinfo{journal}{\emph{Big data}} \bibinfo{volume}{5},
  \bibinfo{number}{2} (\bibinfo{year}{2017}), \bibinfo{pages}{153--163}.
\newblock


\bibitem[\protect\citeauthoryear{Chromik, Eiband, V{\"o}lkel, and
  Buschek}{Chromik et~al\mbox{.}}{2019}]%
        {chromik2019dark}
\bibfield{author}{\bibinfo{person}{Michael Chromik}, \bibinfo{person}{Malin
  Eiband}, \bibinfo{person}{Sarah~Theres V{\"o}lkel}, {and}
  \bibinfo{person}{Daniel Buschek}.} \bibinfo{year}{2019}\natexlab{}.
\newblock \showarticletitle{Dark patterns of explainability, transparency, and
  user control for intelligent systems.}. In \bibinfo{booktitle}{\emph{IUI
  Workshops}}.
\newblock


\bibitem[\protect\citeauthoryear{Chu, Hu, Hu, Wang, and Pei}{Chu
  et~al\mbox{.}}{2018}]%
        {chu2018exact}
\bibfield{author}{\bibinfo{person}{Lingyang Chu}, \bibinfo{person}{Xia Hu},
  \bibinfo{person}{Juhua Hu}, \bibinfo{person}{Lanjun Wang}, {and}
  \bibinfo{person}{Jian Pei}.} \bibinfo{year}{2018}\natexlab{}.
\newblock \showarticletitle{Exact and consistent interpretation for piecewise
  linear neural networks: A closed form solution}. In
  \bibinfo{booktitle}{\emph{Proceedings of the 24th ACM SIGKDD International
  Conference on Knowledge Discovery \& Data Mining}}.
  \bibinfo{pages}{1244--1253}.
\newblock


\bibitem[\protect\citeauthoryear{Clinciu and Hastie}{Clinciu and
  Hastie}{2019}]%
        {clinciu2019survey}
\bibfield{author}{\bibinfo{person}{Miruna-Adriana Clinciu} {and}
  \bibinfo{person}{Helen Hastie}.} \bibinfo{year}{2019}\natexlab{}.
\newblock \showarticletitle{A survey of explainable AI terminology}. In
  \bibinfo{booktitle}{\emph{Proceedings of the 1st Workshop on Interactive
  Natural Language Technology for Explainable Artificial Intelligence (NL4XAI
  2019)}}. \bibinfo{pages}{8--13}.
\newblock


\bibitem[\protect\citeauthoryear{Coppers, Van~den Bergh, Luyten, Coninx,
  Van~der Lek-Ciudin, Vanallemeersch, and Vandeghinste}{Coppers
  et~al\mbox{.}}{2018}]%
        {coppers2018intellingo}
\bibfield{author}{\bibinfo{person}{Sven Coppers}, \bibinfo{person}{Jan Van~den
  Bergh}, \bibinfo{person}{Kris Luyten}, \bibinfo{person}{Karin Coninx},
  \bibinfo{person}{Iulianna Van~der Lek-Ciudin}, \bibinfo{person}{Tom
  Vanallemeersch}, {and} \bibinfo{person}{Vincent Vandeghinste}.}
  \bibinfo{year}{2018}\natexlab{}.
\newblock \showarticletitle{Intellingo: An intelligible translation
  environment}. In \bibinfo{booktitle}{\emph{Proceedings of the 2018 CHI
  Conference on Human Factors in Computing Systems}}. ACM,
  \bibinfo{pages}{524}.
\newblock


\bibitem[\protect\citeauthoryear{Costanza, Fischer, Colley, Rodden, Ramchurn,
  and Jennings}{Costanza et~al\mbox{.}}{2014}]%
        {costanza2014doing}
\bibfield{author}{\bibinfo{person}{Enrico Costanza}, \bibinfo{person}{Joel~E
  Fischer}, \bibinfo{person}{James~A Colley}, \bibinfo{person}{Tom Rodden},
  \bibinfo{person}{Sarvapali~D Ramchurn}, {and} \bibinfo{person}{Nicholas~R
  Jennings}.} \bibinfo{year}{2014}\natexlab{}.
\newblock \showarticletitle{Doing the laundry with agents: a field trial of a
  future smart energy system in the home}. In
  \bibinfo{booktitle}{\emph{Proceedings of the SIGCHI Conference on Human
  Factors in Computing Systems}}. ACM, \bibinfo{pages}{813--822}.
\newblock


\bibitem[\protect\citeauthoryear{Curran, Moore, Kulesza, Wong, Todorovic,
  Stumpf, White, and Burnett}{Curran et~al\mbox{.}}{2012}]%
        {curran2012towards}
\bibfield{author}{\bibinfo{person}{William Curran}, \bibinfo{person}{Travis
  Moore}, \bibinfo{person}{Todd Kulesza}, \bibinfo{person}{Weng-Keen Wong},
  \bibinfo{person}{Sinisa Todorovic}, \bibinfo{person}{Simone Stumpf},
  \bibinfo{person}{Rachel White}, {and} \bibinfo{person}{Margaret Burnett}.}
  \bibinfo{year}{2012}\natexlab{}.
\newblock \showarticletitle{Towards recognizing cool: can end users help
  computer vision recognize subjective attributes of objects in images?}. In
  \bibinfo{booktitle}{\emph{Proceedings of the 2012 ACM International
  Conference on Intelligent User Interfaces}}. ACM, \bibinfo{pages}{285--288}.
\newblock


\bibitem[\protect\citeauthoryear{Das, Agrawal, Zitnick, Parikh, and Batra}{Das
  et~al\mbox{.}}{2016}]%
        {VQA-HAT}
\bibfield{author}{\bibinfo{person}{Abhishek Das}, \bibinfo{person}{Harsh
  Agrawal}, \bibinfo{person}{C.~Lawrence Zitnick}, \bibinfo{person}{Devi
  Parikh}, {and} \bibinfo{person}{Dhruv Batra}.}
  \bibinfo{year}{2016}\natexlab{}.
\newblock \showarticletitle{Human Attention in Visual Question Answering: Do
  Humans and Deep Networks Look at the Same Regions?}. In
  \bibinfo{booktitle}{\emph{Conference on Empirical Methods in Natural Language
  Processing (EMNLP)}}.
\newblock
\urldef\tempurl%
\url{https://computing.ece.vt.edu/~abhshkdz/vqa-hat/}
\showURL{%
\tempurl}


\bibitem[\protect\citeauthoryear{Das, Agrawal, Zitnick, Parikh, and Batra}{Das
  et~al\mbox{.}}{2017}]%
        {das2017human}
\bibfield{author}{\bibinfo{person}{Abhishek Das}, \bibinfo{person}{Harsh
  Agrawal}, \bibinfo{person}{Larry Zitnick}, \bibinfo{person}{Devi Parikh},
  {and} \bibinfo{person}{Dhruv Batra}.} \bibinfo{year}{2017}\natexlab{}.
\newblock \showarticletitle{Human attention in visual question answering: Do
  humans and deep networks look at the same regions?}
\newblock \bibinfo{journal}{\emph{Computer Vision and Image Understanding}}
  \bibinfo{volume}{163} (\bibinfo{year}{2017}), \bibinfo{pages}{90--100}.
\newblock


\bibitem[\protect\citeauthoryear{Datta, Tschantz, and Datta}{Datta
  et~al\mbox{.}}{2015}]%
        {datta2015automated}
\bibfield{author}{\bibinfo{person}{Amit Datta}, \bibinfo{person}{Michael~Carl
  Tschantz}, {and} \bibinfo{person}{Anupam Datta}.}
  \bibinfo{year}{2015}\natexlab{}.
\newblock \showarticletitle{Automated experiments on ad privacy settings}.
\newblock \bibinfo{journal}{\emph{Proceedings on Privacy Enhancing
  Technologies}} \bibinfo{volume}{2015}, \bibinfo{number}{1}
  (\bibinfo{year}{2015}), \bibinfo{pages}{92--112}.
\newblock


\bibitem[\protect\citeauthoryear{Diakopoulos}{Diakopoulos}{2014}]%
        {diakopoulos2014algorithmic}
\bibfield{author}{\bibinfo{person}{Nicholas Diakopoulos}.}
  \bibinfo{year}{2014}\natexlab{}.
\newblock \showarticletitle{Algorithmic-Accountability: the investigation of
  Black Boxes}.
\newblock \bibinfo{journal}{\emph{Tow Center for Digital Journalism}}
  (\bibinfo{year}{2014}).
\newblock


\bibitem[\protect\citeauthoryear{Diakopoulos}{Diakopoulos}{2017}]%
        {diakopoulos2017enabling}
\bibfield{author}{\bibinfo{person}{Nicholas Diakopoulos}.}
  \bibinfo{year}{2017}\natexlab{}.
\newblock \showarticletitle{Enabling Accountability of Algorithmic Media:
  Transparency as a Constructive and Critical Lens}.
\newblock In \bibinfo{booktitle}{\emph{Transparent Data Mining for Big and
  Small Data}}. \bibinfo{publisher}{Springer}, \bibinfo{pages}{25--43}.
\newblock


\bibitem[\protect\citeauthoryear{Dodge, Penney, Anderson, and Burnett}{Dodge
  et~al\mbox{.}}{2018}]%
        {dodge2018should}
\bibfield{author}{\bibinfo{person}{Jonathan Dodge}, \bibinfo{person}{Sean
  Penney}, \bibinfo{person}{Andrew Anderson}, {and} \bibinfo{person}{Margaret~M
  Burnett}.} \bibinfo{year}{2018}\natexlab{}.
\newblock \showarticletitle{What Should Be in an XAI Explanation? What IFT
  Reveals.}. In \bibinfo{booktitle}{\emph{IUI Workshops}}.
\newblock


\bibitem[\protect\citeauthoryear{Doshi-Velez and Kim}{Doshi-Velez and
  Kim}{2017}]%
        {doshi2017towards}
\bibfield{author}{\bibinfo{person}{Finale Doshi-Velez} {and}
  \bibinfo{person}{Been Kim}.} \bibinfo{year}{2017}\natexlab{}.
\newblock \showarticletitle{Towards a rigorous science of interpretable machine
  learning}.
\newblock \bibinfo{journal}{\emph{arXiv preprint arXiv:1702.08608}}
  (\bibinfo{year}{2017}).
\newblock


\bibitem[\protect\citeauthoryear{Doshi-Velez, Kortz, Budish, Bavitz, Gershman,
  O'Brien, Shieber, Waldo, Weinberger, and Wood}{Doshi-Velez
  et~al\mbox{.}}{2017}]%
        {doshi2017accountability}
\bibfield{author}{\bibinfo{person}{Finale Doshi-Velez}, \bibinfo{person}{Mason
  Kortz}, \bibinfo{person}{Ryan Budish}, \bibinfo{person}{Christopher Bavitz},
  \bibinfo{person}{Samuel~J Gershman}, \bibinfo{person}{David O'Brien},
  \bibinfo{person}{Stuart Shieber}, \bibinfo{person}{Jim Waldo},
  \bibinfo{person}{David Weinberger}, {and} \bibinfo{person}{Alexandra Wood}.}
  \bibinfo{year}{2017}\natexlab{}.
\newblock \showarticletitle{Accountability of AI Under the Law: The Role of
  Explanation}.
\newblock \bibinfo{journal}{\emph{Berkman Center Research Publication
  Forthcoming}} (\bibinfo{year}{2017}), \bibinfo{pages}{18--07}.
\newblock


\bibitem[\protect\citeauthoryear{Doyle, Radzicki, and Trees}{Doyle
  et~al\mbox{.}}{2008}]%
        {doyle2008measuring}
\bibfield{author}{\bibinfo{person}{James~K Doyle}, \bibinfo{person}{Michael~J
  Radzicki}, {and} \bibinfo{person}{W~Scott Trees}.}
  \bibinfo{year}{2008}\natexlab{}.
\newblock \showarticletitle{Measuring change in mental models of complex
  dynamic systems}.
\newblock In \bibinfo{booktitle}{\emph{Complex Decision Making}}.
  \bibinfo{publisher}{Springer}, \bibinfo{pages}{269--294}.
\newblock


\bibitem[\protect\citeauthoryear{Du, Plaisant, Spring, Crowley, and
  Shneiderman}{Du et~al\mbox{.}}{2019}]%
        {du2019eventaction}
\bibfield{author}{\bibinfo{person}{Fan Du}, \bibinfo{person}{Catherine
  Plaisant}, \bibinfo{person}{Neil Spring}, \bibinfo{person}{Kenyon Crowley},
  {and} \bibinfo{person}{Ben Shneiderman}.} \bibinfo{year}{2019}\natexlab{}.
\newblock \showarticletitle{EventAction: A Visual Analytics Approach to
  Explainable Recommendation for Event Sequences}.
\newblock \bibinfo{journal}{\emph{ACM Transactions on Interactive Intelligent
  Systems (TiiS)}} \bibinfo{volume}{9}, \bibinfo{number}{4}
  (\bibinfo{year}{2019}), \bibinfo{pages}{1--31}.
\newblock


\bibitem[\protect\citeauthoryear{Du, Liu, Song, and Hu}{Du
  et~al\mbox{.}}{2018}]%
        {du2018towards}
\bibfield{author}{\bibinfo{person}{Mengnan Du}, \bibinfo{person}{Ninghao Liu},
  \bibinfo{person}{Qingquan Song}, {and} \bibinfo{person}{Xia Hu}.}
  \bibinfo{year}{2018}\natexlab{}.
\newblock \showarticletitle{Towards explanation of dnn-based prediction with
  guided feature inversion}. In \bibinfo{booktitle}{\emph{Proceedings of the
  24th ACM SIGKDD International Conference on Knowledge Discovery \& Data
  Mining}}. \bibinfo{pages}{1358--1367}.
\newblock


\bibitem[\protect\citeauthoryear{{Du}, {Liu}, {Yang}, and {Hu}}{{Du}
  et~al\mbox{.}}{2019}]%
        {8970999}
\bibfield{author}{\bibinfo{person}{M. {Du}}, \bibinfo{person}{N. {Liu}},
  \bibinfo{person}{F. {Yang}}, {and} \bibinfo{person}{X. {Hu}}.}
  \bibinfo{year}{2019}\natexlab{}.
\newblock \showarticletitle{Learning credible deep neural networks with
  rationale regularization}. In \bibinfo{booktitle}{\emph{2019 IEEE
  International Conference on Data Mining (ICDM)}}. \bibinfo{pages}{150--159}.
\newblock


\bibitem[\protect\citeauthoryear{Dudley and Kristensson}{Dudley and
  Kristensson}{2018}]%
        {dudley2018review}
\bibfield{author}{\bibinfo{person}{John~J Dudley} {and}
  \bibinfo{person}{Per~Ola Kristensson}.} \bibinfo{year}{2018}\natexlab{}.
\newblock \showarticletitle{A review of user interface design for interactive
  machine learning}.
\newblock \bibinfo{journal}{\emph{ACM Transactions on Interactive Intelligent
  Systems (TiiS)}} \bibinfo{volume}{8}, \bibinfo{number}{2}
  (\bibinfo{year}{2018}), \bibinfo{pages}{8}.
\newblock


\bibitem[\protect\citeauthoryear{Eiband, Buschek, Kremer, and Hussmann}{Eiband
  et~al\mbox{.}}{2019}]%
        {eiband2019impact}
\bibfield{author}{\bibinfo{person}{Malin Eiband}, \bibinfo{person}{Daniel
  Buschek}, \bibinfo{person}{Alexander Kremer}, {and} \bibinfo{person}{Heinrich
  Hussmann}.} \bibinfo{year}{2019}\natexlab{}.
\newblock \showarticletitle{The Impact of Placebic Explanations on Trust in
  Intelligent Systems}. In \bibinfo{booktitle}{\emph{Extended Abstracts of the
  2019 CHI Conference on Human Factors in Computing Systems}}. ACM,
  \bibinfo{pages}{LBW0243}.
\newblock


\bibitem[\protect\citeauthoryear{Eiband, Schneider, Bilandzic, Fazekas-Con,
  Haug, and Hussmann}{Eiband et~al\mbox{.}}{2018}]%
        {Eiband2018practice}
\bibfield{author}{\bibinfo{person}{Malin Eiband}, \bibinfo{person}{Hanna
  Schneider}, \bibinfo{person}{Mark Bilandzic}, \bibinfo{person}{Julian
  Fazekas-Con}, \bibinfo{person}{Mareike Haug}, {and} \bibinfo{person}{Heinrich
  Hussmann}.} \bibinfo{year}{2018}\natexlab{}.
\newblock \showarticletitle{Bringing transparency design into practice}. In
  \bibinfo{booktitle}{\emph{23rd International Conference on Intelligent User
  Interfaces}} \emph{(\bibinfo{series}{IUI '18})}. \bibinfo{publisher}{ACM},
  \bibinfo{address}{New York, NY, USA}, \bibinfo{pages}{211--223}.
\newblock
\showISBNx{978-1-4503-4945-1}
\urldef\tempurl%
\url{https://doi.org/10.1145/3172944.3172961}
\showDOI{\tempurl}


\bibitem[\protect\citeauthoryear{Endert, Ribarsky, Turkay, Wong, Nabney,
  Blanco, and Rossi}{Endert et~al\mbox{.}}{2017}]%
        {endert2017state}
\bibfield{author}{\bibinfo{person}{A Endert}, \bibinfo{person}{W Ribarsky},
  \bibinfo{person}{C Turkay}, \bibinfo{person}{BL Wong}, \bibinfo{person}{Ian
  Nabney}, \bibinfo{person}{I~D{\'\i}az Blanco}, {and} \bibinfo{person}{F
  Rossi}.} \bibinfo{year}{2017}\natexlab{}.
\newblock \showarticletitle{The state of the art in integrating machine
  learning into visual analytics}. In \bibinfo{booktitle}{\emph{Computer
  Graphics Forum}}, Vol.~\bibinfo{volume}{36}. Wiley Online Library,
  \bibinfo{pages}{458--486}.
\newblock


\bibitem[\protect\citeauthoryear{Eslami, Rickman, Vaccaro, Aleyasen, Vuong,
  Karahalios, Hamilton, and Sandvig}{Eslami et~al\mbox{.}}{2015}]%
        {eslami2015always}
\bibfield{author}{\bibinfo{person}{Motahhare Eslami}, \bibinfo{person}{Aimee
  Rickman}, \bibinfo{person}{Kristen Vaccaro}, \bibinfo{person}{Amirhossein
  Aleyasen}, \bibinfo{person}{Andy Vuong}, \bibinfo{person}{Karrie Karahalios},
  \bibinfo{person}{Kevin Hamilton}, {and} \bibinfo{person}{Christian Sandvig}.}
  \bibinfo{year}{2015}\natexlab{}.
\newblock \showarticletitle{I always assumed that I wasn't really that close to
  [her]: Reasoning about Invisible Algorithms in News Feeds}. In
  \bibinfo{booktitle}{\emph{Proceedings of the 33rd Annual ACM Conference on
  Human Factors in Computing Systems}}. ACM, \bibinfo{pages}{153--162}.
\newblock


\bibitem[\protect\citeauthoryear{Eslami, Vaccaro, Karahalios, and
  Hamilton}{Eslami et~al\mbox{.}}{2017}]%
        {eslami2017careful}
\bibfield{author}{\bibinfo{person}{Motahhare Eslami}, \bibinfo{person}{Kristen
  Vaccaro}, \bibinfo{person}{Karrie Karahalios}, {and} \bibinfo{person}{Kevin
  Hamilton}.} \bibinfo{year}{2017}\natexlab{}.
\newblock \showarticletitle{``Be careful; things can be worse than they
  appear'': Understanding Biased Algorithms and Users' Behavior around Them in
  Rating Platforms}. In \bibinfo{booktitle}{\emph{Eleventh International AAAI
  Conference on Web and Social Media}}.
\newblock


\bibitem[\protect\citeauthoryear{Florez-Lopez and Ramon-Jeronimo}{Florez-Lopez
  and Ramon-Jeronimo}{2015}]%
        {florez2015enhancing}
\bibfield{author}{\bibinfo{person}{Raquel Florez-Lopez} {and}
  \bibinfo{person}{Juan~Manuel Ramon-Jeronimo}.}
  \bibinfo{year}{2015}\natexlab{}.
\newblock \showarticletitle{Enhancing accuracy and interpretability of ensemble
  strategies in credit risk assessment. A correlated-adjusted decision forest
  proposal}.
\newblock \bibinfo{journal}{\emph{Expert Systems with Applications}}
  \bibinfo{volume}{42}, \bibinfo{number}{13} (\bibinfo{year}{2015}),
  \bibinfo{pages}{5737--5753}.
\newblock


\bibitem[\protect\citeauthoryear{Fong and Vedaldi}{Fong and Vedaldi}{2017}]%
        {fong2017interpretable}
\bibfield{author}{\bibinfo{person}{Ruth~C Fong} {and} \bibinfo{person}{Andrea
  Vedaldi}.} \bibinfo{year}{2017}\natexlab{}.
\newblock \showarticletitle{Interpretable explanations of black boxes by
  meaningful perturbation}. In \bibinfo{booktitle}{\emph{Proceedings of the
  IEEE International Conference on Computer Vision}}.
  \bibinfo{pages}{3429--3437}.
\newblock


\bibitem[\protect\citeauthoryear{Gedikli, Jannach, and Ge}{Gedikli
  et~al\mbox{.}}{2014}]%
        {gedikli2014should}
\bibfield{author}{\bibinfo{person}{Fatih Gedikli}, \bibinfo{person}{Dietmar
  Jannach}, {and} \bibinfo{person}{Mouzhi Ge}.}
  \bibinfo{year}{2014}\natexlab{}.
\newblock \showarticletitle{How should I explain? A comparison of different
  explanation types for recommender systems}.
\newblock \bibinfo{journal}{\emph{International Journal of Human-Computer
  Studies}} \bibinfo{volume}{72}, \bibinfo{number}{4} (\bibinfo{year}{2014}),
  \bibinfo{pages}{367--382}.
\newblock


\bibitem[\protect\citeauthoryear{Ghorbani, Wexler, Zou, and Kim}{Ghorbani
  et~al\mbox{.}}{2019}]%
        {ghorbani2019towards}
\bibfield{author}{\bibinfo{person}{Amirata Ghorbani}, \bibinfo{person}{James
  Wexler}, \bibinfo{person}{James~Y Zou}, {and} \bibinfo{person}{Been Kim}.}
  \bibinfo{year}{2019}\natexlab{}.
\newblock \showarticletitle{Towards automatic concept-based explanations}. In
  \bibinfo{booktitle}{\emph{Advances in Neural Information Processing
  Systems}}. \bibinfo{pages}{9273--9282}.
\newblock


\bibitem[\protect\citeauthoryear{Gilpin, Bau, Yuan, Bajwa, Specter, and
  Kagal}{Gilpin et~al\mbox{.}}{2018}]%
        {gilpin2018explaining}
\bibfield{author}{\bibinfo{person}{Leilani~H Gilpin}, \bibinfo{person}{David
  Bau}, \bibinfo{person}{Ben~Z Yuan}, \bibinfo{person}{Ayesha Bajwa},
  \bibinfo{person}{Michael Specter}, {and} \bibinfo{person}{Lalana Kagal}.}
  \bibinfo{year}{2018}\natexlab{}.
\newblock \showarticletitle{Explaining Explanations: An Overview of
  Interpretability of Machine Learning}. In \bibinfo{booktitle}{\emph{2018 IEEE
  5th International Conference on Data Science and Advanced Analytics (DSAA)}}.
  IEEE, \bibinfo{pages}{80--89}.
\newblock


\bibitem[\protect\citeauthoryear{Glass, McGuinness, and Wolverton}{Glass
  et~al\mbox{.}}{2008}]%
        {glass2008toward}
\bibfield{author}{\bibinfo{person}{Alyssa Glass}, \bibinfo{person}{Deborah~L
  McGuinness}, {and} \bibinfo{person}{Michael Wolverton}.}
  \bibinfo{year}{2008}\natexlab{}.
\newblock \showarticletitle{Toward establishing trust in adaptive agents}. In
  \bibinfo{booktitle}{\emph{Proceedings of the 13th International Conference on
  Intelligent User Interfaces}}. ACM, \bibinfo{pages}{227--236}.
\newblock


\bibitem[\protect\citeauthoryear{Goodall, Ragan, Steed, Reed, Richardson,
  Huffer, Bridges, and Laska}{Goodall et~al\mbox{.}}{2018}]%
        {goodall2018situ}
\bibfield{author}{\bibinfo{person}{John Goodall}, \bibinfo{person}{Eric~D
  Ragan}, \bibinfo{person}{Chad~A Steed}, \bibinfo{person}{Joel~W Reed},
  \bibinfo{person}{G~David Richardson}, \bibinfo{person}{Kelly~MT Huffer},
  \bibinfo{person}{Robert~A Bridges}, {and} \bibinfo{person}{Jason~A Laska}.}
  \bibinfo{year}{2018}\natexlab{}.
\newblock \showarticletitle{Situ: Identifying and Explaining Suspicious
  Behavior in Networks}.
\newblock \bibinfo{journal}{\emph{IEEE Transactions on Visualization and
  Computer Graphics}} (\bibinfo{year}{2018}).
\newblock


\bibitem[\protect\citeauthoryear{Goodman and Flaxman}{Goodman and
  Flaxman}{2017}]%
        {goodman2016eu}
\bibfield{author}{\bibinfo{person}{Bryce Goodman} {and} \bibinfo{person}{Seth
  Flaxman}.} \bibinfo{year}{2017}\natexlab{}.
\newblock \showarticletitle{European Union regulations on algorithmic
  decision-making and a ``right to explanation''}.
\newblock \bibinfo{journal}{\emph{AI Magazine}} \bibinfo{volume}{38},
  \bibinfo{number}{3} (\bibinfo{year}{2017}), \bibinfo{pages}{50--57}.
\newblock


\bibitem[\protect\citeauthoryear{Gray, Kou, Battles, Hoggatt, and Toombs}{Gray
  et~al\mbox{.}}{2018}]%
        {gray2018dark}
\bibfield{author}{\bibinfo{person}{Colin~M Gray}, \bibinfo{person}{Yubo Kou},
  \bibinfo{person}{Bryan Battles}, \bibinfo{person}{Joseph Hoggatt}, {and}
  \bibinfo{person}{Austin~L Toombs}.} \bibinfo{year}{2018}\natexlab{}.
\newblock \showarticletitle{The dark (patterns) side of UX design}. In
  \bibinfo{booktitle}{\emph{Proceedings of the 2018 CHI Conference on Human
  Factors in Computing Systems}}. ACM, \bibinfo{pages}{534}.
\newblock


\bibitem[\protect\citeauthoryear{Gregor and Benbasat}{Gregor and
  Benbasat}{1999}]%
        {gregor1999explanations}
\bibfield{author}{\bibinfo{person}{Shirley Gregor} {and} \bibinfo{person}{Izak
  Benbasat}.} \bibinfo{year}{1999}\natexlab{}.
\newblock \showarticletitle{Explanations from Intelligent Systems: Theoretical
  Foundations and Implications for Practice}.
\newblock \bibinfo{journal}{\emph{Management Information Systems Quarterly}}
  \bibinfo{volume}{23}, \bibinfo{number}{4} (\bibinfo{year}{1999}),
  \bibinfo{pages}{2}.
\newblock


\bibitem[\protect\citeauthoryear{Groce, Kulesza, Zhang, Shamasunder, Burnett,
  Wong, Stumpf, Das, Shinsel, Bice, et~al\mbox{.}}{Groce et~al\mbox{.}}{2014}]%
        {groce2014you}
\bibfield{author}{\bibinfo{person}{Alex Groce}, \bibinfo{person}{Todd Kulesza},
  \bibinfo{person}{Chaoqiang Zhang}, \bibinfo{person}{Shalini Shamasunder},
  \bibinfo{person}{Margaret Burnett}, \bibinfo{person}{Weng-Keen Wong},
  \bibinfo{person}{Simone Stumpf}, \bibinfo{person}{Shubhomoy Das},
  \bibinfo{person}{Amber Shinsel}, \bibinfo{person}{Forrest Bice},
  {et~al\mbox{.}}} \bibinfo{year}{2014}\natexlab{}.
\newblock \showarticletitle{You are the only possible oracle: Effective test
  selection for end users of interactive machine learning systems}.
\newblock \bibinfo{journal}{\emph{IEEE Transactions on Software Engineering}}
  \bibinfo{volume}{40}, \bibinfo{number}{3} (\bibinfo{year}{2014}),
  \bibinfo{pages}{307--323}.
\newblock


\bibitem[\protect\citeauthoryear{Guidotti, Monreale, Ruggieri, Turini,
  Giannotti, and Pedreschi}{Guidotti et~al\mbox{.}}{2018}]%
        {guidotti2018survey}
\bibfield{author}{\bibinfo{person}{Riccardo Guidotti}, \bibinfo{person}{Anna
  Monreale}, \bibinfo{person}{Salvatore Ruggieri}, \bibinfo{person}{Franco
  Turini}, \bibinfo{person}{Fosca Giannotti}, {and} \bibinfo{person}{Dino
  Pedreschi}.} \bibinfo{year}{2018}\natexlab{}.
\newblock \showarticletitle{A survey of methods for explaining black box
  models}.
\newblock \bibinfo{journal}{\emph{ACM Computing Surveys (CSUR)}}
  \bibinfo{volume}{51}, \bibinfo{number}{5} (\bibinfo{year}{2018}),
  \bibinfo{pages}{93}.
\newblock


\bibitem[\protect\citeauthoryear{Gunning}{Gunning}{2017}]%
        {gunning2017explainable}
\bibfield{author}{\bibinfo{person}{David Gunning}.}
  \bibinfo{year}{2017}\natexlab{}.
\newblock \showarticletitle{Explainable artificial intelligence ({XAI})}.
\newblock \bibinfo{journal}{\emph{Defense Advanced Research Projects Agency
  (DARPA)}} (\bibinfo{year}{2017}).
\newblock


\bibitem[\protect\citeauthoryear{Hannak, Sapiezynski, Molavi~Kakhki,
  Krishnamurthy, Lazer, Mislove, and Wilson}{Hannak et~al\mbox{.}}{2013}]%
        {hannak2013measuring}
\bibfield{author}{\bibinfo{person}{Aniko Hannak}, \bibinfo{person}{Piotr
  Sapiezynski}, \bibinfo{person}{Arash Molavi~Kakhki},
  \bibinfo{person}{Balachander Krishnamurthy}, \bibinfo{person}{David Lazer},
  \bibinfo{person}{Alan Mislove}, {and} \bibinfo{person}{Christo Wilson}.}
  \bibinfo{year}{2013}\natexlab{}.
\newblock \showarticletitle{Measuring personalization of web search}. In
  \bibinfo{booktitle}{\emph{Proceedings of the 22nd International Conference on
  World Wide Web}}. ACM, \bibinfo{pages}{527--538}.
\newblock


\bibitem[\protect\citeauthoryear{Haynes, Cohen, and Ritter}{Haynes
  et~al\mbox{.}}{2009}]%
        {haynes2009designs}
\bibfield{author}{\bibinfo{person}{Steven~R Haynes}, \bibinfo{person}{Mark~A
  Cohen}, {and} \bibinfo{person}{Frank~E Ritter}.}
  \bibinfo{year}{2009}\natexlab{}.
\newblock \showarticletitle{Designs for explaining intelligent agents}.
\newblock \bibinfo{journal}{\emph{International Journal of Human-Computer
  Studies}} \bibinfo{volume}{67}, \bibinfo{number}{1} (\bibinfo{year}{2009}),
  \bibinfo{pages}{90--110}.
\newblock


\bibitem[\protect\citeauthoryear{Heer}{Heer}{2019}]%
        {heer2019agency}
\bibfield{author}{\bibinfo{person}{Jeffrey Heer}.}
  \bibinfo{year}{2019}\natexlab{}.
\newblock \showarticletitle{Agency plus automation: Designing artificial
  intelligence into interactive systems}.
\newblock \bibinfo{journal}{\emph{Proceedings of the National Academy of
  Sciences}} \bibinfo{volume}{116}, \bibinfo{number}{6} (\bibinfo{year}{2019}),
  \bibinfo{pages}{1844--1850}.
\newblock


\bibitem[\protect\citeauthoryear{Hendricks, Burns, Saenko, Darrell, and
  Rohrbach}{Hendricks et~al\mbox{.}}{2018}]%
        {hendricks2018women}
\bibfield{author}{\bibinfo{person}{Lisa~Anne Hendricks},
  \bibinfo{person}{Kaylee Burns}, \bibinfo{person}{Kate Saenko},
  \bibinfo{person}{Trevor Darrell}, {and} \bibinfo{person}{Anna Rohrbach}.}
  \bibinfo{year}{2018}\natexlab{}.
\newblock \showarticletitle{Women also snowboard: Overcoming bias in captioning
  models}. In \bibinfo{booktitle}{\emph{European Conference on Computer
  Vision}}. Springer, \bibinfo{pages}{793--811}.
\newblock


\bibitem[\protect\citeauthoryear{Herlocker, Konstan, and Riedl}{Herlocker
  et~al\mbox{.}}{2000}]%
        {herlocker2000explaining}
\bibfield{author}{\bibinfo{person}{Jonathan~L Herlocker},
  \bibinfo{person}{Joseph~A Konstan}, {and} \bibinfo{person}{John Riedl}.}
  \bibinfo{year}{2000}\natexlab{}.
\newblock \showarticletitle{Explaining collaborative filtering
  recommendations}. In \bibinfo{booktitle}{\emph{Proceedings of the 2000 ACM
  Conference on Computer Supported Cooperative Work}}. ACM,
  \bibinfo{pages}{241--250}.
\newblock


\bibitem[\protect\citeauthoryear{Herman}{Herman}{2017}]%
        {herman2017promise}
\bibfield{author}{\bibinfo{person}{Bernease Herman}.}
  \bibinfo{year}{2017}\natexlab{}.
\newblock \showarticletitle{The Promise and Peril of Human Evaluation for Model
  Interpretability}.
\newblock \bibinfo{journal}{\emph{arXiv preprint arXiv:1711.07414}}
  (\bibinfo{year}{2017}).
\newblock


\bibitem[\protect\citeauthoryear{Hoffman, Miller, Mueller, Klein, and
  Clancey}{Hoffman et~al\mbox{.}}{2018a}]%
        {hoffman2018part4explaining}
\bibfield{author}{\bibinfo{person}{Robert Hoffman}, \bibinfo{person}{Tim
  Miller}, \bibinfo{person}{Shane~T Mueller}, \bibinfo{person}{Gary Klein},
  {and} \bibinfo{person}{William~J Clancey}.} \bibinfo{year}{2018}\natexlab{a}.
\newblock \showarticletitle{Explaining explanation, part 4: a deep dive on deep
  nets}.
\newblock \bibinfo{journal}{\emph{IEEE Intelligent Systems}}
  \bibinfo{volume}{33}, \bibinfo{number}{3} (\bibinfo{year}{2018}),
  \bibinfo{pages}{87--95}.
\newblock


\bibitem[\protect\citeauthoryear{Hoffman}{Hoffman}{2017}]%
        {hoffman2017theory}
\bibfield{author}{\bibinfo{person}{Robert~R Hoffman}.}
  \bibinfo{year}{2017}\natexlab{}.
\newblock \showarticletitle{Theory concepts measures but policies metrics}.
\newblock In \bibinfo{booktitle}{\emph{Macrocognition Metrics and Scenarios}}.
  \bibinfo{publisher}{CRC Press}, \bibinfo{pages}{35--42}.
\newblock


\bibitem[\protect\citeauthoryear{Hoffman, Hawley, and Bradshaw}{Hoffman
  et~al\mbox{.}}{2014}]%
        {hoffman2014myths}
\bibfield{author}{\bibinfo{person}{Robert~R Hoffman}, \bibinfo{person}{John~K
  Hawley}, {and} \bibinfo{person}{Jeffrey~M Bradshaw}.}
  \bibinfo{year}{2014}\natexlab{}.
\newblock \showarticletitle{Myths of automation, part 2: Some very human
  consequences}.
\newblock \bibinfo{journal}{\emph{IEEE Intelligent Systems}}
  \bibinfo{volume}{29}, \bibinfo{number}{2} (\bibinfo{year}{2014}),
  \bibinfo{pages}{82--85}.
\newblock


\bibitem[\protect\citeauthoryear{Hoffman, Johnson, Bradshaw, and
  Underbrink}{Hoffman et~al\mbox{.}}{2013}]%
        {hoffman2013trust}
\bibfield{author}{\bibinfo{person}{Robert~R Hoffman}, \bibinfo{person}{Matthew
  Johnson}, \bibinfo{person}{Jeffrey~M Bradshaw}, {and} \bibinfo{person}{Al
  Underbrink}.} \bibinfo{year}{2013}\natexlab{}.
\newblock \showarticletitle{Trust in automation}.
\newblock \bibinfo{journal}{\emph{IEEE Intelligent Systems}}
  \bibinfo{volume}{28}, \bibinfo{number}{1} (\bibinfo{year}{2013}),
  \bibinfo{pages}{84--88}.
\newblock


\bibitem[\protect\citeauthoryear{Hoffman and Klein}{Hoffman and Klein}{2017}]%
        {hoffman2017part1explaining}
\bibfield{author}{\bibinfo{person}{Robert~R Hoffman} {and}
  \bibinfo{person}{Gary Klein}.} \bibinfo{year}{2017}\natexlab{}.
\newblock \showarticletitle{Explaining explanation, part 1: theoretical
  foundations}.
\newblock \bibinfo{journal}{\emph{IEEE Intelligent Systems}}
  \bibinfo{volume}{32}, \bibinfo{number}{3} (\bibinfo{year}{2017}),
  \bibinfo{pages}{68--73}.
\newblock


\bibitem[\protect\citeauthoryear{Hoffman, Mueller, and Klein}{Hoffman
  et~al\mbox{.}}{2017}]%
        {hoffman2017part2explaining}
\bibfield{author}{\bibinfo{person}{Robert~R Hoffman}, \bibinfo{person}{Shane~T
  Mueller}, {and} \bibinfo{person}{Gary Klein}.}
  \bibinfo{year}{2017}\natexlab{}.
\newblock \showarticletitle{Explaining explanation, part 2: empirical
  foundations}.
\newblock \bibinfo{journal}{\emph{IEEE Intelligent Systems}}
  \bibinfo{volume}{32}, \bibinfo{number}{4} (\bibinfo{year}{2017}),
  \bibinfo{pages}{78--86}.
\newblock


\bibitem[\protect\citeauthoryear{Hoffman, Mueller, Klein, and Litman}{Hoffman
  et~al\mbox{.}}{2018b}]%
        {hoffman2018metrics}
\bibfield{author}{\bibinfo{person}{Robert~R Hoffman}, \bibinfo{person}{Shane~T
  Mueller}, \bibinfo{person}{Gary Klein}, {and} \bibinfo{person}{Jordan
  Litman}.} \bibinfo{year}{2018}\natexlab{b}.
\newblock \showarticletitle{Metrics for explainable AI: challenges and
  prospects}.
\newblock \bibinfo{journal}{\emph{arXiv preprint arXiv:1812.04608}}
  (\bibinfo{year}{2018}).
\newblock


\bibitem[\protect\citeauthoryear{Hohman, Park, Robinson, and Chau}{Hohman
  et~al\mbox{.}}{2019a}]%
        {hohman2019summit}
\bibfield{author}{\bibinfo{person}{Fred Hohman}, \bibinfo{person}{Haekyu Park},
  \bibinfo{person}{Caleb Robinson}, {and} \bibinfo{person}{Duen Horng~Polo
  Chau}.} \bibinfo{year}{2019}\natexlab{a}.
\newblock \showarticletitle{Summit: scaling deep learning interpretability by
  visualizing activation and attribution summarizations}.
\newblock \bibinfo{journal}{\emph{IEEE Transactions on Visualization and
  Computer Graphics}} \bibinfo{volume}{26}, \bibinfo{number}{1}
  (\bibinfo{year}{2019}), \bibinfo{pages}{1096--1106}.
\newblock


\bibitem[\protect\citeauthoryear{Hohman, Srinivasan, and Drucker}{Hohman
  et~al\mbox{.}}{2019b}]%
        {hohman2019telegam}
\bibfield{author}{\bibinfo{person}{Fred Hohman}, \bibinfo{person}{Arjun
  Srinivasan}, {and} \bibinfo{person}{Steven~M. Drucker}.}
  \bibinfo{year}{2019}\natexlab{b}.
\newblock \showarticletitle{TeleGam: combining visualization and verbalization
  for interpretable machine learning}.
\newblock \bibinfo{journal}{\emph{IEEE Visualization Conference (VIS)}}
  (\bibinfo{year}{2019}).
\newblock


\bibitem[\protect\citeauthoryear{Hohman, Kahng, Pienta, and Chau}{Hohman
  et~al\mbox{.}}{2018}]%
        {hohman2018visual}
\bibfield{author}{\bibinfo{person}{Fred~Matthew Hohman},
  \bibinfo{person}{Minsuk Kahng}, \bibinfo{person}{Robert Pienta}, {and}
  \bibinfo{person}{Duen~Horng Chau}.} \bibinfo{year}{2018}\natexlab{}.
\newblock \showarticletitle{Visual Analytics in Deep Learning: An Interrogative
  Survey for the Next Frontiers}.
\newblock \bibinfo{journal}{\emph{IEEE Transactions on Visualization and
  Computer Graphics}} (\bibinfo{year}{2018}).
\newblock


\bibitem[\protect\citeauthoryear{Holliday, Wilson, and Stumpf}{Holliday
  et~al\mbox{.}}{2016}]%
        {holliday2016user}
\bibfield{author}{\bibinfo{person}{Daniel Holliday}, \bibinfo{person}{Stephanie
  Wilson}, {and} \bibinfo{person}{Simone Stumpf}.}
  \bibinfo{year}{2016}\natexlab{}.
\newblock \showarticletitle{User trust in intelligent systems: A journey over
  time}. In \bibinfo{booktitle}{\emph{Proceedings of the 21st International
  Conference on Intelligent User Interfaces}}. ACM, \bibinfo{pages}{164--168}.
\newblock


\bibitem[\protect\citeauthoryear{H{\"o}{\"o}k}{H{\"o}{\"o}k}{2000}]%
        {hook2000steps}
\bibfield{author}{\bibinfo{person}{Kristina H{\"o}{\"o}k}.}
  \bibinfo{year}{2000}\natexlab{}.
\newblock \showarticletitle{Steps to take before intelligent user interfaces
  become real}.
\newblock \bibinfo{journal}{\emph{Interacting with Computers}}
  \bibinfo{volume}{12}, \bibinfo{number}{4} (\bibinfo{year}{2000}),
  \bibinfo{pages}{409--426}.
\newblock


\bibitem[\protect\citeauthoryear{Howard and Kollanyi}{Howard and
  Kollanyi}{2016}]%
        {howard2016bots}
\bibfield{author}{\bibinfo{person}{Philip~N Howard} {and}
  \bibinfo{person}{Bence Kollanyi}.} \bibinfo{year}{2016}\natexlab{}.
\newblock \showarticletitle{Bots, \#StrongerIn, and \#Brexit: computational
  propaganda during the UK-EU referendum}.
\newblock  (\bibinfo{year}{2016}).
\newblock


\bibitem[\protect\citeauthoryear{Hu, Boyd-Graber, Satinoff, and Smith}{Hu
  et~al\mbox{.}}{2014}]%
        {hu2014interactive}
\bibfield{author}{\bibinfo{person}{Yuening Hu}, \bibinfo{person}{Jordan
  Boyd-Graber}, \bibinfo{person}{Brianna Satinoff}, {and}
  \bibinfo{person}{Alison Smith}.} \bibinfo{year}{2014}\natexlab{}.
\newblock \showarticletitle{Interactive topic modeling}.
\newblock \bibinfo{journal}{\emph{Machine Learning}} \bibinfo{volume}{95},
  \bibinfo{number}{3} (\bibinfo{year}{2014}), \bibinfo{pages}{423--469}.
\newblock


\bibitem[\protect\citeauthoryear{Jhaver, Karpfen, and Antin}{Jhaver
  et~al\mbox{.}}{2018}]%
        {jhaver2018algorithmic}
\bibfield{author}{\bibinfo{person}{Shagun Jhaver}, \bibinfo{person}{Yoni
  Karpfen}, {and} \bibinfo{person}{Judd Antin}.}
  \bibinfo{year}{2018}\natexlab{}.
\newblock \showarticletitle{Algorithmic anxiety and coping strategies of Airbnb
  hosts}. In \bibinfo{booktitle}{\emph{Proceedings of the 2018 CHI Conference
  on Human Factors in Computing Systems}}. ACM, \bibinfo{pages}{421}.
\newblock


\bibitem[\protect\citeauthoryear{Jian, Bisantz, and Drury}{Jian
  et~al\mbox{.}}{2000}]%
        {jian2000foundations}
\bibfield{author}{\bibinfo{person}{Jiun-Yin Jian}, \bibinfo{person}{Ann~M
  Bisantz}, {and} \bibinfo{person}{Colin~G Drury}.}
  \bibinfo{year}{2000}\natexlab{}.
\newblock \showarticletitle{Foundations for an empirically determined scale of
  trust in automated systems}.
\newblock \bibinfo{journal}{\emph{International Journal of Cognitive
  Ergonomics}} \bibinfo{volume}{4}, \bibinfo{number}{1} (\bibinfo{year}{2000}),
  \bibinfo{pages}{53--71}.
\newblock


\bibitem[\protect\citeauthoryear{Kahng, Andrews, Kalro, and Chau}{Kahng
  et~al\mbox{.}}{2018}]%
        {kahng2018cti}
\bibfield{author}{\bibinfo{person}{Minsuk Kahng}, \bibinfo{person}{Pierre~Y
  Andrews}, \bibinfo{person}{Aditya Kalro}, {and} \bibinfo{person}{Duen
  Horng~Polo Chau}.} \bibinfo{year}{2018}\natexlab{}.
\newblock \showarticletitle{ActiVis: visual exploration of industry-scale deep
  neural network models}.
\newblock \bibinfo{journal}{\emph{IEEE Transactions on Visualization and
  Computer Graphics}} \bibinfo{volume}{24}, \bibinfo{number}{1}
  (\bibinfo{year}{2018}), \bibinfo{pages}{88--97}.
\newblock


\bibitem[\protect\citeauthoryear{Kay, Kola, Hullman, and Munson}{Kay
  et~al\mbox{.}}{2016}]%
        {kay2016ish}
\bibfield{author}{\bibinfo{person}{Matthew Kay}, \bibinfo{person}{Tara Kola},
  \bibinfo{person}{Jessica~R Hullman}, {and} \bibinfo{person}{Sean~A Munson}.}
  \bibinfo{year}{2016}\natexlab{}.
\newblock \showarticletitle{When (ish) is my bus?: User-centered visualizations
  of uncertainty in everyday, mobile predictive systems}. In
  \bibinfo{booktitle}{\emph{Proceedings of the 2016 CHI Conference on Human
  Factors in Computing Systems}}. ACM, \bibinfo{pages}{5092--5103}.
\newblock


\bibitem[\protect\citeauthoryear{Keil}{Keil}{2006}]%
        {keil2006explanation}
\bibfield{author}{\bibinfo{person}{Frank~C Keil}.}
  \bibinfo{year}{2006}\natexlab{}.
\newblock \showarticletitle{Explanation and understanding}.
\newblock \bibinfo{journal}{\emph{Annu. Rev. Psychol.}}  \bibinfo{volume}{57}
  (\bibinfo{year}{2006}), \bibinfo{pages}{227--254}.
\newblock


\bibitem[\protect\citeauthoryear{Kim, Khanna, and Koyejo}{Kim
  et~al\mbox{.}}{2016}]%
        {kim2016examples}
\bibfield{author}{\bibinfo{person}{Been Kim}, \bibinfo{person}{Rajiv Khanna},
  {and} \bibinfo{person}{Oluwasanmi~O Koyejo}.}
  \bibinfo{year}{2016}\natexlab{}.
\newblock \showarticletitle{Examples are not enough, learn to criticize!
  criticism for interpretability}. In \bibinfo{booktitle}{\emph{Advances in
  Neural Information Processing Systems}}. \bibinfo{pages}{2280--2288}.
\newblock


\bibitem[\protect\citeauthoryear{Kim, Wattenberg, Gilmer, Cai, Wexler, Viegas,
  et~al\mbox{.}}{Kim et~al\mbox{.}}{2018}]%
        {kim2018interpretability}
\bibfield{author}{\bibinfo{person}{Been Kim}, \bibinfo{person}{Martin
  Wattenberg}, \bibinfo{person}{Justin Gilmer}, \bibinfo{person}{Carrie Cai},
  \bibinfo{person}{James Wexler}, \bibinfo{person}{Fernanda Viegas},
  {et~al\mbox{.}}} \bibinfo{year}{2018}\natexlab{}.
\newblock \showarticletitle{Interpretability beyond Feature Attribution:
  Quantitative Testing with Concept Activation Vectors (TCAV)}. In
  \bibinfo{booktitle}{\emph{International Conference on Machine Learning}}.
  \bibinfo{pages}{2673--2682}.
\newblock


\bibitem[\protect\citeauthoryear{Kim and Seo}{Kim and Seo}{2017}]%
        {kim2017human}
\bibfield{author}{\bibinfo{person}{Jaedeok Kim} {and} \bibinfo{person}{Jingoo
  Seo}.} \bibinfo{year}{2017}\natexlab{}.
\newblock \showarticletitle{Human understandable explanation extraction for
  black-box classification models based on matrix factorization}.
\newblock \bibinfo{journal}{\emph{arXiv preprint arXiv:1709.06201}}
  (\bibinfo{year}{2017}).
\newblock


\bibitem[\protect\citeauthoryear{Kindermans, Hooker, Adebayo, Alber,
  Sch{\"u}tt, D{\"a}hne, Erhan, and Kim}{Kindermans et~al\mbox{.}}{2019}]%
        {kindermans2017reliability}
\bibfield{author}{\bibinfo{person}{Pieter-Jan Kindermans},
  \bibinfo{person}{Sara Hooker}, \bibinfo{person}{Julius Adebayo},
  \bibinfo{person}{Maximilian Alber}, \bibinfo{person}{Kristof~T Sch{\"u}tt},
  \bibinfo{person}{Sven D{\"a}hne}, \bibinfo{person}{Dumitru Erhan}, {and}
  \bibinfo{person}{Been Kim}.} \bibinfo{year}{2019}\natexlab{}.
\newblock \showarticletitle{The (un) reliability of saliency methods}.
\newblock In \bibinfo{booktitle}{\emph{Explainable AI: Interpreting, Explaining
  and Visualizing Deep Learning}}. \bibinfo{publisher}{Springer},
  \bibinfo{pages}{267--280}.
\newblock


\bibitem[\protect\citeauthoryear{Klein}{Klein}{2018}]%
        {klein2018part3explaining}
\bibfield{author}{\bibinfo{person}{Gary Klein}.}
  \bibinfo{year}{2018}\natexlab{}.
\newblock \showarticletitle{Explaining explanation, part 3: The causal
  landscape}.
\newblock \bibinfo{journal}{\emph{IEEE Intelligent Systems}}
  \bibinfo{volume}{33}, \bibinfo{number}{2} (\bibinfo{year}{2018}),
  \bibinfo{pages}{83--88}.
\newblock


\bibitem[\protect\citeauthoryear{Kocielnik, Amershi, and Bennett}{Kocielnik
  et~al\mbox{.}}{2019}]%
        {kocielnik2019will}
\bibfield{author}{\bibinfo{person}{Rafal Kocielnik}, \bibinfo{person}{Saleema
  Amershi}, {and} \bibinfo{person}{Paul~N Bennett}.}
  \bibinfo{year}{2019}\natexlab{}.
\newblock \showarticletitle{Will you accept an imperfect ai? exploring designs
  for adjusting end-user expectations of ai systems}. In
  \bibinfo{booktitle}{\emph{Proceedings of the 2019 CHI Conference on Human
  Factors in Computing Systems}}. \bibinfo{pages}{1--14}.
\newblock


\bibitem[\protect\citeauthoryear{Kraus, Scholz, Stiegemeier, and Baumann}{Kraus
  et~al\mbox{.}}{2019}]%
        {kraus2019more}
\bibfield{author}{\bibinfo{person}{Johannes Kraus}, \bibinfo{person}{David
  Scholz}, \bibinfo{person}{Dina Stiegemeier}, {and} \bibinfo{person}{Martin
  Baumann}.} \bibinfo{year}{2019}\natexlab{}.
\newblock \showarticletitle{The more you know: Trust dynamics and calibration
  in highly automated driving and the effects of take-overs, system
  malfunction, and system transparency}.
\newblock \bibinfo{journal}{\emph{Human Factors}} (\bibinfo{year}{2019}),
  \bibinfo{pages}{0018720819853686}.
\newblock


\bibitem[\protect\citeauthoryear{Krause, Dasgupta, Swartz, Aphinyanaphongs, and
  Bertini}{Krause et~al\mbox{.}}{2017}]%
        {Krause2017workflow}
\bibfield{author}{\bibinfo{person}{Josua Krause}, \bibinfo{person}{Aritra
  Dasgupta}, \bibinfo{person}{Jordan Swartz}, \bibinfo{person}{Yindalon
  Aphinyanaphongs}, {and} \bibinfo{person}{Enrico Bertini}.}
  \bibinfo{year}{2017}\natexlab{}.
\newblock \showarticletitle{A workflow for visual diagnostics of binary
  classifiers using instance-level explanations}. In
  \bibinfo{booktitle}{\emph{2017 IEEE Conference on Visual Analytics Science
  and Technology (VAST)}}. IEEE, \bibinfo{pages}{162--172}.
\newblock


\bibitem[\protect\citeauthoryear{Krause, Perer, and Bertini}{Krause
  et~al\mbox{.}}{2014}]%
        {Krause2014}
\bibfield{author}{\bibinfo{person}{Josua Krause}, \bibinfo{person}{Adam Perer},
  {and} \bibinfo{person}{Enrico Bertini}.} \bibinfo{year}{2014}\natexlab{}.
\newblock \showarticletitle{INFUSE: interactive feature selection for
  predictive modeling of high dimensional data}.
\newblock \bibinfo{journal}{\emph{IEEE Transactions on Visualization and
  Computer Graphics}} \bibinfo{volume}{20}, \bibinfo{number}{12}
  (\bibinfo{year}{2014}), \bibinfo{pages}{1614--1623}.
\newblock


\bibitem[\protect\citeauthoryear{Krause, Perer, and Ng}{Krause
  et~al\mbox{.}}{2016}]%
        {krause2016interacting}
\bibfield{author}{\bibinfo{person}{Josua Krause}, \bibinfo{person}{Adam Perer},
  {and} \bibinfo{person}{Kenney Ng}.} \bibinfo{year}{2016}\natexlab{}.
\newblock \showarticletitle{Interacting with predictions: Visual inspection of
  black-box machine learning models}. In \bibinfo{booktitle}{\emph{Proceedings
  of the 2016 CHI Conference on Human Factors in Computing Systems}}. ACM,
  \bibinfo{pages}{5686--5697}.
\newblock


\bibitem[\protect\citeauthoryear{Kulesza, Burnett, Wong, and Stumpf}{Kulesza
  et~al\mbox{.}}{2015}]%
        {kulesza2015principles}
\bibfield{author}{\bibinfo{person}{Todd Kulesza}, \bibinfo{person}{Margaret
  Burnett}, \bibinfo{person}{Weng-Keen Wong}, {and} \bibinfo{person}{Simone
  Stumpf}.} \bibinfo{year}{2015}\natexlab{}.
\newblock \showarticletitle{Principles of explanatory debugging to personalize
  interactive machine learning}. In \bibinfo{booktitle}{\emph{Proceedings of
  the 20th International Conference on Intelligent User Interfaces}}. ACM,
  \bibinfo{pages}{126--137}.
\newblock


\bibitem[\protect\citeauthoryear{Kulesza, Stumpf, Burnett, and Kwan}{Kulesza
  et~al\mbox{.}}{2012}]%
        {Kulesza2012Tell}
\bibfield{author}{\bibinfo{person}{Todd Kulesza}, \bibinfo{person}{Simone
  Stumpf}, \bibinfo{person}{Margaret Burnett}, {and} \bibinfo{person}{Irwin
  Kwan}.} \bibinfo{year}{2012}\natexlab{}.
\newblock \showarticletitle{Tell Me More?: The Effects of Mental Model
  Soundness on Personalizing an Intelligent Agent}. In
  \bibinfo{booktitle}{\emph{Proceedings of the SIGCHI Conference on Human
  Factors in Computing Systems}} \emph{(\bibinfo{series}{CHI '12})}.
  \bibinfo{publisher}{ACM}, \bibinfo{address}{New York, NY, USA},
  \bibinfo{pages}{1--10}.
\newblock
\showISBNx{978-1-4503-1015-4}


\bibitem[\protect\citeauthoryear{Kulesza, Stumpf, Burnett, Wong, Riche, Moore,
  Oberst, Shinsel, and McIntosh}{Kulesza et~al\mbox{.}}{2010}]%
        {kulesza2010explanatory}
\bibfield{author}{\bibinfo{person}{Todd Kulesza}, \bibinfo{person}{Simone
  Stumpf}, \bibinfo{person}{Margaret Burnett}, \bibinfo{person}{Weng-Keen
  Wong}, \bibinfo{person}{Yann Riche}, \bibinfo{person}{Travis Moore},
  \bibinfo{person}{Ian Oberst}, \bibinfo{person}{Amber Shinsel}, {and}
  \bibinfo{person}{Kevin McIntosh}.} \bibinfo{year}{2010}\natexlab{}.
\newblock \showarticletitle{Explanatory debugging: Supporting end-user
  debugging of machine-learned programs}. In \bibinfo{booktitle}{\emph{Visual
  Languages and Human-Centric Computing (VL/HCC), 2010 IEEE Symposium on}}.
  IEEE, \bibinfo{pages}{41--48}.
\newblock


\bibitem[\protect\citeauthoryear{Kulesza, Stumpf, Burnett, Yang, Kwan, and
  Wong}{Kulesza et~al\mbox{.}}{2013}]%
        {kulesza2013too}
\bibfield{author}{\bibinfo{person}{Todd Kulesza}, \bibinfo{person}{Simone
  Stumpf}, \bibinfo{person}{Margaret Burnett}, \bibinfo{person}{Sherry Yang},
  \bibinfo{person}{Irwin Kwan}, {and} \bibinfo{person}{Weng-Keen Wong}.}
  \bibinfo{year}{2013}\natexlab{}.
\newblock \showarticletitle{Too much, too little, or just right? Ways
  explanations impact end users' mental models}. In
  \bibinfo{booktitle}{\emph{Visual Languages and Human-Centric Computing
  (VL/HCC), 2013 IEEE Symposium on}}. IEEE, \bibinfo{pages}{3--10}.
\newblock


\bibitem[\protect\citeauthoryear{Lage, Chen, He, Narayanan, Kim, Gershman, and
  Doshi-Velez}{Lage et~al\mbox{.}}{2019}]%
        {lage2019human}
\bibfield{author}{\bibinfo{person}{Isaac Lage}, \bibinfo{person}{Emily Chen},
  \bibinfo{person}{Jeffrey He}, \bibinfo{person}{Menaka Narayanan},
  \bibinfo{person}{Been Kim}, \bibinfo{person}{Samuel~J Gershman}, {and}
  \bibinfo{person}{Finale Doshi-Velez}.} \bibinfo{year}{2019}\natexlab{}.
\newblock \showarticletitle{Human evaluation of models built for
  interpretability}. In \bibinfo{booktitle}{\emph{Proceedings of the AAAI
  Conference on Human Computation and Crowdsourcing}},
  Vol.~\bibinfo{volume}{7}. \bibinfo{pages}{59--67}.
\newblock


\bibitem[\protect\citeauthoryear{Lakkaraju, Bach, and Leskovec}{Lakkaraju
  et~al\mbox{.}}{2016}]%
        {lakkaraju2016interpretable}
\bibfield{author}{\bibinfo{person}{Himabindu Lakkaraju},
  \bibinfo{person}{Stephen~H Bach}, {and} \bibinfo{person}{Jure Leskovec}.}
  \bibinfo{year}{2016}\natexlab{}.
\newblock \showarticletitle{Interpretable decision sets: A joint framework for
  description and prediction}. In \bibinfo{booktitle}{\emph{Proceedings of the
  22nd ACM SIGKDD International Conference on Knowledge Discovery and Data
  Mining}}. ACM, \bibinfo{pages}{1675--1684}.
\newblock


\bibitem[\protect\citeauthoryear{Langer, Blank, and Chanowitz}{Langer
  et~al\mbox{.}}{1978}]%
        {langer1978mindlessness}
\bibfield{author}{\bibinfo{person}{Ellen~J Langer}, \bibinfo{person}{Arthur
  Blank}, {and} \bibinfo{person}{Benzion Chanowitz}.}
  \bibinfo{year}{1978}\natexlab{}.
\newblock \showarticletitle{The mindlessness of ostensibly thoughtful action:
  The role of ``placebic'' information in interpersonal interaction.}
\newblock \bibinfo{journal}{\emph{Journal of Personality and Social
  Psychology}} \bibinfo{volume}{36}, \bibinfo{number}{6}
  (\bibinfo{year}{1978}), \bibinfo{pages}{635}.
\newblock


\bibitem[\protect\citeauthoryear{Lee, Jain, Cha, Ojha, and Kusbit}{Lee
  et~al\mbox{.}}{2019}]%
        {lee2019procedural}
\bibfield{author}{\bibinfo{person}{Min~Kyung Lee}, \bibinfo{person}{Anuraag
  Jain}, \bibinfo{person}{Hea~Jin Cha}, \bibinfo{person}{Shashank Ojha}, {and}
  \bibinfo{person}{Daniel Kusbit}.} \bibinfo{year}{2019}\natexlab{}.
\newblock \showarticletitle{Procedural justice in algorithmic fairness:
  Leveraging transparency and outcome control for fair algorithmic mediation}.
\newblock \bibinfo{journal}{\emph{Proceedings of the ACM on Human-Computer
  Interaction}} \bibinfo{volume}{3}, \bibinfo{number}{CSCW}
  (\bibinfo{year}{2019}), \bibinfo{pages}{182}.
\newblock


\bibitem[\protect\citeauthoryear{Lee, Kusbit, Metsky, and Dabbish}{Lee
  et~al\mbox{.}}{2015}]%
        {lee2015working}
\bibfield{author}{\bibinfo{person}{Min~Kyung Lee}, \bibinfo{person}{Daniel
  Kusbit}, \bibinfo{person}{Evan Metsky}, {and} \bibinfo{person}{Laura
  Dabbish}.} \bibinfo{year}{2015}\natexlab{}.
\newblock \showarticletitle{Working with machines: The impact of algorithmic
  and data-driven management on human workers}. In
  \bibinfo{booktitle}{\emph{Proceedings of the 33rd Annual ACM Conference on
  Human Factors in Computing Systems}}. ACM, \bibinfo{pages}{1603--1612}.
\newblock


\bibitem[\protect\citeauthoryear{Lepri, Oliver, Letouz{\'e}, Pentland, and
  Vinck}{Lepri et~al\mbox{.}}{2017}]%
        {lepri2017fair}
\bibfield{author}{\bibinfo{person}{Bruno Lepri}, \bibinfo{person}{Nuria
  Oliver}, \bibinfo{person}{Emmanuel Letouz{\'e}}, \bibinfo{person}{Alex
  Pentland}, {and} \bibinfo{person}{Patrick Vinck}.}
  \bibinfo{year}{2017}\natexlab{}.
\newblock \showarticletitle{Fair, Transparent, and Accountable Algorithmic
  Decision-making Processes}.
\newblock \bibinfo{journal}{\emph{Philosophy \& Technology}}
  (\bibinfo{year}{2017}), \bibinfo{pages}{1--17}.
\newblock


\bibitem[\protect\citeauthoryear{Lertvittayakumjorn and
  Toni}{Lertvittayakumjorn and Toni}{2019}]%
        {Lertvittayakumjorn2019human}
\bibfield{author}{\bibinfo{person}{Piyawat Lertvittayakumjorn} {and}
  \bibinfo{person}{Francesca Toni}.} \bibinfo{year}{2019}\natexlab{}.
\newblock \showarticletitle{Human-grounded Evaluations of Explanation Methods
  for Text Classification}. In \bibinfo{booktitle}{\emph{Proceedings of the
  2019 Conference on Empirical Methods in Natural Language Processing and the
  9th International Joint Conference on Natural Language Processing
  (EMNLP-IJCNLP)}}. \bibinfo{pages}{5198--5208}.
\newblock


\bibitem[\protect\citeauthoryear{Letham, Rudin, McCormick, Madigan,
  et~al\mbox{.}}{Letham et~al\mbox{.}}{2015}]%
        {letham2015interpretable}
\bibfield{author}{\bibinfo{person}{Benjamin Letham}, \bibinfo{person}{Cynthia
  Rudin}, \bibinfo{person}{Tyler~H McCormick}, \bibinfo{person}{David Madigan},
  {et~al\mbox{.}}} \bibinfo{year}{2015}\natexlab{}.
\newblock \showarticletitle{Interpretable classifiers using rules and bayesian
  analysis: Building a better stroke prediction model}.
\newblock \bibinfo{journal}{\emph{The Annals of Applied Statistics}}
  \bibinfo{volume}{9}, \bibinfo{number}{3} (\bibinfo{year}{2015}),
  \bibinfo{pages}{1350--1371}.
\newblock


\bibitem[\protect\citeauthoryear{Lex, Streit, Schulz, Partl, Schmalstieg, Park,
  and Gehlenborg}{Lex et~al\mbox{.}}{2012}]%
        {Lex2012}
\bibfield{author}{\bibinfo{person}{Alexander Lex}, \bibinfo{person}{Marc
  Streit}, \bibinfo{person}{H-J Schulz}, \bibinfo{person}{Christian Partl},
  \bibinfo{person}{Dieter Schmalstieg}, \bibinfo{person}{Peter~J Park}, {and}
  \bibinfo{person}{Nils Gehlenborg}.} \bibinfo{year}{2012}\natexlab{}.
\newblock \showarticletitle{StratomeX: Visual Analysis of Large-Scale
  Heterogeneous Genomics Data for Cancer Subtype Characterization}. In
  \bibinfo{booktitle}{\emph{Computer Graphics Forum}},
  Vol.~\bibinfo{volume}{31}. Wiley Online Library, \bibinfo{pages}{1175--1184}.
\newblock


\bibitem[\protect\citeauthoryear{Li, Wu, Peng, Ernst, and Fu}{Li
  et~al\mbox{.}}{2018}]%
        {li2018tell}
\bibfield{author}{\bibinfo{person}{Kunpeng Li}, \bibinfo{person}{Ziyan Wu},
  \bibinfo{person}{Kuan-Chuan Peng}, \bibinfo{person}{Jan Ernst}, {and}
  \bibinfo{person}{Yun Fu}.} \bibinfo{year}{2018}\natexlab{}.
\newblock \showarticletitle{Tell me where to look: Guided attention inference
  network}. In \bibinfo{booktitle}{\emph{Proceedings of the IEEE Conference on
  Computer Vision and Pattern Recognition}}. \bibinfo{pages}{9215--9223}.
\newblock


\bibitem[\protect\citeauthoryear{Li, Zhang, Li, Li, and Fu}{Li
  et~al\mbox{.}}{2019}]%
        {li2019attention}
\bibfield{author}{\bibinfo{person}{Kunpeng Li}, \bibinfo{person}{Yulun Zhang},
  \bibinfo{person}{Kai Li}, \bibinfo{person}{Yuanyuan Li}, {and}
  \bibinfo{person}{Yun Fu}.} \bibinfo{year}{2019}\natexlab{}.
\newblock \showarticletitle{Attention bridging network for knowledge transfer}.
  In \bibinfo{booktitle}{\emph{Proceedings of the IEEE International Conference
  on Computer Vision}}. \bibinfo{pages}{5198--5207}.
\newblock


\bibitem[\protect\citeauthoryear{Lim}{Lim}{2011}]%
        {lim2011improving}
\bibfield{author}{\bibinfo{person}{Brian Lim}.}
  \bibinfo{year}{2011}\natexlab{}.
\newblock \showarticletitle{Improving Understanding, Trust, and Control with
  Intelligibility in Context-Aware Applications}.
\newblock \bibinfo{journal}{\emph{Human-Computer Interaction}}
  (\bibinfo{year}{2011}).
\newblock


\bibitem[\protect\citeauthoryear{Lim and Dey}{Lim and Dey}{2009}]%
        {lim2009assessing}
\bibfield{author}{\bibinfo{person}{Brian~Y Lim} {and} \bibinfo{person}{Anind~K
  Dey}.} \bibinfo{year}{2009}\natexlab{}.
\newblock \showarticletitle{Assessing demand for intelligibility in
  context-aware applications}. In \bibinfo{booktitle}{\emph{Proceedings of the
  11th International Conference on Ubiquitous Computing}}. ACM,
  \bibinfo{pages}{195--204}.
\newblock


\bibitem[\protect\citeauthoryear{Lim, Dey, and Avrahami}{Lim
  et~al\mbox{.}}{2009}]%
        {lim2009and}
\bibfield{author}{\bibinfo{person}{Brian~Y Lim}, \bibinfo{person}{Anind~K Dey},
  {and} \bibinfo{person}{Daniel Avrahami}.} \bibinfo{year}{2009}\natexlab{}.
\newblock \showarticletitle{Why and why not explanations improve the
  intelligibility of context-aware intelligent systems}. In
  \bibinfo{booktitle}{\emph{Proceedings of the SIGCHI Conference on Human
  Factors in Computing Systems}}. ACM, \bibinfo{pages}{2119--2128}.
\newblock


\bibitem[\protect\citeauthoryear{Lim, Yang, Abdul, and Wang}{Lim
  et~al\mbox{.}}{2019}]%
        {lim2019these}
\bibfield{author}{\bibinfo{person}{Brian~Y Lim}, \bibinfo{person}{Qian Yang},
  \bibinfo{person}{Ashraf~M Abdul}, {and} \bibinfo{person}{Danding Wang}.}
  \bibinfo{year}{2019}\natexlab{}.
\newblock \showarticletitle{Why these explanations? Selecting intelligibility
  types for explanation Goals}. In \bibinfo{booktitle}{\emph{IUI Workshops}}.
\newblock


\bibitem[\protect\citeauthoryear{Lipton}{Lipton}{2016}]%
        {lipton2016mythos}
\bibfield{author}{\bibinfo{person}{Zachary~C Lipton}.}
  \bibinfo{year}{2016}\natexlab{}.
\newblock \showarticletitle{The mythos of model interpretability}.
\newblock \bibinfo{journal}{\emph{arXiv preprint arXiv:1606.03490}}
  (\bibinfo{year}{2016}).
\newblock


\bibitem[\protect\citeauthoryear{Liu, Liu, Zhu, Liao, Wei, and Pan}{Liu
  et~al\mbox{.}}{2016}]%
        {liu2016uncertainty}
\bibfield{author}{\bibinfo{person}{Mengchen Liu}, \bibinfo{person}{Shixia Liu},
  \bibinfo{person}{Xizhou Zhu}, \bibinfo{person}{Qinying Liao},
  \bibinfo{person}{Furu Wei}, {and} \bibinfo{person}{Shimei Pan}.}
  \bibinfo{year}{2016}\natexlab{}.
\newblock \showarticletitle{An uncertainty-aware approach for exploratory
  microblog retrieval}.
\newblock \bibinfo{journal}{\emph{IEEE Transactions on Visualization and
  Computer Graphics}} \bibinfo{volume}{22}, \bibinfo{number}{1}
  (\bibinfo{year}{2016}), \bibinfo{pages}{250--259}.
\newblock


\bibitem[\protect\citeauthoryear{Liu, Shi, Cao, Zhu, and Liu}{Liu
  et~al\mbox{.}}{2018}]%
        {liu2018analyzing}
\bibfield{author}{\bibinfo{person}{Mengchen Liu}, \bibinfo{person}{Jiaxin Shi},
  \bibinfo{person}{Kelei Cao}, \bibinfo{person}{Jun Zhu}, {and}
  \bibinfo{person}{Shixia Liu}.} \bibinfo{year}{2018}\natexlab{}.
\newblock \showarticletitle{Analyzing the training processes of deep generative
  models}.
\newblock \bibinfo{journal}{\emph{IEEE Transactions on Visualization and
  Computer Graphics}} \bibinfo{volume}{24}, \bibinfo{number}{1}
  (\bibinfo{year}{2018}), \bibinfo{pages}{77--87}.
\newblock


\bibitem[\protect\citeauthoryear{Liu, Shi, Li, Li, Zhu, and Liu}{Liu
  et~al\mbox{.}}{2017}]%
        {liu2017towards}
\bibfield{author}{\bibinfo{person}{Mengchen Liu}, \bibinfo{person}{Jiaxin Shi},
  \bibinfo{person}{Zhen Li}, \bibinfo{person}{Chongxuan Li},
  \bibinfo{person}{Jun Zhu}, {and} \bibinfo{person}{Shixia Liu}.}
  \bibinfo{year}{2017}\natexlab{}.
\newblock \showarticletitle{Towards better analysis of deep convolutional
  neural networks}.
\newblock \bibinfo{journal}{\emph{IEEE Transactions on Visualization and
  Computer Graphics}} \bibinfo{volume}{23}, \bibinfo{number}{1}
  (\bibinfo{year}{2017}), \bibinfo{pages}{91--100}.
\newblock


\bibitem[\protect\citeauthoryear{Liu, Wang, Chen, Zhu, and Guo}{Liu
  et~al\mbox{.}}{2014}]%
        {liu2014topicpanorama}
\bibfield{author}{\bibinfo{person}{Shixia Liu}, \bibinfo{person}{Xiting Wang},
  \bibinfo{person}{Jianfei Chen}, \bibinfo{person}{Jim Zhu}, {and}
  \bibinfo{person}{Baining Guo}.} \bibinfo{year}{2014}\natexlab{}.
\newblock \showarticletitle{TopicPanorama: A full picture of relevant topics}.
  In \bibinfo{booktitle}{\emph{Visual Analytics Science and Technology (VAST),
  2014 IEEE Conference on}}. IEEE, \bibinfo{pages}{183--192}.
\newblock


\bibitem[\protect\citeauthoryear{Lombrozo}{Lombrozo}{2006}]%
        {lombrozo2006structure}
\bibfield{author}{\bibinfo{person}{Tania Lombrozo}.}
  \bibinfo{year}{2006}\natexlab{}.
\newblock \showarticletitle{The structure and function of explanations}.
\newblock \bibinfo{journal}{\emph{Trends in Cognitive Sciences}}
  \bibinfo{volume}{10}, \bibinfo{number}{10} (\bibinfo{year}{2006}),
  \bibinfo{pages}{464--470}.
\newblock


\bibitem[\protect\citeauthoryear{Lombrozo}{Lombrozo}{2009}]%
        {lombrozo2009explanation}
\bibfield{author}{\bibinfo{person}{Tania Lombrozo}.}
  \bibinfo{year}{2009}\natexlab{}.
\newblock \showarticletitle{Explanation and categorization: How ``why?''
  informs ``what?''}.
\newblock \bibinfo{journal}{\emph{Cognition}} \bibinfo{volume}{110},
  \bibinfo{number}{2} (\bibinfo{year}{2009}), \bibinfo{pages}{248--253}.
\newblock


\bibitem[\protect\citeauthoryear{Lundberg and Lee}{Lundberg and Lee}{2017}]%
        {lundberg2017unified}
\bibfield{author}{\bibinfo{person}{Scott~M Lundberg} {and}
  \bibinfo{person}{Su-In Lee}.} \bibinfo{year}{2017}\natexlab{}.
\newblock \showarticletitle{A unified approach to interpreting model
  predictions}. In \bibinfo{booktitle}{\emph{Advances in Neural Information
  Processing Systems}}. \bibinfo{pages}{4765--4774}.
\newblock


\bibitem[\protect\citeauthoryear{Maaten and Hinton}{Maaten and Hinton}{2008}]%
        {maaten2008visualizing}
\bibfield{author}{\bibinfo{person}{Laurens van~der Maaten} {and}
  \bibinfo{person}{Geoffrey Hinton}.} \bibinfo{year}{2008}\natexlab{}.
\newblock \showarticletitle{Visualizing data using t-SNE}.
\newblock \bibinfo{journal}{\emph{Journal of Machine Learning Research}}
  \bibinfo{volume}{9}, \bibinfo{number}{Nov} (\bibinfo{year}{2008}),
  \bibinfo{pages}{2579--2605}.
\newblock


\bibitem[\protect\citeauthoryear{Madsen and Gregor}{Madsen and Gregor}{2000}]%
        {madsen2000measuring}
\bibfield{author}{\bibinfo{person}{Maria Madsen} {and} \bibinfo{person}{Shirley
  Gregor}.} \bibinfo{year}{2000}\natexlab{}.
\newblock \showarticletitle{Measuring human-computer trust}. In
  \bibinfo{booktitle}{\emph{11th Australasian Conference on Information
  Systems}}, Vol.~\bibinfo{volume}{53}. Citeseer, \bibinfo{pages}{6--8}.
\newblock


\bibitem[\protect\citeauthoryear{Mehrabi, Morstatter, Saxena, Lerman, and
  Galstyan}{Mehrabi et~al\mbox{.}}{2019}]%
        {mehrabi2019survey}
\bibfield{author}{\bibinfo{person}{Ninareh Mehrabi}, \bibinfo{person}{Fred
  Morstatter}, \bibinfo{person}{Nripsuta Saxena}, \bibinfo{person}{Kristina
  Lerman}, {and} \bibinfo{person}{Aram Galstyan}.}
  \bibinfo{year}{2019}\natexlab{}.
\newblock \showarticletitle{A survey on bias and fairness in machine learning}.
\newblock \bibinfo{journal}{\emph{arXiv preprint arXiv:1908.09635}}
  (\bibinfo{year}{2019}).
\newblock


\bibitem[\protect\citeauthoryear{Mennicken, Vermeulen, and Huang}{Mennicken
  et~al\mbox{.}}{2014}]%
        {mennicken2014today}
\bibfield{author}{\bibinfo{person}{Sarah Mennicken}, \bibinfo{person}{Jo
  Vermeulen}, {and} \bibinfo{person}{Elaine~M Huang}.}
  \bibinfo{year}{2014}\natexlab{}.
\newblock \showarticletitle{From today's augmented houses to tomorrow's smart
  homes: new directions for home automation research}. In
  \bibinfo{booktitle}{\emph{Proceedings of the 2014 ACM International Joint
  Conference on Pervasive and Ubiquitous Computing}}. ACM,
  \bibinfo{pages}{105--115}.
\newblock


\bibitem[\protect\citeauthoryear{Merritt, Heimbaugh, LaChapell, and
  Lee}{Merritt et~al\mbox{.}}{2013}]%
        {merritt2013trust}
\bibfield{author}{\bibinfo{person}{Stephanie~M Merritt},
  \bibinfo{person}{Heather Heimbaugh}, \bibinfo{person}{Jennifer LaChapell},
  {and} \bibinfo{person}{Deborah Lee}.} \bibinfo{year}{2013}\natexlab{}.
\newblock \showarticletitle{I trust it, but I don't know why: Effects of
  implicit attitudes toward automation on trust in an automated system}.
\newblock \bibinfo{journal}{\emph{Human Factors}} \bibinfo{volume}{55},
  \bibinfo{number}{3} (\bibinfo{year}{2013}), \bibinfo{pages}{520--534}.
\newblock


\bibitem[\protect\citeauthoryear{Meyer, Sedlmair, Quinan, and Munzner}{Meyer
  et~al\mbox{.}}{2015}]%
        {meyer2015nested}
\bibfield{author}{\bibinfo{person}{Miriah Meyer}, \bibinfo{person}{Michael
  Sedlmair}, \bibinfo{person}{P~Samuel Quinan}, {and} \bibinfo{person}{Tamara
  Munzner}.} \bibinfo{year}{2015}\natexlab{}.
\newblock \showarticletitle{The nested blocks and guidelines model}.
\newblock \bibinfo{journal}{\emph{Information Visualization}}
  \bibinfo{volume}{14}, \bibinfo{number}{3} (\bibinfo{year}{2015}),
  \bibinfo{pages}{234--249}.
\newblock


\bibitem[\protect\citeauthoryear{Meyerson, Weick, and Kramer}{Meyerson
  et~al\mbox{.}}{1996}]%
        {meyerson1996swift}
\bibfield{author}{\bibinfo{person}{Debra Meyerson}, \bibinfo{person}{Karl~E
  Weick}, {and} \bibinfo{person}{Roderick~M Kramer}.}
  \bibinfo{year}{1996}\natexlab{}.
\newblock \showarticletitle{Swift trust and temporary groups}.
\newblock \bibinfo{journal}{\emph{Trust in organizations: Frontiers of theory
  and research}}  \bibinfo{volume}{166} (\bibinfo{year}{1996}),
  \bibinfo{pages}{195}.
\newblock


\bibitem[\protect\citeauthoryear{Miller}{Miller}{2019}]%
        {miller2017explanation}
\bibfield{author}{\bibinfo{person}{Tim Miller}.}
  \bibinfo{year}{2019}\natexlab{}.
\newblock \showarticletitle{Explanation in artificial intelligence: Insights
  from the social sciences}.
\newblock \bibinfo{journal}{\emph{Artificial Intelligence}}
  \bibinfo{volume}{267} (\bibinfo{year}{2019}), \bibinfo{pages}{1--38}.
\newblock


\bibitem[\protect\citeauthoryear{Ming, Cao, Zhang, Li, Chen, Song, and Qu}{Ming
  et~al\mbox{.}}{2017}]%
        {ming2017understanding}
\bibfield{author}{\bibinfo{person}{Yao Ming}, \bibinfo{person}{Shaozu Cao},
  \bibinfo{person}{Ruixiang Zhang}, \bibinfo{person}{Zhen Li},
  \bibinfo{person}{Yuanzhe Chen}, \bibinfo{person}{Yangqiu Song}, {and}
  \bibinfo{person}{Huamin Qu}.} \bibinfo{year}{2017}\natexlab{}.
\newblock \showarticletitle{Understanding hidden memories of recurrent neural
  networks}. In \bibinfo{booktitle}{\emph{2017 IEEE Conference on Visual
  Analytics Science and Technology (VAST)}}. IEEE, \bibinfo{pages}{13--24}.
\newblock


\bibitem[\protect\citeauthoryear{Ming, Qu, and Bertini}{Ming
  et~al\mbox{.}}{2018}]%
        {ming2018rulematrix}
\bibfield{author}{\bibinfo{person}{Yao Ming}, \bibinfo{person}{Huamin Qu},
  {and} \bibinfo{person}{Enrico Bertini}.} \bibinfo{year}{2018}\natexlab{}.
\newblock \showarticletitle{Rulematrix: Visualizing and understanding
  classifiers with rules}.
\newblock \bibinfo{journal}{\emph{IEEE Transactions on Visualization and
  Computer Graphics}} \bibinfo{volume}{25}, \bibinfo{number}{1}
  (\bibinfo{year}{2018}), \bibinfo{pages}{342--352}.
\newblock


\bibitem[\protect\citeauthoryear{Mittelstadt}{Mittelstadt}{2016}]%
        {mittelstadt2016automation}
\bibfield{author}{\bibinfo{person}{Brent Mittelstadt}.}
  \bibinfo{year}{2016}\natexlab{}.
\newblock \showarticletitle{Automation, algorithms, and politics: Auditing for
  transparency in content personalization systems}.
\newblock \bibinfo{journal}{\emph{International Journal of Communication}}
  \bibinfo{volume}{10} (\bibinfo{year}{2016}), \bibinfo{pages}{12}.
\newblock


\bibitem[\protect\citeauthoryear{Mohseni, Jagadeesh, and Wang}{Mohseni
  et~al\mbox{.}}{2019a}]%
        {mohseni2019predicting}
\bibfield{author}{\bibinfo{person}{Sina Mohseni}, \bibinfo{person}{Akshay
  Jagadeesh}, {and} \bibinfo{person}{Zhangyang Wang}.}
  \bibinfo{year}{2019}\natexlab{a}.
\newblock \showarticletitle{Predicting model failure using saliency maps in
  autonomous driving systems}.
\newblock \bibinfo{journal}{\emph{ICML Workshop on Uncertainty \& Robustness in
  Deep Learning}} (\bibinfo{year}{2019}).
\newblock


\bibitem[\protect\citeauthoryear{Mohseni, Pitale, Singh, and Wang}{Mohseni
  et~al\mbox{.}}{2020a}]%
        {mohseni2020practical}
\bibfield{author}{\bibinfo{person}{Sina Mohseni}, \bibinfo{person}{Mandar
  Pitale}, \bibinfo{person}{Vasu Singh}, {and} \bibinfo{person}{Zhangyang
  Wang}.} \bibinfo{year}{2020}\natexlab{a}.
\newblock \showarticletitle{Practical solutions for machine learning safety in
  autonomous vehicles}. In \bibinfo{booktitle}{\emph{The AAAI Workshop on
  Artificial Intelligence Safety (Safe AI)}}.
\newblock


\bibitem[\protect\citeauthoryear{Mohseni, Ragan, and Hu}{Mohseni
  et~al\mbox{.}}{2019b}]%
        {mohseni2019open}
\bibfield{author}{\bibinfo{person}{Sina Mohseni}, \bibinfo{person}{Eric Ragan},
  {and} \bibinfo{person}{Xia Hu}.} \bibinfo{year}{2019}\natexlab{b}.
\newblock \showarticletitle{Open issues in combating fake news:
  Interpretability as an opportunity}.
\newblock \bibinfo{journal}{\emph{arXiv preprint arXiv:1904.03016}}
  (\bibinfo{year}{2019}).
\newblock


\bibitem[\protect\citeauthoryear{Mohseni and Ragan}{Mohseni and Ragan}{2018}]%
        {mohseni2018human}
\bibfield{author}{\bibinfo{person}{Sina Mohseni} {and} \bibinfo{person}{Eric~D
  Ragan}.} \bibinfo{year}{2018}\natexlab{}.
\newblock \showarticletitle{A human-grounded evaluation benchmark for local
  explanations of machine learning}.
\newblock \bibinfo{journal}{\emph{arXiv preprint arXiv:1801.05075}}
  (\bibinfo{year}{2018}).
\newblock


\bibitem[\protect\citeauthoryear{Mohseni, Yang, Pentyala, Du, Liu, Lupfer, Hu,
  Ji, and Ragan}{Mohseni et~al\mbox{.}}{2020b}]%
        {mohseni2020trust}
\bibfield{author}{\bibinfo{person}{Sina Mohseni}, \bibinfo{person}{Fan Yang},
  \bibinfo{person}{Shiva Pentyala}, \bibinfo{person}{Mengnan Du},
  \bibinfo{person}{Yi Liu}, \bibinfo{person}{Nic Lupfer}, \bibinfo{person}{Xia
  Hu}, \bibinfo{person}{Shuiwang Ji}, {and} \bibinfo{person}{Eric Ragan}.}
  \bibinfo{year}{2020}\natexlab{b}.
\newblock \showarticletitle{Trust evolution over time in explainable AI for
  fake news detection}.
\newblock \bibinfo{journal}{\emph{Fair \& Responsible AI Workshop at CHI 2020}}
  (\bibinfo{year}{2020}).
\newblock


\bibitem[\protect\citeauthoryear{Molnar}{Molnar}{2019}]%
        {molnar2018interpretable}
\bibfield{author}{\bibinfo{person}{Christoph Molnar}.}
  \bibinfo{year}{2019}\natexlab{}.
\newblock \bibinfo{booktitle}{\emph{Interpretable machine learning}}.
\newblock \bibinfo{publisher}{Lulu. com}.
\newblock


\bibitem[\protect\citeauthoryear{Montavon, Samek, and M{\"u}ller}{Montavon
  et~al\mbox{.}}{2017}]%
        {montavon2017methods}
\bibfield{author}{\bibinfo{person}{Gr{\'e}goire Montavon},
  \bibinfo{person}{Wojciech Samek}, {and} \bibinfo{person}{Klaus-Robert
  M{\"u}ller}.} \bibinfo{year}{2017}\natexlab{}.
\newblock \showarticletitle{Methods for interpreting and understanding deep
  neural networks}.
\newblock \bibinfo{journal}{\emph{Digital Signal Processing}}
  (\bibinfo{year}{2017}).
\newblock


\bibitem[\protect\citeauthoryear{Mueller and Klein}{Mueller and Klein}{2011}]%
        {mueller2011improving}
\bibfield{author}{\bibinfo{person}{Shane~T Mueller} {and} \bibinfo{person}{Gary
  Klein}.} \bibinfo{year}{2011}\natexlab{}.
\newblock \showarticletitle{Improving users' mental models of intelligent
  software tools}.
\newblock \bibinfo{journal}{\emph{IEEE Intelligent Systems}}
  \bibinfo{volume}{26}, \bibinfo{number}{2} (\bibinfo{year}{2011}),
  \bibinfo{pages}{77--83}.
\newblock


\bibitem[\protect\citeauthoryear{Muir}{Muir}{1987}]%
        {muir1987trust}
\bibfield{author}{\bibinfo{person}{Bonnie~M Muir}.}
  \bibinfo{year}{1987}\natexlab{}.
\newblock \showarticletitle{Trust between humans and machines, and the design
  of decision aids}.
\newblock \bibinfo{journal}{\emph{International Journal of Man-Machine
  Studies}} \bibinfo{volume}{27}, \bibinfo{number}{5-6} (\bibinfo{year}{1987}),
  \bibinfo{pages}{527--539}.
\newblock


\bibitem[\protect\citeauthoryear{Munzner}{Munzner}{2009}]%
        {munzner2009nested}
\bibfield{author}{\bibinfo{person}{Tamara Munzner}.}
  \bibinfo{year}{2009}\natexlab{}.
\newblock \showarticletitle{A nested process model for visualization design and
  validation}.
\newblock \bibinfo{journal}{\emph{IEEE Transactions on Visualization and
  Computer Graphics}} \bibinfo{number}{6} (\bibinfo{year}{2009}),
  \bibinfo{pages}{921--928}.
\newblock


\bibitem[\protect\citeauthoryear{Myers, Weitzman, Ko, and Chau}{Myers
  et~al\mbox{.}}{2006}]%
        {myers2006answering}
\bibfield{author}{\bibinfo{person}{Brad~A Myers}, \bibinfo{person}{David~A
  Weitzman}, \bibinfo{person}{Andrew~J Ko}, {and} \bibinfo{person}{Duen~H
  Chau}.} \bibinfo{year}{2006}\natexlab{}.
\newblock \showarticletitle{Answering why and why not questions in user
  interfaces}. In \bibinfo{booktitle}{\emph{Proceedings of the SIGCHI
  conference on Human Factors in computing systems}}. ACM,
  \bibinfo{pages}{397--406}.
\newblock


\bibitem[\protect\citeauthoryear{Norton and Qi}{Norton and Qi}{2017}]%
        {norton2017adversarial}
\bibfield{author}{\bibinfo{person}{Andrew~P Norton} {and}
  \bibinfo{person}{Yanjun Qi}.} \bibinfo{year}{2017}\natexlab{}.
\newblock \showarticletitle{Adversarial-playground: A visualization suite
  showing how adversarial examples fool deep learning}. In
  \bibinfo{booktitle}{\emph{Visualization for Cyber Security (VizSec), 2017
  IEEE Symposium on}}. IEEE, \bibinfo{pages}{1--4}.
\newblock


\bibitem[\protect\citeauthoryear{Nothdurft, Richter, and Minker}{Nothdurft
  et~al\mbox{.}}{2014}]%
        {nothdurft2014probabilistic}
\bibfield{author}{\bibinfo{person}{Florian Nothdurft}, \bibinfo{person}{Felix
  Richter}, {and} \bibinfo{person}{Wolfgang Minker}.}
  \bibinfo{year}{2014}\natexlab{}.
\newblock \showarticletitle{Probabilistic human-computer trust handling}. In
  \bibinfo{booktitle}{\emph{Proceedings of the 15th Annual Meeting of the
  Special Interest Group on Discourse and Dialogue (SIGDIAL)}}.
  \bibinfo{pages}{51--59}.
\newblock


\bibitem[\protect\citeauthoryear{Nourani, Honeycutt, Block, Roy, Rahman, Ragan,
  and Gogate}{Nourani et~al\mbox{.}}{2020}]%
        {nourani2020time}
\bibfield{author}{\bibinfo{person}{Mahsan Nourani}, \bibinfo{person}{Dondald
  Honeycutt}, \bibinfo{person}{Jeremy Block}, \bibinfo{person}{Chiradeep Roy},
  \bibinfo{person}{Tahrima Rahman}, \bibinfo{person}{Eric~D. Ragan}, {and}
  \bibinfo{person}{Vibhav Gogate}.} \bibinfo{year}{2020}\natexlab{}.
\newblock \showarticletitle{Investigating the importance of first impressions
  and explainable AI with interactive video analysis}. In
  \bibinfo{booktitle}{\emph{Extended Abstracts of the 2019 CHI Conference on
  Human Factors in Computing Systems}}. ACM.
\newblock


\bibitem[\protect\citeauthoryear{Nourani, Kabir, Mohseni, and Ragan}{Nourani
  et~al\mbox{.}}{2019}]%
        {nourani2019effects}
\bibfield{author}{\bibinfo{person}{Mahsan Nourani}, \bibinfo{person}{Samia
  Kabir}, \bibinfo{person}{Sina Mohseni}, {and} \bibinfo{person}{Eric~D
  Ragan}.} \bibinfo{year}{2019}\natexlab{}.
\newblock \showarticletitle{The effects of meaningful and meaningless
  explanations on trust and perceived system accuracy in intelligent systems}.
  In \bibinfo{booktitle}{\emph{Proceedings of the AAAI Conference on Human
  Computation and Crowdsourcing}}, Vol.~\bibinfo{volume}{7}.
  \bibinfo{pages}{97--105}.
\newblock


\bibitem[\protect\citeauthoryear{Nushi, Kamar, and Horvitz}{Nushi
  et~al\mbox{.}}{2018}]%
        {nushi2018towards}
\bibfield{author}{\bibinfo{person}{Besmira Nushi}, \bibinfo{person}{Ece Kamar},
  {and} \bibinfo{person}{Eric Horvitz}.} \bibinfo{year}{2018}\natexlab{}.
\newblock \showarticletitle{Towards accountable AI: Hybrid human-machine
  analyses for characterizing system failure}. In
  \bibinfo{booktitle}{\emph{Sixth AAAI Conference on Human Computation and
  Crowdsourcing}}.
\newblock


\bibitem[\protect\citeauthoryear{Olah, Satyanarayan, Johnson, Carter, Schubert,
  Ye, and Mordvintsev}{Olah et~al\mbox{.}}{2018}]%
        {olah2018the}
\bibfield{author}{\bibinfo{person}{Chris Olah}, \bibinfo{person}{Arvind
  Satyanarayan}, \bibinfo{person}{Ian Johnson}, \bibinfo{person}{Shan Carter},
  \bibinfo{person}{Ludwig Schubert}, \bibinfo{person}{Katherine Ye}, {and}
  \bibinfo{person}{Alexander Mordvintsev}.} \bibinfo{year}{2018}\natexlab{}.
\newblock \showarticletitle{The Building Blocks of Interpretability}.
\newblock \bibinfo{journal}{\emph{Distill}} (\bibinfo{year}{2018}).
\newblock
\urldef\tempurl%
\url{https://doi.org/10.23915/distill.00010}
\showDOI{\tempurl}
\newblock
\shownote{https://distill.pub/2018/building-blocks.}


\bibitem[\protect\citeauthoryear{O'Neil}{O'Neil}{2016}]%
        {o2016weapons}
\bibfield{author}{\bibinfo{person}{Cathy O'Neil}.}
  \bibinfo{year}{2016}\natexlab{}.
\newblock \bibinfo{booktitle}{\emph{Weapons of math destruction: How big data
  increases inequality and threatens democracy}}.
\newblock \bibinfo{publisher}{Broadway Books}.
\newblock


\bibitem[\protect\citeauthoryear{Penney, Dodge, Hilderbrand, Anderson, Simpson,
  and Burnett}{Penney et~al\mbox{.}}{2018}]%
        {Penney:2018:TFU:3172944.3172946}
\bibfield{author}{\bibinfo{person}{Sean Penney}, \bibinfo{person}{Jonathan
  Dodge}, \bibinfo{person}{Claudia Hilderbrand}, \bibinfo{person}{Andrew
  Anderson}, \bibinfo{person}{Logan Simpson}, {and} \bibinfo{person}{Margaret
  Burnett}.} \bibinfo{year}{2018}\natexlab{}.
\newblock \showarticletitle{Toward foraging for understanding of StarCraft
  agents: An empirical study}. In \bibinfo{booktitle}{\emph{23rd International
  Conference on Intelligent User Interfaces}} \emph{(\bibinfo{series}{IUI
  '18})}. \bibinfo{publisher}{ACM}, \bibinfo{address}{New York, NY, USA},
  \bibinfo{pages}{225--237}.
\newblock
\showISBNx{978-1-4503-4945-1}
\urldef\tempurl%
\url{https://doi.org/10.1145/3172944.3172946}
\showDOI{\tempurl}


\bibitem[\protect\citeauthoryear{Pezzotti, H{\"o}llt, Van~Gemert, Lelieveldt,
  Eisemann, and Vilanova}{Pezzotti et~al\mbox{.}}{2018}]%
        {pezzotti2018deepeyes}
\bibfield{author}{\bibinfo{person}{Nicola Pezzotti}, \bibinfo{person}{Thomas
  H{\"o}llt}, \bibinfo{person}{Jan Van~Gemert}, \bibinfo{person}{Boudewijn~PF
  Lelieveldt}, \bibinfo{person}{Elmar Eisemann}, {and} \bibinfo{person}{Anna
  Vilanova}.} \bibinfo{year}{2018}\natexlab{}.
\newblock \showarticletitle{DeepEyes: Progressive visual analytics for
  designing deep neural networks}.
\newblock \bibinfo{journal}{\emph{IEEE Transactions on Visualization and
  Computer Graphics}} \bibinfo{volume}{24}, \bibinfo{number}{1}
  (\bibinfo{year}{2018}), \bibinfo{pages}{98--108}.
\newblock


\bibitem[\protect\citeauthoryear{Poerner, Sch{\"u}tze, and Roth}{Poerner
  et~al\mbox{.}}{2018}]%
        {poerner2018evaluating}
\bibfield{author}{\bibinfo{person}{Nina Poerner}, \bibinfo{person}{Hinrich
  Sch{\"u}tze}, {and} \bibinfo{person}{Benjamin Roth}.}
  \bibinfo{year}{2018}\natexlab{}.
\newblock \showarticletitle{Evaluating neural network explanation methods using
  hybrid documents and morphological prediction}. In
  \bibinfo{booktitle}{\emph{56th Annual Meeting of the Association for
  Computational Linguistics (ACL)}}.
\newblock


\bibitem[\protect\citeauthoryear{Poulin, Eisner, Szafron, Lu, Greiner, Wishart,
  Fyshe, Pearcy, MacDonell, and Anvik}{Poulin et~al\mbox{.}}{2006}]%
        {poulin2006visual}
\bibfield{author}{\bibinfo{person}{Brett Poulin}, \bibinfo{person}{Roman
  Eisner}, \bibinfo{person}{Duane Szafron}, \bibinfo{person}{Paul Lu},
  \bibinfo{person}{Russell Greiner}, \bibinfo{person}{David~S Wishart},
  \bibinfo{person}{Alona Fyshe}, \bibinfo{person}{Brandon Pearcy},
  \bibinfo{person}{Cam MacDonell}, {and} \bibinfo{person}{John Anvik}.}
  \bibinfo{year}{2006}\natexlab{}.
\newblock \showarticletitle{Visual explanation of evidence with additive
  classifiers}. In \bibinfo{booktitle}{\emph{Proceedings Of The National
  Conference On Artificial Intelligence}}, Vol.~\bibinfo{volume}{21}. Menlo
  Park, CA; Cambridge, MA; London; AAAI Press; MIT Press; 1999,
  \bibinfo{pages}{1822}.
\newblock


\bibitem[\protect\citeauthoryear{Poursabzi-Sangdeh, Goldstein, Hofman, Vaughan,
  and Wallach}{Poursabzi-Sangdeh et~al\mbox{.}}{2018}]%
        {poursabzi2018manipulating}
\bibfield{author}{\bibinfo{person}{Forough Poursabzi-Sangdeh},
  \bibinfo{person}{Daniel~G Goldstein}, \bibinfo{person}{Jake~M Hofman},
  \bibinfo{person}{Jennifer~Wortman Vaughan}, {and} \bibinfo{person}{Hanna
  Wallach}.} \bibinfo{year}{2018}\natexlab{}.
\newblock \showarticletitle{Manipulating and measuring model interpretability}.
\newblock \bibinfo{journal}{\emph{arXiv preprint arXiv:1802.07810}}
  (\bibinfo{year}{2018}).
\newblock


\bibitem[\protect\citeauthoryear{Pu and Chen}{Pu and Chen}{2006}]%
        {pu2006trust}
\bibfield{author}{\bibinfo{person}{Pearl Pu} {and} \bibinfo{person}{Li Chen}.}
  \bibinfo{year}{2006}\natexlab{}.
\newblock \showarticletitle{Trust building with explanation interfaces}. In
  \bibinfo{booktitle}{\emph{Proceedings of the 11th International Conference on
  Intelligent User Interfaces}}. ACM, \bibinfo{pages}{93--100}.
\newblock


\bibitem[\protect\citeauthoryear{Rader, Cotter, and Cho}{Rader
  et~al\mbox{.}}{2018}]%
        {rader2018explanations}
\bibfield{author}{\bibinfo{person}{Emilee Rader}, \bibinfo{person}{Kelley
  Cotter}, {and} \bibinfo{person}{Janghee Cho}.}
  \bibinfo{year}{2018}\natexlab{}.
\newblock \showarticletitle{Explanations as mechanisms for supporting
  algorithmic transparency}. In \bibinfo{booktitle}{\emph{Proceedings of the
  2018 CHI Conference on Human Factors in Computing Systems}}. ACM,
  \bibinfo{pages}{103}.
\newblock


\bibitem[\protect\citeauthoryear{Rader and Gray}{Rader and Gray}{2015}]%
        {rader2015understanding}
\bibfield{author}{\bibinfo{person}{Emilee Rader} {and} \bibinfo{person}{Rebecca
  Gray}.} \bibinfo{year}{2015}\natexlab{}.
\newblock \showarticletitle{Understanding user beliefs about algorithmic
  curation in the Facebook news feed}. In \bibinfo{booktitle}{\emph{Proceedings
  of the 33rd Annual ACM Conference on Human Factors in Computing Systems}}.
  ACM, \bibinfo{pages}{173--182}.
\newblock


\bibitem[\protect\citeauthoryear{Ribeiro, Singh, and Guestrin}{Ribeiro
  et~al\mbox{.}}{2016}]%
        {ribeiro2016should}
\bibfield{author}{\bibinfo{person}{Marco~Tulio Ribeiro},
  \bibinfo{person}{Sameer Singh}, {and} \bibinfo{person}{Carlos Guestrin}.}
  \bibinfo{year}{2016}\natexlab{}.
\newblock \showarticletitle{Why should i you? Explaining the predictions of any
  classifier}. In \bibinfo{booktitle}{\emph{Proceedings of the 22nd ACM SIGKDD
  International Conference on Knowledge Discovery and Data Mining}}. ACM,
  \bibinfo{pages}{1135--1144}.
\newblock


\bibitem[\protect\citeauthoryear{Ribeiro, Singh, and Guestrin}{Ribeiro
  et~al\mbox{.}}{2018}]%
        {ribeiro2018anchors}
\bibfield{author}{\bibinfo{person}{Marco~Tulio Ribeiro},
  \bibinfo{person}{Sameer Singh}, {and} \bibinfo{person}{Carlos Guestrin}.}
  \bibinfo{year}{2018}\natexlab{}.
\newblock \showarticletitle{Anchors: High-precision model-agnostic
  explanations}. In \bibinfo{booktitle}{\emph{AAAI Conference on Artificial
  Intelligence}}.
\newblock


\bibitem[\protect\citeauthoryear{Robinson, Hohman, and Dilkina}{Robinson
  et~al\mbox{.}}{2017}]%
        {robinson2017deep}
\bibfield{author}{\bibinfo{person}{Caleb Robinson}, \bibinfo{person}{Fred
  Hohman}, {and} \bibinfo{person}{Bistra Dilkina}.}
  \bibinfo{year}{2017}\natexlab{}.
\newblock \showarticletitle{A deep learning approach for population estimation
  from satellite imagery}. In \bibinfo{booktitle}{\emph{Proceedings of the 1st
  ACM SIGSPATIAL Workshop on Geospatial Humanities}}. ACM,
  \bibinfo{pages}{47--54}.
\newblock


\bibitem[\protect\citeauthoryear{Robnik-{\v{S}}ikonja and
  Bohanec}{Robnik-{\v{S}}ikonja and Bohanec}{2018}]%
        {robnik2018perturbation}
\bibfield{author}{\bibinfo{person}{Marko Robnik-{\v{S}}ikonja} {and}
  \bibinfo{person}{Marko Bohanec}.} \bibinfo{year}{2018}\natexlab{}.
\newblock \showarticletitle{Perturbation-based explanations of prediction
  models}.
\newblock In \bibinfo{booktitle}{\emph{Human and Machine Learning}}.
  \bibinfo{publisher}{Springer}, \bibinfo{pages}{159--175}.
\newblock


\bibitem[\protect\citeauthoryear{Rosenthal, Selvaraj, and Veloso}{Rosenthal
  et~al\mbox{.}}{2016}]%
        {rosenthal2016verbalization}
\bibfield{author}{\bibinfo{person}{Stephanie Rosenthal}, \bibinfo{person}{Sai~P
  Selvaraj}, {and} \bibinfo{person}{Manuela Veloso}.}
  \bibinfo{year}{2016}\natexlab{}.
\newblock \showarticletitle{Verbalization: narration of autonomous robot
  experience}. In \bibinfo{booktitle}{\emph{Proceedings of the Twenty-Fifth
  International Joint Conference on Artificial Intelligence}}.
  \bibinfo{pages}{862--868}.
\newblock


\bibitem[\protect\citeauthoryear{Ross and Doshi-Velez}{Ross and
  Doshi-Velez}{2018}]%
        {ross2017improving}
\bibfield{author}{\bibinfo{person}{Andrew~Slavin Ross} {and}
  \bibinfo{person}{Finale Doshi-Velez}.} \bibinfo{year}{2018}\natexlab{}.
\newblock \showarticletitle{Improving the adversarial robustness and
  interpretability of deep neural networks by regularizing their input
  gradients}. In \bibinfo{booktitle}{\emph{Thirty-second AAAI Conference on
  Artificial Intelligence}}.
\newblock


\bibitem[\protect\citeauthoryear{Ross, Hughes, and Doshi-Velez}{Ross
  et~al\mbox{.}}{2017}]%
        {ijcai2017-371}
\bibfield{author}{\bibinfo{person}{Andrew~Slavin Ross},
  \bibinfo{person}{Michael~C. Hughes}, {and} \bibinfo{person}{Finale
  Doshi-Velez}.} \bibinfo{year}{2017}\natexlab{}.
\newblock \showarticletitle{Right for the right reasons: Training
  differentiable models by constraining their explanations}. In
  \bibinfo{booktitle}{\emph{Proceedings of the Twenty-Sixth International Joint
  Conference on Artificial Intelligence, {IJCAI-17}}}.
  \bibinfo{pages}{2662--2670}.
\newblock
\urldef\tempurl%
\url{https://doi.org/10.24963/ijcai.2017/371}
\showDOI{\tempurl}


\bibitem[\protect\citeauthoryear{Rudolph, Savikhin, and Ebert}{Rudolph
  et~al\mbox{.}}{2009}]%
        {rudolph2009finvis}
\bibfield{author}{\bibinfo{person}{Stephen Rudolph}, \bibinfo{person}{Anya
  Savikhin}, {and} \bibinfo{person}{David~S Ebert}.}
  \bibinfo{year}{2009}\natexlab{}.
\newblock \showarticletitle{Finvis: Applied visual analytics for personal
  financial planning}. In \bibinfo{booktitle}{\emph{Visual Analytics Science
  and Technology, 2009. IEEE Symposium on}}. Citeseer,
  \bibinfo{pages}{195--202}.
\newblock


\bibitem[\protect\citeauthoryear{Sacha, Sedlmair, Zhang, Lee, Weiskopf, North,
  and Keim}{Sacha et~al\mbox{.}}{2016a}]%
        {sacha2016human}
\bibfield{author}{\bibinfo{person}{Dominik Sacha}, \bibinfo{person}{Michael
  Sedlmair}, \bibinfo{person}{Leishi Zhang}, \bibinfo{person}{John~Aldo Lee},
  \bibinfo{person}{Daniel Weiskopf}, \bibinfo{person}{Stephen North}, {and}
  \bibinfo{person}{Daniel Keim}.} \bibinfo{year}{2016}\natexlab{a}.
\newblock \showarticletitle{Human-centered machine learning through interactive
  visualization}. In \bibinfo{booktitle}{\emph{24th European Symposium on
  Artificial Neural Networks, Computational Intelligence and Machine
  Learning}}. \bibinfo{pages}{641--646}.
\newblock


\bibitem[\protect\citeauthoryear{Sacha, Senaratne, Kwon, Ellis, and Keim}{Sacha
  et~al\mbox{.}}{2016b}]%
        {sacha2016role}
\bibfield{author}{\bibinfo{person}{Dominik Sacha}, \bibinfo{person}{Hansi
  Senaratne}, \bibinfo{person}{Bum~Chul Kwon}, \bibinfo{person}{Geoffrey
  Ellis}, {and} \bibinfo{person}{Daniel~A Keim}.}
  \bibinfo{year}{2016}\natexlab{b}.
\newblock \showarticletitle{The role of uncertainty, awareness, and trust in
  visual analytics}.
\newblock \bibinfo{journal}{\emph{IEEE Transactions on Visualization and
  Computer Graphics}} \bibinfo{volume}{22}, \bibinfo{number}{1}
  (\bibinfo{year}{2016}), \bibinfo{pages}{240--249}.
\newblock


\bibitem[\protect\citeauthoryear{Saket, Srinivasan, Ragan, and Endert}{Saket
  et~al\mbox{.}}{2017}]%
        {saket2017evaluating}
\bibfield{author}{\bibinfo{person}{Bahador Saket}, \bibinfo{person}{Arjun
  Srinivasan}, \bibinfo{person}{Eric~D Ragan}, {and} \bibinfo{person}{Alex
  Endert}.} \bibinfo{year}{2017}\natexlab{}.
\newblock \showarticletitle{Evaluating interactive graphical encodings for data
  visualization}.
\newblock \bibinfo{journal}{\emph{IEEE Transactions on Visualization and
  Computer Graphics}} \bibinfo{volume}{24}, \bibinfo{number}{3}
  (\bibinfo{year}{2017}), \bibinfo{pages}{1316--1330}.
\newblock


\bibitem[\protect\citeauthoryear{Samek, Binder, Montavon, Lapuschkin, and
  M{\"u}ller}{Samek et~al\mbox{.}}{2017}]%
        {samek2017evaluating}
\bibfield{author}{\bibinfo{person}{Wojciech Samek}, \bibinfo{person}{Alexander
  Binder}, \bibinfo{person}{Gr{\'e}goire Montavon}, \bibinfo{person}{Sebastian
  Lapuschkin}, {and} \bibinfo{person}{Klaus-Robert M{\"u}ller}.}
  \bibinfo{year}{2017}\natexlab{}.
\newblock \showarticletitle{Evaluating the visualization of what a deep neural
  network has learned}.
\newblock \bibinfo{journal}{\emph{IEEE Transactions on Neural Networks and
  Learning Systems}} \bibinfo{volume}{28}, \bibinfo{number}{11}
  (\bibinfo{year}{2017}), \bibinfo{pages}{2660--2673}.
\newblock


\bibitem[\protect\citeauthoryear{Sandvig, Hamilton, Karahalios, and
  Langbort}{Sandvig et~al\mbox{.}}{2014}]%
        {sandvig2014auditing}
\bibfield{author}{\bibinfo{person}{Christian Sandvig}, \bibinfo{person}{Kevin
  Hamilton}, \bibinfo{person}{Karrie Karahalios}, {and} \bibinfo{person}{Cedric
  Langbort}.} \bibinfo{year}{2014}\natexlab{}.
\newblock \showarticletitle{Auditing algorithms: Research methods for detecting
  discrimination on internet platforms}.
\newblock \bibinfo{journal}{\emph{Data and Discrimination: Converting Critical
  Concerns Into Productive Inquiry}} (\bibinfo{year}{2014}),
  \bibinfo{pages}{1--23}.
\newblock


\bibitem[\protect\citeauthoryear{Schaffernicht and Groesser}{Schaffernicht and
  Groesser}{2011}]%
        {schaffernicht2011comprehensive}
\bibfield{author}{\bibinfo{person}{Martin Schaffernicht} {and}
  \bibinfo{person}{Stefan~N Groesser}.} \bibinfo{year}{2011}\natexlab{}.
\newblock \showarticletitle{A comprehensive method for comparing mental models
  of dynamic systems}.
\newblock \bibinfo{journal}{\emph{European Journal of Operational Research}}
  \bibinfo{volume}{210}, \bibinfo{number}{1} (\bibinfo{year}{2011}),
  \bibinfo{pages}{57--67}.
\newblock


\bibitem[\protect\citeauthoryear{Schmid, Zeller, Besold, Tamaddoni-Nezhad, and
  Muggleton}{Schmid et~al\mbox{.}}{2016}]%
        {schmid2016does}
\bibfield{author}{\bibinfo{person}{Ute Schmid}, \bibinfo{person}{Christina
  Zeller}, \bibinfo{person}{Tarek Besold}, \bibinfo{person}{Alireza
  Tamaddoni-Nezhad}, {and} \bibinfo{person}{Stephen Muggleton}.}
  \bibinfo{year}{2016}\natexlab{}.
\newblock \showarticletitle{How does predicate invention affect human
  comprehensibility?}. In \bibinfo{booktitle}{\emph{International Conference on
  Inductive Logic Programming}}. Springer, \bibinfo{pages}{52--67}.
\newblock


\bibitem[\protect\citeauthoryear{Schmidt and Biessmann}{Schmidt and
  Biessmann}{2019}]%
        {schmidt2019quantifying}
\bibfield{author}{\bibinfo{person}{Philipp Schmidt} {and}
  \bibinfo{person}{Felix Biessmann}.} \bibinfo{year}{2019}\natexlab{}.
\newblock \showarticletitle{Quantifying interpretability and trust in machine
  learning systems}.
\newblock \bibinfo{journal}{\emph{arXiv preprint arXiv:1901.08558}}
  (\bibinfo{year}{2019}).
\newblock


\bibitem[\protect\citeauthoryear{Selvaraju, Cogswell, Das, Vedantam, Parikh,
  and Batra}{Selvaraju et~al\mbox{.}}{2017}]%
        {selvaraju2017grad}
\bibfield{author}{\bibinfo{person}{Ramprasaath~R Selvaraju},
  \bibinfo{person}{Michael Cogswell}, \bibinfo{person}{Abhishek Das},
  \bibinfo{person}{Ramakrishna Vedantam}, \bibinfo{person}{Devi Parikh}, {and}
  \bibinfo{person}{Dhruv Batra}.} \bibinfo{year}{2017}\natexlab{}.
\newblock \showarticletitle{Grad-cam: Visual explanations from deep networks
  via gradient-based localization}. In \bibinfo{booktitle}{\emph{Proceedings of
  the IEEE international conference on computer vision}}.
  \bibinfo{pages}{618--626}.
\newblock


\bibitem[\protect\citeauthoryear{Shrikumar, Greenside, and Kundaje}{Shrikumar
  et~al\mbox{.}}{2017}]%
        {shrikumar2017learning}
\bibfield{author}{\bibinfo{person}{Avanti Shrikumar}, \bibinfo{person}{Peyton
  Greenside}, {and} \bibinfo{person}{Anshul Kundaje}.}
  \bibinfo{year}{2017}\natexlab{}.
\newblock \showarticletitle{Learning important features through propagating
  activation differences}. In \bibinfo{booktitle}{\emph{Proceedings of the 34th
  International Conference on Machine Learning-Volume 70}}. JMLR. org,
  \bibinfo{pages}{3145--3153}.
\newblock


\bibitem[\protect\citeauthoryear{Simonyan, Vedaldi, and Zisserman}{Simonyan
  et~al\mbox{.}}{2013}]%
        {simonyan2013deep}
\bibfield{author}{\bibinfo{person}{Karen Simonyan}, \bibinfo{person}{Andrea
  Vedaldi}, {and} \bibinfo{person}{Andrew Zisserman}.}
  \bibinfo{year}{2013}\natexlab{}.
\newblock \showarticletitle{Deep inside convolutional networks: Visualising
  image classification models and saliency maps}.
\newblock \bibinfo{journal}{\emph{arXiv preprint arXiv:1312.6034}}
  (\bibinfo{year}{2013}).
\newblock


\bibitem[\protect\citeauthoryear{Smilkov, Carter, Sculley, Vi{\'e}gas, and
  Wattenberg}{Smilkov et~al\mbox{.}}{2017}]%
        {smilkov2017direct}
\bibfield{author}{\bibinfo{person}{Daniel Smilkov}, \bibinfo{person}{Shan
  Carter}, \bibinfo{person}{D Sculley}, \bibinfo{person}{Fernanda~B
  Vi{\'e}gas}, {and} \bibinfo{person}{Martin Wattenberg}.}
  \bibinfo{year}{2017}\natexlab{}.
\newblock \showarticletitle{Direct-manipulation visualization of deep
  networks}.
\newblock \bibinfo{journal}{\emph{arXiv preprint arXiv:1708.03788}}
  (\bibinfo{year}{2017}).
\newblock


\bibitem[\protect\citeauthoryear{Spinner, Schlegel, Sch{\"a}fer, and
  El-Assady}{Spinner et~al\mbox{.}}{2019}]%
        {spinner2019explainer}
\bibfield{author}{\bibinfo{person}{Thilo Spinner}, \bibinfo{person}{Udo
  Schlegel}, \bibinfo{person}{Hanna Sch{\"a}fer}, {and}
  \bibinfo{person}{Mennatallah El-Assady}.} \bibinfo{year}{2019}\natexlab{}.
\newblock \showarticletitle{explAIner: A visual analytics framework for
  interactive and explainable machine learning}.
\newblock \bibinfo{journal}{\emph{IEEE Transactions on Visualization and
  Computer Graphics}} (\bibinfo{year}{2019}).
\newblock


\bibitem[\protect\citeauthoryear{Strobelt, Gehrmann, Pfister, and
  Rush}{Strobelt et~al\mbox{.}}{2018}]%
        {strobelt2018lstmvis}
\bibfield{author}{\bibinfo{person}{Hendrik Strobelt},
  \bibinfo{person}{Sebastian Gehrmann}, \bibinfo{person}{Hanspeter Pfister},
  {and} \bibinfo{person}{Alexander~M Rush}.} \bibinfo{year}{2018}\natexlab{}.
\newblock \showarticletitle{LstmVis: A tool for visual analysis of hidden state
  dynamics in recurrent neural networks}.
\newblock \bibinfo{journal}{\emph{IEEE Transactions on Visualization and
  Computer Graphics}} \bibinfo{volume}{24}, \bibinfo{number}{1}
  (\bibinfo{year}{2018}), \bibinfo{pages}{667--676}.
\newblock


\bibitem[\protect\citeauthoryear{Stumpf, Rajaram, Li, Wong, Burnett,
  Dietterich, Sullivan, and Herlocker}{Stumpf et~al\mbox{.}}{2009}]%
        {stumpf2009interacting}
\bibfield{author}{\bibinfo{person}{Simone Stumpf}, \bibinfo{person}{Vidya
  Rajaram}, \bibinfo{person}{Lida Li}, \bibinfo{person}{Weng-Keen Wong},
  \bibinfo{person}{Margaret Burnett}, \bibinfo{person}{Thomas Dietterich},
  \bibinfo{person}{Erin Sullivan}, {and} \bibinfo{person}{Jonathan Herlocker}.}
  \bibinfo{year}{2009}\natexlab{}.
\newblock \showarticletitle{Interacting meaningfully with machine learning
  systems: Three experiments}.
\newblock \bibinfo{journal}{\emph{International Journal of Human-Computer
  Studies}} \bibinfo{volume}{67}, \bibinfo{number}{8} (\bibinfo{year}{2009}),
  \bibinfo{pages}{639--662}.
\newblock


\bibitem[\protect\citeauthoryear{Stumpf, Skrebe, Aymer, and Hobson}{Stumpf
  et~al\mbox{.}}{2018}]%
        {stumpf2018explaining}
\bibfield{author}{\bibinfo{person}{Simone Stumpf}, \bibinfo{person}{Simonas
  Skrebe}, \bibinfo{person}{Graeme Aymer}, {and} \bibinfo{person}{Julie
  Hobson}.} \bibinfo{year}{2018}\natexlab{}.
\newblock \showarticletitle{Explaining smart heating systems to discourage
  fiddling with optimized behavior}.
\newblock  (\bibinfo{year}{2018}).
\newblock


\bibitem[\protect\citeauthoryear{Sweeney}{Sweeney}{2013}]%
        {sweeney2013discrimination}
\bibfield{author}{\bibinfo{person}{Latanya Sweeney}.}
  \bibinfo{year}{2013}\natexlab{}.
\newblock \showarticletitle{Discrimination in online ad delivery}.
\newblock \bibinfo{journal}{\emph{Commun. ACM}} \bibinfo{volume}{56},
  \bibinfo{number}{5} (\bibinfo{year}{2013}), \bibinfo{pages}{44--54}.
\newblock


\bibitem[\protect\citeauthoryear{Tang, Gao, Liu, and Das~Sarma}{Tang
  et~al\mbox{.}}{2012}]%
        {tang2012etrust}
\bibfield{author}{\bibinfo{person}{Jiliang Tang}, \bibinfo{person}{Huiji Gao},
  \bibinfo{person}{Huan Liu}, {and} \bibinfo{person}{Atish Das~Sarma}.}
  \bibinfo{year}{2012}\natexlab{}.
\newblock \showarticletitle{eTrust: Understanding trust evolution in an online
  world}. In \bibinfo{booktitle}{\emph{Proceedings of the 18th ACM SIGKDD
  International Conference on Knowledge Discovery and Data Mining}}. ACM,
  \bibinfo{pages}{253--261}.
\newblock


\bibitem[\protect\citeauthoryear{Tikkinen-Piri, Rohunen, and
  Markkula}{Tikkinen-Piri et~al\mbox{.}}{2018}]%
        {tikkinen2018eu}
\bibfield{author}{\bibinfo{person}{Christina Tikkinen-Piri},
  \bibinfo{person}{Anna Rohunen}, {and} \bibinfo{person}{Jouni Markkula}.}
  \bibinfo{year}{2018}\natexlab{}.
\newblock \showarticletitle{EU general data protection regulation: Changes and
  implications for personal data collecting companies}.
\newblock \bibinfo{journal}{\emph{Computer Law \& Security Review}}
  \bibinfo{volume}{34}, \bibinfo{number}{1} (\bibinfo{year}{2018}),
  \bibinfo{pages}{134--153}.
\newblock


\bibitem[\protect\citeauthoryear{Tintarev and Masthoff}{Tintarev and
  Masthoff}{2011}]%
        {tintarev2011designing}
\bibfield{author}{\bibinfo{person}{Nava Tintarev} {and} \bibinfo{person}{Judith
  Masthoff}.} \bibinfo{year}{2011}\natexlab{}.
\newblock \showarticletitle{Designing and evaluating explanations for
  recommender systems}.
\newblock In \bibinfo{booktitle}{\emph{Recommender Systems Handbook}}.
  \bibinfo{publisher}{Springer}, \bibinfo{pages}{479--510}.
\newblock


\bibitem[\protect\citeauthoryear{Tomsett, Braines, Harborne, Preece, and
  Chakraborty}{Tomsett et~al\mbox{.}}{2018}]%
        {tomsett2018interpretable}
\bibfield{author}{\bibinfo{person}{Richard Tomsett}, \bibinfo{person}{Dave
  Braines}, \bibinfo{person}{Dan Harborne}, \bibinfo{person}{Alun Preece},
  {and} \bibinfo{person}{Supriyo Chakraborty}.}
  \bibinfo{year}{2018}\natexlab{}.
\newblock \showarticletitle{Interpretable to whom? A role-based model for
  analyzing interpretable machine learning systems}.
\newblock \bibinfo{journal}{\emph{arXiv preprint arXiv:1806.07552}}
  (\bibinfo{year}{2018}).
\newblock


\bibitem[\protect\citeauthoryear{Turilli and Floridi}{Turilli and
  Floridi}{2009}]%
        {turilli2009ethics}
\bibfield{author}{\bibinfo{person}{Matteo Turilli} {and}
  \bibinfo{person}{Luciano Floridi}.} \bibinfo{year}{2009}\natexlab{}.
\newblock \showarticletitle{The ethics of information transparency}.
\newblock \bibinfo{journal}{\emph{Ethics and Information Technology}}
  \bibinfo{volume}{11}, \bibinfo{number}{2} (\bibinfo{year}{2009}),
  \bibinfo{pages}{105--112}.
\newblock


\bibitem[\protect\citeauthoryear{Vermeulen, Vanderhulst, Luyten, and
  Coninx}{Vermeulen et~al\mbox{.}}{2010}]%
        {vermeulen2010pervasivecrystal}
\bibfield{author}{\bibinfo{person}{Jo Vermeulen}, \bibinfo{person}{Geert
  Vanderhulst}, \bibinfo{person}{Kris Luyten}, {and} \bibinfo{person}{Karin
  Coninx}.} \bibinfo{year}{2010}\natexlab{}.
\newblock \showarticletitle{PervasiveCrystal: Asking and answering why and why
  not questions about pervasive computing applications}. In
  \bibinfo{booktitle}{\emph{Intelligent Environments (IE), 2010 Sixth
  International Conference on}}. IEEE, \bibinfo{pages}{271--276}.
\newblock


\bibitem[\protect\citeauthoryear{Wachter, Mittelstadt, and Russell}{Wachter
  et~al\mbox{.}}{2017}]%
        {wachter2017counterfactual}
\bibfield{author}{\bibinfo{person}{Sandra Wachter}, \bibinfo{person}{Brent
  Mittelstadt}, {and} \bibinfo{person}{Chris Russell}.}
  \bibinfo{year}{2017}\natexlab{}.
\newblock \showarticletitle{Counterfactual explanations without opening the
  black box: Automated decisions and the GDPR}.
\newblock \bibinfo{journal}{\emph{Harvard Journal of Law \& Technology}}
  \bibinfo{volume}{31} (\bibinfo{year}{2017}), \bibinfo{pages}{841}.
\newblock


\bibitem[\protect\citeauthoryear{Wang, Yang, Abdul, and Lim}{Wang
  et~al\mbox{.}}{2019a}]%
        {Wang2019Theory}
\bibfield{author}{\bibinfo{person}{Danding Wang}, \bibinfo{person}{Qian Yang},
  \bibinfo{person}{Ashraf Abdul}, {and} \bibinfo{person}{Brian~Y. Lim}.}
  \bibinfo{year}{2019}\natexlab{a}.
\newblock \showarticletitle{Designing theory-driven user-centric explainable
  AI}. In \bibinfo{booktitle}{\emph{Proceedings of the 2019 CHI Conference on
  Human Factors in Computing Systems}} \emph{(\bibinfo{series}{CHI '19})}.
  \bibinfo{publisher}{ACM}, \bibinfo{address}{New York, NY, USA}, Article
  \bibinfo{articleno}{601}, \bibinfo{numpages}{15}~pages.
\newblock
\showISBNx{978-1-4503-5970-2}


\bibitem[\protect\citeauthoryear{Wang and Rudin}{Wang and Rudin}{2015}]%
        {wang2015falling}
\bibfield{author}{\bibinfo{person}{Fulton Wang} {and} \bibinfo{person}{Cynthia
  Rudin}.} \bibinfo{year}{2015}\natexlab{}.
\newblock \showarticletitle{Falling rule lists}. In
  \bibinfo{booktitle}{\emph{Artificial Intelligence and Statistics}}.
  \bibinfo{pages}{1013--1022}.
\newblock


\bibitem[\protect\citeauthoryear{Wang, Yuan, Chen, Su, Qu, and Liu}{Wang
  et~al\mbox{.}}{2019b}]%
        {wang2019visual}
\bibfield{author}{\bibinfo{person}{Qianwen Wang}, \bibinfo{person}{Jun Yuan},
  \bibinfo{person}{Shuxin Chen}, \bibinfo{person}{Hang Su},
  \bibinfo{person}{Huamin Qu}, {and} \bibinfo{person}{Shixia Liu}.}
  \bibinfo{year}{2019}\natexlab{b}.
\newblock \showarticletitle{Visual genealogy of deep neural networks}.
\newblock \bibinfo{journal}{\emph{IEEE Transactions on Visualization and
  Computer Graphics}} (\bibinfo{year}{2019}).
\newblock


\bibitem[\protect\citeauthoryear{Weld and Bansal}{Weld and Bansal}{2019}]%
        {Daniel2019challenge}
\bibfield{author}{\bibinfo{person}{Daniel~S. Weld} {and} \bibinfo{person}{Gagan
  Bansal}.} \bibinfo{year}{2019}\natexlab{}.
\newblock \showarticletitle{The challenge of crafting intelligible
  intelligence}.
\newblock \bibinfo{journal}{\emph{Commun. ACM}} \bibinfo{volume}{62},
  \bibinfo{number}{6} (\bibinfo{date}{May} \bibinfo{year}{2019}),
  \bibinfo{pages}{70–79}.
\newblock
\showISSN{0001-0782}


\bibitem[\protect\citeauthoryear{Weller}{Weller}{2017}]%
        {weller2017challenges}
\bibfield{author}{\bibinfo{person}{Adrian Weller}.}
  \bibinfo{year}{2017}\natexlab{}.
\newblock \showarticletitle{Challenges for transparency}.
\newblock \bibinfo{journal}{\emph{arXiv preprint arXiv:1708.01870}}
  (\bibinfo{year}{2017}).
\newblock


\bibitem[\protect\citeauthoryear{Wiegand, Schmidmaier, Weber, Liu, and
  Hussmann}{Wiegand et~al\mbox{.}}{2019}]%
        {wiegand2019drive}
\bibfield{author}{\bibinfo{person}{Gesa Wiegand}, \bibinfo{person}{Matthias
  Schmidmaier}, \bibinfo{person}{Thomas Weber}, \bibinfo{person}{Yuanting Liu},
  {and} \bibinfo{person}{Heinrich Hussmann}.} \bibinfo{year}{2019}\natexlab{}.
\newblock \showarticletitle{I drive-you trust: Explaining driving behavior of
  autonomous cars}. In \bibinfo{booktitle}{\emph{Extended Abstracts of the 2019
  CHI Conference on Human Factors in Computing Systems}}. ACM,
  \bibinfo{pages}{LBW0163}.
\newblock


\bibitem[\protect\citeauthoryear{Wise, Thomas, Pennock, Lantrip, Pottier,
  Schur, and Crow}{Wise et~al\mbox{.}}{1995}]%
        {wise1995visualizing}
\bibfield{author}{\bibinfo{person}{James~A Wise}, \bibinfo{person}{James~J
  Thomas}, \bibinfo{person}{Kelly Pennock}, \bibinfo{person}{David Lantrip},
  \bibinfo{person}{Marc Pottier}, \bibinfo{person}{Anne Schur}, {and}
  \bibinfo{person}{Vern Crow}.} \bibinfo{year}{1995}\natexlab{}.
\newblock \showarticletitle{Visualizing the non-visual: Spatial analysis and
  interaction with information from text documents}. In
  \bibinfo{booktitle}{\emph{Information Visualization, 1995. Proceedings.}}
  IEEE, \bibinfo{pages}{51--58}.
\newblock


\bibitem[\protect\citeauthoryear{Wongsuphasawat, Smilkov, Wexler, Wilson, Mane,
  Fritz, Krishnan, Vi{\'e}gas, and Wattenberg}{Wongsuphasawat
  et~al\mbox{.}}{2017}]%
        {wongsuphasawat2017visualizing}
\bibfield{author}{\bibinfo{person}{Kanit Wongsuphasawat},
  \bibinfo{person}{Daniel Smilkov}, \bibinfo{person}{James Wexler},
  \bibinfo{person}{Jimbo Wilson}, \bibinfo{person}{Dandelion Mane},
  \bibinfo{person}{Doug Fritz}, \bibinfo{person}{Dilip Krishnan},
  \bibinfo{person}{Fernanda~B Vi{\'e}gas}, {and} \bibinfo{person}{Martin
  Wattenberg}.} \bibinfo{year}{2017}\natexlab{}.
\newblock \showarticletitle{Visualizing dataflow graphs of deep learning models
  in tensorflow}.
\newblock \bibinfo{journal}{\emph{IEEE Transactions on Visualization and
  Computer Graphics}} \bibinfo{volume}{24}, \bibinfo{number}{1}
  (\bibinfo{year}{2017}), \bibinfo{pages}{1--12}.
\newblock


\bibitem[\protect\citeauthoryear{Woolley}{Woolley}{2016}]%
        {woolley2016automating}
\bibfield{author}{\bibinfo{person}{Samuel~C Woolley}.}
  \bibinfo{year}{2016}\natexlab{}.
\newblock \showarticletitle{Automating power: Social bot interference in global
  politics}.
\newblock \bibinfo{journal}{\emph{First Monday}} \bibinfo{volume}{21},
  \bibinfo{number}{4} (\bibinfo{year}{2016}).
\newblock


\bibitem[\protect\citeauthoryear{Wu, Hughes, Parbhoo, Zazzi, Roth, and
  Doshi-Velez}{Wu et~al\mbox{.}}{2018}]%
        {wu2018beyond}
\bibfield{author}{\bibinfo{person}{Mike Wu}, \bibinfo{person}{Michael~C
  Hughes}, \bibinfo{person}{Sonali Parbhoo}, \bibinfo{person}{Maurizio Zazzi},
  \bibinfo{person}{Volker Roth}, {and} \bibinfo{person}{Finale Doshi-Velez}.}
  \bibinfo{year}{2018}\natexlab{}.
\newblock \showarticletitle{Beyond sparsity: Tree regularization of deep models
  for interpretability}. In \bibinfo{booktitle}{\emph{Thirty-Second AAAI
  Conference on Artificial Intelligence}}.
\newblock


\bibitem[\protect\citeauthoryear{Yin, Wortman~Vaughan, and Wallach}{Yin
  et~al\mbox{.}}{2019}]%
        {yin2019understanding}
\bibfield{author}{\bibinfo{person}{Ming Yin}, \bibinfo{person}{Jennifer
  Wortman~Vaughan}, {and} \bibinfo{person}{Hanna Wallach}.}
  \bibinfo{year}{2019}\natexlab{}.
\newblock \showarticletitle{Understanding the effect of accuracy on trust in
  machine learning models}. In \bibinfo{booktitle}{\emph{Proceedings of the
  2019 CHI Conference on Human Factors in Computing Systems}}.
  \bibinfo{pages}{1--12}.
\newblock


\bibitem[\protect\citeauthoryear{Yosinski, Clune, Nguyen, Fuchs, and
  Lipson}{Yosinski et~al\mbox{.}}{2015}]%
        {yosinski2015understanding}
\bibfield{author}{\bibinfo{person}{Jason Yosinski}, \bibinfo{person}{Jeff
  Clune}, \bibinfo{person}{Anh Nguyen}, \bibinfo{person}{Thomas Fuchs}, {and}
  \bibinfo{person}{Hod Lipson}.} \bibinfo{year}{2015}\natexlab{}.
\newblock \showarticletitle{Understanding neural networks through deep
  visualization}.
\newblock \bibinfo{journal}{\emph{In ICML Deep Learning Workshop 2015}}
  (\bibinfo{year}{2015}).
\newblock


\bibitem[\protect\citeauthoryear{Yu and Shi}{Yu and Shi}{2018}]%
        {yu2018user}
\bibfield{author}{\bibinfo{person}{Rulei Yu} {and} \bibinfo{person}{Lei Shi}.}
  \bibinfo{year}{2018}\natexlab{}.
\newblock \showarticletitle{A user-based taxonomy for deep learning
  visualization}.
\newblock \bibinfo{journal}{\emph{Visual Informatics}} \bibinfo{volume}{2},
  \bibinfo{number}{3} (\bibinfo{year}{2018}), \bibinfo{pages}{147--154}.
\newblock


\bibitem[\protect\citeauthoryear{Zahavy, Ben-Zrihem, and Mannor}{Zahavy
  et~al\mbox{.}}{2016}]%
        {zahavy2016graying}
\bibfield{author}{\bibinfo{person}{Tom Zahavy}, \bibinfo{person}{Nir
  Ben-Zrihem}, {and} \bibinfo{person}{Shie Mannor}.}
  \bibinfo{year}{2016}\natexlab{}.
\newblock \showarticletitle{Graying the black box: Understanding DQNs}. In
  \bibinfo{booktitle}{\emph{International Conference on Machine Learning}}.
  \bibinfo{pages}{1899--1908}.
\newblock


\bibitem[\protect\citeauthoryear{Zarsky}{Zarsky}{2016}]%
        {zarsky2016trouble}
\bibfield{author}{\bibinfo{person}{Tal Zarsky}.}
  \bibinfo{year}{2016}\natexlab{}.
\newblock \showarticletitle{The trouble with algorithmic decisions: An analytic
  road map to examine efficiency and fairness in automated and opaque decision
  making}.
\newblock \bibinfo{journal}{\emph{Science, Technology, \& Human Values}}
  \bibinfo{volume}{41}, \bibinfo{number}{1} (\bibinfo{year}{2016}),
  \bibinfo{pages}{118--132}.
\newblock


\bibitem[\protect\citeauthoryear{Zeiler and Fergus}{Zeiler and Fergus}{2014}]%
        {zeiler2014visualizing}
\bibfield{author}{\bibinfo{person}{Matthew~D Zeiler} {and} \bibinfo{person}{Rob
  Fergus}.} \bibinfo{year}{2014}\natexlab{}.
\newblock \showarticletitle{Visualizing and understanding convolutional
  networks}. In \bibinfo{booktitle}{\emph{European Conference on Computer
  Vision}}. Springer, \bibinfo{pages}{818--833}.
\newblock


\bibitem[\protect\citeauthoryear{Zhang, Wang, and Zhu}{Zhang
  et~al\mbox{.}}{2018}]%
        {zhang2018examining}
\bibfield{author}{\bibinfo{person}{Quanshi Zhang}, \bibinfo{person}{Wenguan
  Wang}, {and} \bibinfo{person}{Song-Chun Zhu}.}
  \bibinfo{year}{2018}\natexlab{}.
\newblock \showarticletitle{Examining cnn representations with respect to
  dataset bias}. In \bibinfo{booktitle}{\emph{Thirty-Second AAAI Conference on
  Artificial Intelligence}}.
\newblock


\bibitem[\protect\citeauthoryear{Zhang and Zhu}{Zhang and Zhu}{2018}]%
        {zhang2018visual}
\bibfield{author}{\bibinfo{person}{Quan-shi Zhang} {and}
  \bibinfo{person}{Song-Chun Zhu}.} \bibinfo{year}{2018}\natexlab{}.
\newblock \showarticletitle{Visual interpretability for deep learning: a
  survey}.
\newblock \bibinfo{journal}{\emph{Frontiers of Information Technology \&
  Electronic Engineering}} \bibinfo{volume}{19}, \bibinfo{number}{1}
  (\bibinfo{year}{2018}), \bibinfo{pages}{27--39}.
\newblock


\bibitem[\protect\citeauthoryear{Zhang, Liao, and Bellamy}{Zhang
  et~al\mbox{.}}{2020}]%
        {zhang2020effect}
\bibfield{author}{\bibinfo{person}{Yunfeng Zhang}, \bibinfo{person}{Q.~Vera
  Liao}, {and} \bibinfo{person}{Rachel K.~E. Bellamy}.}
  \bibinfo{year}{2020}\natexlab{}.
\newblock \showarticletitle{Effect of confidence and explanation on accuracy
  and trust calibration in AI-assisted decision making}. In
  \bibinfo{booktitle}{\emph{Proceedings of the 2020 Conference on Fairness,
  Accountability, and Transparency}} \emph{(\bibinfo{series}{FAT* '20})}.
\newblock


\bibitem[\protect\citeauthoryear{Zhang, Singh, Gadiraju, and Anand}{Zhang
  et~al\mbox{.}}{2019}]%
        {zhang2019dissonance}
\bibfield{author}{\bibinfo{person}{Zijian Zhang}, \bibinfo{person}{Jaspreet
  Singh}, \bibinfo{person}{Ujwal Gadiraju}, {and} \bibinfo{person}{Avishek
  Anand}.} \bibinfo{year}{2019}\natexlab{}.
\newblock \showarticletitle{Dissonance between human and machine
  understanding}.
\newblock \bibinfo{journal}{\emph{Proceedings of the ACM on Human-Computer
  Interaction}} \bibinfo{volume}{3}, \bibinfo{number}{CSCW}
  (\bibinfo{year}{2019}), \bibinfo{pages}{56}.
\newblock


\bibitem[\protect\citeauthoryear{Zhong, Xie, Zhong, Wang, Xu, Cheng, and
  Mueller}{Zhong et~al\mbox{.}}{2017}]%
        {zhong2017evolutionary}
\bibfield{author}{\bibinfo{person}{Wen Zhong}, \bibinfo{person}{Cong Xie},
  \bibinfo{person}{Yuan Zhong}, \bibinfo{person}{Yang Wang},
  \bibinfo{person}{Wei Xu}, \bibinfo{person}{Shenghui Cheng}, {and}
  \bibinfo{person}{Klaus Mueller}.} \bibinfo{year}{2017}\natexlab{}.
\newblock \showarticletitle{Evolutionary visual analysis of deep neural
  networks}. In \bibinfo{booktitle}{\emph{ICML Workshop on Visualization for
  Deep Learning}}.
\newblock


\bibitem[\protect\citeauthoryear{Zhu, Liapis, Risi, Bidarra, and
  Youngblood}{Zhu et~al\mbox{.}}{2018}]%
        {zhu2018explainable}
\bibfield{author}{\bibinfo{person}{Jichen Zhu}, \bibinfo{person}{Antonios
  Liapis}, \bibinfo{person}{Sebastian Risi}, \bibinfo{person}{Rafael Bidarra},
  {and} \bibinfo{person}{G~Michael Youngblood}.}
  \bibinfo{year}{2018}\natexlab{}.
\newblock \showarticletitle{Explainable AI for designers: A human-centered
  perspective on mixed-initiative co-creation}. In
  \bibinfo{booktitle}{\emph{2018 IEEE Conference on Computational Intelligence
  and Games (CIG)}}. IEEE, \bibinfo{pages}{1--8}.
\newblock


\bibitem[\protect\citeauthoryear{Zintgraf, Cohen, Adel, and Welling}{Zintgraf
  et~al\mbox{.}}{2017}]%
        {zintgraf2017visualizing}
\bibfield{author}{\bibinfo{person}{Luisa~M Zintgraf}, \bibinfo{person}{Taco~S
  Cohen}, \bibinfo{person}{Tameem Adel}, {and} \bibinfo{person}{Max Welling}.}
  \bibinfo{year}{2017}\natexlab{}.
\newblock \showarticletitle{Visualizing deep neural network decisions:
  Prediction difference analysis}.
\newblock \bibinfo{journal}{\emph{arXiv preprint arXiv:1702.04595}}
  (\bibinfo{year}{2017}).
\newblock


\end{thebibliography}

%
\appendix

\end{document}